

\input jytex.tex

\font\tenBbb=msym10
\font\twelveBbb=msym10 scaled \magstep1
\font\nineBbb=msym9
\font\sevenBbb=msym7
\newfam\Bbbfam
\textfont\Bbbfam=\twelveBbb
\scriptfont\Bbbfam=\nineBbb
\scriptscriptfont\Bbbfam=\sevenBbb
\def\Bbb{\fam\Bbbfam\twelveBbb}
  \def\C{{\Bbb C}}

 \def\R{{\Bbb R}}  
   
 \def\Z{{\Bbb Z}}

\baselinestretch=1500
\pagenumstyle{roman}
\footnotenumstyle{arabic}
\footnotenum= 0
\head={\hfil\normalfonts\numstyle\pagenum\hfil}
\foot={\hfil}
\topmargin= 1truein\vsize=9truein
\leftmargin=1.5truein\hsize=6truein


%
%
\def\h{\hbox to .5cm{\hfill}}
\def\hof{\hbox to .15cm{\hfill}}
\def\hi{\hbox to .2cm{\hfill}}
\def\htf{\hbox to .35cm{\hfill}}
\def\hso{\hskip 1cm}
\def\hsf{\hskip .7cm}
\def\ha#1{\n\hbox to .6cm{\n {#1}\hfill}}
\def\hb#1{\indent\hbox to .7cm{\n {#1}\hfill}}
\def\hc#1{\indent\hbox to .7cm{\hfill}\hbox to 1.1cm{\n {#1}\hfill}}
\def\hd#1{\indent\hbox to 1.8cm{\hfill}\hbox to 1.4cm{\n {#1}\hfill}}
\def\vone{\vskip 1cm}
\def\vhalf{\vskip .5cm}
\def\eq#1{\eqno\eqnlabel{#1}}

\def\mpr#1{\markup{[\putref{#1}]}}
\def\pr#1{[\putref{#1}]}
\def\pe#1{(\puteqn{#1})}
\def\n{\noindent}
\def\no{\noindent}
\def\ref{\reference}

\def\subon{\subequationnumstyle{alphabetic}}
\def\suboff{\subequationnumstyle{blank}}
\def\undbib{\underbar {\hbox to 2cm{\hfill}}, }
%
%

\def\al{\alpha}
\def\balpha{{\bmit \alpha}}
\def\bal{{\bmit \alpha}}
\def\bbe{{\bmit \beta}}

\def\bA{{{\bmit A}^{\rm KM}}}
\def\cc#1#2{c^{#1}_{#2}}
\def\dag{ \dagger }
\def\dc{\tilde h}

\def\desc{:\phi^1_0 (z)\phi^1_0 (z):}

\def\e{{\rm e}}
\def\ep{\epsilon}

\def\Om{\Omega_{\delta}}
\def\ra{\rightarrow  }

\def\lal{{\cal L}}
\def\kml{{\tilde {\cal L}}}

\def\t{\vartheta}
\def\theta{\vartheta}
\def\bt{\bar\vartheta}
\def\Kt{\tilde K}
%
%
\def\half{{1\over 2}}

%
%

\def\ie{{\it i.e.}}
\def\KM{KM }
\def\KMA{KM algebra }
\def\KMAS{KM algebras }

\def\LAS{Lie algebras }

%
%


{\bigfonts
\pagenumstyle{blank}
\vskip 2.5 truecm
\centertext{\bf KA\v C-MOODY ALGEBRAS and STRING THEORY}
\vskip 6 truecm
\centertext{Dissertation by\\ \bf Gerald B. Cleaver}
\vskip 6 truecm
\centertext{In Partial Fulfillment of the Requirements\\
for the Degree of\\
Doctor of Philosophy}
\vskip 2 truecm
\centertext{California Institute of Technology\\
Pasadena, California}
\vskip .4truecm
\centertext{1993}
\centertext{(Submitted May 14, 1993)}
\hfill\vfill\eject}
\pagenum=1
\pagenumstyle{roman}
\hfill\vfill
\centertext{$\copyright$ 1993\\
Gerald Bryan Cleaver\\
All Rights Reserved}
\newpage
\centertext{\bf\bigfonts Acknowledgements}
\vskip 2 truecm
I want to thank my fiancee, Lisa, for her understanding and support
these past months while this dissertation was written.  I also
thank my family for their support and encouragement these many years.
My good friend and coauthor of two papers, Phil Rosenthal, deserves
special recognition
for his contributions to our discussions and research
in both fractional superstrings and stringy cosmology.
The others in my advisor's research group,
Dimitris Kallifatides, Christoph Schmidhuber, and Malik Rakhmanov,
additionally contributed much to the stimulating and enjoyable
research atmosphere at
Caltech, as did all the other graduate students, postdoc's and senior
faculty on the fourth floor.
I also thank David Lewellen for spurring my interest
in higher level \KM  algebras with his 1989 paper,\mpr{lewellen}
and for the opportunity of working together on the paper\mpr{cleaver93a}
that forms much of chapter three of this thesis.
I also want to recognize those people outside
of Caltech, with whom discussions, on issues relevant to this thesis,
were very rewarding.
This list includes David Lewellen, Henry Tye, Phil Argyres, and
Keith Dienes. Several others merit thanks for discussions on
various issues of string cosmology:
Mark Bowick, Joe Polchinski, Jaques Distler, Chris Pope,
and Cumrum Vafa.
My advisor, John H. Schwarz,
deserves special thanks for his guidance of my research.
I thank him for the opportunity to carry on
an apprenticeship into elementary particle physics and string theory
under his supervision.
Lastly, I want to recognize Helen Tuck for the help she so kindly
gives in her role as secretary of the  particle theory group.

\hfill\vfill\eject
\hfill\\
\vskip 6.0 cm
\centertext{\bigfonts To Lisa}
\hfill\vfill\eject

\hfill\\
\vskip 5.0cm
\n With regard to strings:
\vskip 1.0cm
\begin{narrow}[.75in]
For a theory may be formed by the mind, and
therefore be shaped by its structures. However, it frequently happens that
the descriptive power of a good theory so transcends the meager
understanding of the original conception, that is is very hard to believe
that the theory only reflects our cognitive structures.  It would then
appear to be in much closer contact with nature than with our minds; our
minds are continually agasped at its applicability.\\
\vskip .2cm
\hfill --Albert Einstein
\end{narrow}
\hfill\vfill\eject

\chapternumstyle{blank}
\n\chapter{\bf Abstract}

The focus of this thesis
is on (1) the role of Ka\v c-Moody algebras in string theory and the
development of techniques for
systematically building  string theory models based on higher
level ($K\geq 2$) \KMAS and (2) fractional superstrings, a new class of
solutions based on $SU(2)_K/U(1)$ conformal field theories. The content of
this thesis is as follows.

In chapter two we review \KMAS and their role in string theory.
In the next chapter,
we present two results concerning the construction of modular invariant
partition functions for conformal field theories built by tensoring
together other conformal field theories.
This is based upon our research in ref.~\pr{cleaver93a}.
First we show how the possible modular invariants for the tensor product
theory are constrained if the allowed modular
invariants of the individual conformal field theory factors have been
classified. We illustrate the use of these constraints for theories of
the type $SU(2)_{K_A}\otimes SU(2)_{K_B}$,
finding all consistent theories for $K_A$ and $K_B$ odd. Second we show how
known diagonal
modular invariants can be used to construct inherently asymmetric invariants
where the holomorphic and anti-holomorphic theories do not share the
same chiral algebra. Explicit examples are given.

Next, in chapter four
we investigate some issues relating to recently
proposed fractional superstring theories with $D_{\rm critical}<10$.
Using the factorization
approach of Gepner and Qiu, we systematically rederive
the partition functions of the $K=4,\, 8,$ and $16$ theories
and examine their spacetime
supersymmetry. Generalized GSO projection operators for the $K=4$ model
are found. Uniqueness of the twist field, $\phi^{K/4}_{K/4}$,
as source of spacetime fermions, is demonstrated. Our research
was originally presented in
refs.~[\putref{cleaver92a}, \putref{cleaver93b}]
\hfill\vfill\eject

{\parindent=.6cm
\n{\bf\bigfonts Table of Contents}\hfill\\
\vskip .5cm
\no
{\bf Acknowledgements}\quad\dotfill\quad iii\\
{\bf Dedication}\quad\dotfill\quad iv\\
{\bf Quotation}\quad\dotfill\quad v\\
{\bf Abstract}\quad\dotfill\quad vi\\
{\bf Table of Contents}\quad\dotfill\quad vii\\
{\bf List of Figures and Tables}\quad\dotfill\quad ix\\
{\ha{\bf 1.}{\bf Introduction}\quad\dotfill\quad 2\\}
{\hb{1.1}Reasons for String Theory Research\quad\dotfill\quad 2\\}
{\hb{1.2}Status of String Theory Phenomenology\quad\dotfill\quad 3\\}
{\hc{1.2.a} Phenomenological Restrictions\quad\dotfill\quad 7\\}
{\ha{\bf 2.}{\bf Ka\v c-Moody Algebras and Superstring
Theory}\quad\dotfill\quad 9\\}
{\hb{2.1}Review of Ka\v c-Moody Algebras\quad\dotfill\quad 9\\}
{\hc{2.1.a}Categories of Ka\v c-Moody Algebras\quad\dotfill\quad 12\\}
{\hc{2.1.b}Affine Algebras\quad\dotfill\quad 15\\}
{\hb{2.2}Application to String Theory\quad\dotfill\quad 18\\}
{\ha{\bf 3.}{\bf Modular Invariant Partition Functions}\quad\dotfill\quad 24\\}
{\hb{3.1}Review of Modular Invariance\quad\dotfill 24\\}
{\hb{3.2}Complications for Models Based on General Ka\v c-Moody
Models\quad\dotfill\quad 31\\}
{\hb{3.3}Constraints on Tensor Product Modular Invariants\quad\dotfill\quad
36\\}
{\hc{3.3.a} Example: SU(2)$_{K_A}\otimes$SU(2)$_{K_B}$ Tensor Product
Theories\quad\dotfill\quad 38\\}
{\hb{3.4}Left-Right Asymmetric Modular Invariants\quad\dotfill\quad 42\\}
{\hb{3.5}Concluding Comments on MIPFs\quad\dotfill\quad 46\\}
{\ha{\bf 4.}{\bf Fractional Superstrings}\quad\dotfill\quad 47\\}
{\hb{4.1}Introduction to Fractional Superstrings\quad\dotfill\quad 47\\}
{\hc{4.1.a}Parafermion Characters\quad\dotfill\quad 49\\}
{\hb{4.2}Fractional Superstring Partition Functions\quad\dotfill\quad 51\\}
{\hc{4.2.a}New Derivation of the Partition Functions\quad\dotfill\quad 54\\}
{\hc{4.2.b}Affine Factor and ``W'' Partition Function\quad\dotfill\quad 57\\}
{\hc{4.2.c}Theta-Function Factor and the ``V'' Partition
Function\quad\dotfill\quad 60\\}
{\hb{4.3}Beyond the Partition Function: Additional
Comments\quad\dotfill\quad 69\\}
{\hc{4.3.a}Bosonization of the $K=4$ Theory\quad\dotfill\quad 69\\}
{\hc{4.3.b}Generalized Commutation Relations and the GSO
Projection\quad\dotfill\quad 75\\}
{\hc{4.3.c}Unique Role of Twist Field
$(\phi^{K/4}_{K/4})^{D-2}$\quad\dotfill\quad 85\\}
{\hb{4.4}Concluding Discussion\quad\dotfill\quad 92\\}
{{\bf Appendix A:}  Dynken Diagrams for Lie Algebras
and Ka\v c-Moody Algebras\quad\dotfill\quad 94\\}
\n {{\bf Appendix B:} Proof that Completeness of the A-D-E Classification
of\hfill\\}
\n{\phantom{\bf Appendix B:} Modular Invariant Partition Functions
for $SU(2)_K$ is Unrelated\hfill\\}
\n{\phantom{\bf Appendix B:} to Uniqueness of the
Vacuum\quad\dotfill\quad 97\\}
{{\bf References}\quad\dotfill\quad 106\\}}
\hfill\vfill\eject

{\parindent=.6cm
\n{\bf\bigfonts List of Figures and Tables}\hfill\\
\vskip .5cm
\no
\hbox to 2.1cm{Figure 3.1\hfill}Two Conformally
Inequivalent Tori\quad\dotfill\quad 28\\
\hbox to 2.1cm{Figure 3.2\hfill}Lattice Representation
of a Two-Dimensional Torus Defined by\hfill\\
\hbox to 2.1cm{\hfill}Complex Number $\tau$\quad\dotfill\quad 28\\
\hbox to 2.1cm{Figure 3.3\hfill}Lattice Representation of a
Two-Dimensional Torus Defined by\hfill\\
\hbox to 2.1cm{\hfill}Complex Numbers $\lambda_1$
and $\lambda_2$\quad\dotfill\quad 28\\
\hbox to 2.1cm{Figure 3.4\hfill}The Two Independent Cycles on a
Torus\quad\dotfill\quad 29\\
\hbox to 2.1cm{Figure 3.5\hfill}Transformation of $\tau$ from Dehn Twist
around the $a$ Cycle\quad\dotfill\quad 29\\
\hbox to 2.1cm{Figure 3.6\hfill}Transformation of $\tau$ from Dehn Twist
around the $b$ Cycle\quad\dotfill\quad 29\\
\hbox to 2.1cm{Figure 3.7\hfill}Fundamental Domain $\cal F$ in Moduli Space
and Its Images under $S$\hfill\\
\hbox to 2.1cm{\hfill}and $T$\quad\dotfill\quad 30\\
\hbox to 2.1cm{Figure 4.1\hfill}Supersymmetry of Lowest Mass States of
Fractional Open String\quad\dotfill\quad 41\\
\hbox to 2.1cm{Figure A.1\hfill}Generalized Dynkin Diagrams of the Finite
KM Algebras\quad\dotfill\quad 94\\
\hbox to 2.1cm{Figure A.2\hfill}Generalized Dynkin Diagrams of the
Untwisted Affine KM\hfill\\
\hbox to 2.1cm{\hfill}Algebras\quad\dotfill\quad 95\\
\hbox to 2.1cm{Figure A.3\hfill}Generalized Dynkin Diagrams of the Twisted
Affine KM\hfill\\
\hbox to 2.1cm{\hfill}Algebras\quad\dotfill\quad 96\\
\hbox to 2.1cm{Figure B.1\hfill}The Integers (mod $N$)
Mapped to a Circle of Radius $N/2\pi$}\quad\dotfill\quad 98\\
\hbox to 2.1cm{\hfill}\hfill\\
\n\hbox to 1.9cm{Table 4.1\hfill}$\Z_4$ Primary Fields\quad\dotfill\quad 70\\
\hbox to 1.9cm{Table 4.2\hfill}Primary Field Representation from Orbifold
Bosonization\quad\dotfill\quad 72\\
\hbox to 1.9cm{Table 4.3\hfill}Primary Field Representation from $R={\sqrt 6}$
Bosonization\quad\dotfill\quad 74\\
\hbox to 1.9cm{Table 4.4\hfill}Masses of $K=4$ Highest Weight
States\quad\dotfill\quad 83\\
\hbox to 1.9cm{Table 4.5\hfill}Mass Sectors as Function of $\Z_3$
Charge\quad\dotfill\quad 84\\
\hbox to 1.9cm{Table 4.6\hfill}Fields with $\phi^{j_1}_{m_1}\neq
\phi^1_0$ with Conformal Dimensions in Integer Ratio\hfill\\
\hbox to 1.9cm{\hfill}with
$h(\phi^{K/4}_{K/4})$\quad\dotfill\quad 88\\
\hbox to 1.9cm{Table 4.7\hfill}Potential Alternatives, $\phi^{j_3}_{\pm j_3}$,
to $\phi^{K/4}_{K/4}$ for Spin Fields\quad\dotfill\quad 90\\
\hbox to 1.9cm{Table B.1\hfill}A--D--E Classification in Terms of
$\Omega_{\delta}$ Basis Set\quad\dotfill\quad 98
\hfill\vfill\eject

\n Figures 3.1-7 are from ref.~\pr{lust89}.

\n Figures A.1-3 are from ref.~\pr{cornwell89}.

\hfill\vfill\eject


\pagenum=0\pagenumstyle{arabic}

\pagenumstyle{arabic}\pagenum=0
\hbox to 1in{\hfill}
\vskip 3in
\centertext{\bf Ka\v c-Moody Algebras and String Theory}
\hfill\vfill\eject

\chapternumstyle{blank}
\n {\bf Chapter 1: Introduction}\vskip .8cm
{\hb{1.1}{\bfs Reasons for String Theory Research}}\vhalf
\chapternumstyle{arabic}\sectionnumstyle{arabic}
\chapternum=1\sectionnum=1\equationnum=0

Elementary particle physics has achieved phenomenal success in recent
decades, resulting in the Standard Model (SM),
$SU(3)_C\times SU(2)_L\times U(1)_Y$,
and the verification, to high
precision, of many of its predictions. However, there are still several
shortcomings or unsatisfying aspects of the theory. Consider, for example,
the following:\mpr{langacker92}\vone
\item {1.} The SM is very complicated, requiring measurement of some 21 free
parameters, such as the masses the quarks and leptons
and the coupling constants.  We should expect the true fundamental theory to
have at most one free parameter.
\item {2.} The SM has a complicated gauge structure.  A gauge group
that is the direct product of three gauge groups with independent couplings
does not seem fundamental.
\item {3.} There seems to be a naturalness problem concerning
 the scale at which the
electroweak (EW) symmetry, $SU(2)_L\times U(1)_Y$, breaks to the
electromagnetic
$U(1)_{EM}$.  Why is this scale of 100 GeV so much smaller than the Planck
scale of $10^{19}$ GeV?  Although this is ``explained'' by the scale of the
Higgs mass, fine-tuning is required in renormalization theory to keep
the Higgs mass on the order of the symmetry breaking scale.  This seems to
suggest the need for supersymmetry at a higher scale.
\item{4.} Fine-tuning is also required to solve the strong CP problem.
\item{5.} The SM provides no unification with gravity, {\ie}, no means of
forming a consistent theory of quantum gravity.
\item{6.} The cosmological constant resulting from EW symmetry breaking
should be approximately 50 orders of magnitude higher than the experimental
limit. Solving this problem from the SM viewpoint again requires a
fine-tuned
cancellation.
\vone

These shortcomings have
motivated a search for phenomenologically
viable
Grand Unified Theories (GUT's) that would unify
SM physics through a single force and even for a Theory of Everything
(TOE) that
could consistently combine the SM with gravity.  In the last decade,
this pursuit has resulted in an intensive study of string theory, which
 involves only one truly
elementary ``particle,'' a (closed) string-like (rather than point-like)
object with a length on the order of the Planck scale,
$l_{\rm Pl}=10^{-33}$cm.  In this theory all particles ordinarily
regarded as
``elementary'' are explained as vibrational or
internal modes of this fundamental string.
One of the advantages of string theory
is that it removes the infinities resulting from high-energy
interactions of point-like
particles, without requiring renormalization techniques.  The supersymmetric
version of string theory contains no ultraviolet divergences.

String theory is the first theory to successfully
combine the SM forces with gravity.
Any string theory with $D>2$ contains an infinite tower of
vibrational/internal
excitation modes.  Included in the closed (super)string spectrum is a massless
spin-2 (and spin-3/2) state which can be identified with the graviton
(and its supersymmetric partner, the gravitino).

\sectionnumstyle{blank}\vhalf
{\hb{1.2}{\bfs Status of String Theory Phenomenology}}\vhalf
\sectionnumstyle{arabic}\sectionnum=2\equationnum=0

In a sense string theory has been too successful following the explosion
of interest in the mid-80's.  The (super)string
theory is inherently a (10) 26 dimensional spacetime theory. Although
in both cases there
are only a very few solutions for the theories when all
spacetime dimensions are uncompactified, for each dimension that is
compactified, there arise many more possible solutions.
With only four uncompactified
spacetime dimensions, there is a plethora (on the order of several
million) of distinct solutions to the superstring theory.
Many different approaches to ``compactification,'' {\it e.g.},
bosonic lattices and orbifolds, free fermions, Calabi-Yau manifolds,
and $N=2$ minimal models, have been devised.
(Often there is, however, much overlap and
sometimes even complete equivalence between alternative methods of
compactification.)  Four-dimensional solutions
can be classified into two broad categories: (1) those solutions
involving an actual geometrical
compactification from ten uncompactified dimensions,
and (2) those with
internal degrees of freedom having no equivalent representation
in terms of six well-defined compactified dimensions.

There is a
potential problem with solutions in the first class: such models with
$N=1$ spacetime supersymmetry (SUSY) and/or chiral fermions cannot
contain massless spacetime scalar fields in the adjoint or higher dimensional
representations of the gauge group.\mpr{lewellen,font90,ellis90}
This presents a possible difficulty for string theory, because typical
GUT's  depend  upon scalars in these representations to break the
gauge symmetry down to the  SM. In the usual approach,
spontaneous symmetry breaking is brought about
by vacuum expectation values (VEV's) of these scalars.
Thus, the gauge groups of these string models either must break to
the standard model near the string (Planck) scale or a non-standard
Higgs breaking is required.  An example of the first method is symmetry
breaking by Wilson lines in Calabi-Yau vacua.\mpr{candelas85}
Flipped $SU(5)$ is the primary example of the second
approach.\mpr{antoniadis87b}
However, standard GUT's such as $SU(5)$ or $SO(10)$
are excluded from this class of string theory models.

In the first class of models,
the absence of spacetime scalars in higher representations results
from the association of geometrical compactification with level-one
\KM algebras.
In other words,
the connection of these models to level-one \KMAS is basically a byproduct of
the classical idea of ``compactification.''
Because of this,
basing a model on level-one \KMAS has been
the standard approach to string theory phenomenology.
Starting from
either the ten dimensional type-II or heterotic superstrings,
four-dimensional spacetime has most often been derived through
``spontaneous compactification'' of the extra six dimensions.
In ten uncompactified
dimensions the only modular invariant heterotic
string models with spacetime SUSY and gauge symmetry are the
level-one $E_8\otimes E_8$ and level-one $SO(32)$ solutions.
(In ten uncompactified dimensions, the type-II string has $N=2$ SUSY,
but no gauge group.)
Compactification of the extra six dimensions
on a Calabi-Yau manifold or symmetric orbifold,
naturally keeps
the \KMA at level-one. The resulting gauge group
$g$, is a subgroup of either $E_8\otimes E_8$ or $SO(32)$,
and the representations of the gauge group that appear are
determined by the level of the algebra.
Models using bosonic lattice compactification, or equivalently
complex world sheet fermions,\mpr{kawai87a,antoniadis87,antoniadis88}
likewise have level-one
\KM algebras, with the associated gauge group being a subgroup of
either $SO(12)\otimes E_8\otimes E_8$ or $SO(44)$.

Models can be based on higher-level \KM algebras, if the demand for
a classical interpretation of compactification is relaxed.
Such models fall into the second general
class of string solutions and can contain scalars in the adjoint or higher
representations.
These states can exist in the spectrum if
their gauge group arises from a level-$K\geq 2$ \KMA on
the world sheet.
Examples are given in \pr{lewellen}, where the approach to
such models is via {\it real} fermions.

The unitary representations of a level-$K$ \KMA in a string model
are required to satisfy (see section 2.2.)
$$ K\geq \sum_{i=1}^{{\rm rank}\, \lal}
n_i m_i\,\, , \eqno\eqnlabel{unitaryrep}$$
where $n_i$ are the Dynkin labels of the highest weight representation of the
associated Lie algebra, $\lal$,
and $m_i$ are the related co-marks.
Based on this unitarity constraint, at level-one only the singlet, spinor,
conjugate spinor and vector representations of $SO(4n+2)$ can
appear. For $SU(N)$ level-one, only
the ${N\choose 0}$ ({\it i.e.,} the singlet),
${N\choose 1}$, ${N\choose 2}$, $\dots$, ${N\choose N-1}$
representations are allowed,\footnote{For $SU(N)$ at level-$K$,
the rule for determining all unitary representations
is the following:
Only those representations that correspond to a Young tableau
with $K$ or fewer columns are allowed. Henceforth
a level-K \KM algebra, $\kml$, based on a Lie algebra, $\lal$, will often be
denoted by ${\lal}_K$, with the exception of those Lie algebras that already
carry a subscript denoting the rank, {\it e.g.,} $E_6$.}
while for $E_6$ level-one the $1$, $27$ and $\overline{27}$
representations can be present.

Until geometric ``compactification'' from ten dimensions is sacrificed,
higher-level models cannot be reached.
However, when the basic strategy is generalized,
level-one models become much less special.
If one starts with all spatial dimensions initially
compactified, and not well defined spatially,
the occurrence of level-one \KMAS is not necessarily favored.
That is, after ``decompactification'' of three spatial dimensions,
a gauge group in the $(3+1)$-dimensional
space based on higher level algebras
becomes possible (and not unlikely).

The difference between the two classes of ``four-dimensional'' string
models relates
to the question of  how valid it is to think geometrically about
physics at the Planck scale.
Are the lattice, free fermion, and Calabi-Yau approaches
to compactification
too classical for Planck scale physics?
Going beyond the classical notion of spatial dimensions
was one reason that Gepner considered $N=2$ minimal
models,\mpr{gepner87b} (even though
Calabi-Yau manifolds and $N=2$
minimal models were eventually found to be
equivalent\mpr{gepner87b}).\footnote{We suggest that
this indicates more than just the
mathematical equivalences of the appraoches
as demonstrated through Landau-Ginsburg potentials.
It is another example that makes us question the meaningfulness of the
concept of well-defined compactified
spatial dimensions of Plank scale length.}
Many considerations suggest investigating  string models
based on higher level K-M algebras, even though the
degrees of freedom of the models generally cannot be
expressed in terms of compactified spatial dimensions.

The central charge of the
level-$K$ algebra
(which measures the contribution to the conformal anomaly of the world
sheet theory) is
$$c_{\kml}= {K\, {\rm dim}\, \lal\over K+ \half C_A} \,\, ,
\eqno\eqnlabel{centralcharge}$$
 where $C_A$ is the quadratic Casimir for the adjoint representation.
For simply-laced groups the central charge equals the rank of the
group at level one and monotonically approaches the dimension of the group
as $K\rightarrow\infty$. This demonstrates that heterotic models constructed
from free real world-sheet bosons, $\partial X^i$, compactified on a lattice
(or equivalently free complex fermions)
include only simply-laced level-one
algebras.\footnote{The rank of the group is at least equal to the
number of bosons, since each bosonic operator $\partial X^i$ generates
a $U(1)$ \KM algebra. Also note that we have specified simply-laced,
because, as eqs.~(2.2.3) and (2.2.9) together indicate,
the central charge for non-simply-laced algebras
is greater than the rank of the group even at level-one.}
Hence, as stated above, both the $E_8\otimes E_8$ and $SO(32)$ ten-dimensional
models are level-one. With
compactification that treats left-moving and right-moving modes of the
string symmetrically the level remains one.

Construction of models with higher-level gauge groups requires
asymmetry between the left- and right-moving fields on the
world-sheet.\mpr{lewellen}
Associated with this property of the fields there
are asymmetric modular invariants.
Systematically constructing asymmetric modular invariants has proven
very difficult, except for the special case of models based
on free bosons or fermions.
However, even for asymmetric models,
use of lattice bosons (or equivalently complex fermions) limits
the possibilities to level one-models.
The first and simplest alternative is
to use real fermions instead.\mpr{lewellen} However, to date, no
phenomenologically viable model has been found using this
approach. A more general method for  constructing
(asymmetric) modular invariant tensor products of \KMAS (and of
conformal field theories) has not been developed. Several years of
research has shown that an enormous collection of consistent
free fermion models exist, of which only a small percentage are
actually left-right
symmetric. Perhaps, a systematic approach to developing
asymmetric modular invariants for tensor products of higher-level \KMAS
could produce a new class of string models with viable
phenomenology. Steps toward developing this approach
is the focus of chapter 3 of this thesis.

\vhalf
{\hc{1.2.a}{\sl Phenomenological Restrictions}}
\vhalf

Recent results from LEP have resulted in tighter constraints for viable
string models.
Using renormalization group equations (RGE), the measured high precision
values of
the standard model coupling constants have been extrapolated
from $M_Z$ to near the Planck scale.  It was found that the RGE for the
minimal supersymmetric standard model with just two Higgs doublets predict
a unification of the three coupling constants $g_3$, $g_2$ and $g_1$ for
$SU(3)\times SU(2)_L\times U(1)_Y$, respectively,
at about $10^{16}$ GeV.  For string theory
this naively poses a problem since the string unification scale is generally
required, at tree level,  to be near the Planck scale
(around $10^{18-19}$ GeV).
Three classes of solutions have been proposed for resolving
the potential inconsistency between these  extrapolations and string
theory.\markup{[\putref{bailin92}]}

The first proposal is to regard the unification of the couplings at
$10^{16}$ GeV using the minimal SUSY standard model RGE as a coincidence,
 and to allow additional states  between the electroweak scale and
the string unification scale that raise the RGE unification scale.
A second suggestion is that string threshold effects could
significantly lower the
string scale down to the  minimal SUSY standard model RGE unification
scale.
The third possibility is that a
 grand unified gauge group results from
a Ka\v c-Moody algebra at level $K\geq 2$.
As we have discussed,
adjoint (and higher) representations for Lorentz scalars become possible
when  the level of the \KM algebra is greater than one.
These adjoint scalars might allow $SU(5)$ or $SO(10)$ grand
unification.  Thus, the SUSY standard model couplings could
unify at $10^{16}$
GeV and  run upward from there
with a common value to the string unification scale.

The last proposal appears most natural and
appealing. The concept of a grand unified gauge group
fits well with the idea of successive levels of increasing
symmetry much better
than does
going directly from the symmetry of the standard model to the symmetry of
the string.  It seems far more natural for the strong force to merge
with the electroweak significantly
 below the string scale, rather than where the gravitational
coupling
(and, additionally, all hidden sector gauge couplings) finally merge.

Thus, we will examine various aspects of higher-level Ka\v c-Moody
algebras in string models. In chapter 2 we review Ka\v c-Moody
algebras in greater depth
and discuss their applications to string theory, including
general properties of and restrictions on higher level models.
In chapter 3 we develop
tools for systematically constructing string
models containing (asymmetric) higher-level \KM algebras.
Chapter 4 heads along a different direction as we
investigate aspects of a potentially new class of string models
with spacetime SUSY and critical dimensions below ten.
These models seem to have a local world sheet symmetry that pairs
the world sheet boson not with a fermion, but rather with
a primary field of a higher-level $SU(2)\over U(1)$
conformal field theory.
\hfill\vfill\eject

\chapternumstyle{blank}
\n {\bf Chapter 2: Ka\v c-Moody Algebras and String Theory}\vskip .8cm
\chapternumstyle{arabic}\sectionnumstyle{arabic}
\chapternum=2\sectionnum=1\equationnum=0
{\hb{2.1}{\bfs Review of Ka\v c-Moody Algebras}}
\sectionnumstyle{arabic}\sectionnum=1\equationnum=0
\vhalf

At the heart of the gauge symmetries of string theory are not only Lie
algebras, but the more complicated Ka\v c-Moody (KM) algebras,\mpr{kac83}
for which the former are subalgebras.
Because of the
importance of \KMAS in string theory, we review them in this chapter
before proceeding in the next chapter
with our study of modular invariant partition functions
for tensor products of \KM algebras.
Often in string theory the terms ``affine algebra,''
``affine Ka\v c-Moody algebra,'' and ``Ka\v c-Moody algebra''
are used interchangeably. The imprecise use of these
terms can be confusing, since there are actually three distinct
classes of \KM algebras, only
one of which is ``affine.''\mpr{cornwell89}
The basic step required to progress
from Lie algebras\footnote{Specifically, compact simple or compact
semi-simple Lie algebras, which is henceforth implied.}
 to \KMAS is to relax the
{\it finite}-dimension restriction on Lie algebras and
consider {\it infinite}-dimensional generalizations.
As we shall show, many of the features of semi-simple
\LAS reappear in \KM algebras. In fact, Lie
algebras can be regarded as particular cases of \KMAS with the
special property of being finite-dimensional.

Analogous to Lie algebras, \KMAS are defined by
generalized Cartan matrices (or equivalently by
Dynkin diagrams).
We will discuss \KM algebras in terms of these matrices.
After lifting the finite-dimension restriction, examination of
generalized Cartan matrices shows that \KM algebras can be grouped into
three distinct classes, called the ``finite'' (corresponding to
standard Lie algebras), ``affine,''
and ``indefinite'' types. Within the affine class are two subclasses,
denoted as ``twisted'' and ``untwisted.''

Recall that the elements of an $l\times
l$-dimensional Cartan matrix, $\bmit A$, for a Lie algebra, $\lal$,
of rank-$l$ are defined by
$$ A_{jk}= {2\langle \bal_j,\bal_k\rangle\over
\langle \bal_j,\bal_j\rangle} \,\, ,
\quad j,k\in I^{\lal}\equiv \{ 1,2,\dots ,l\}\, ,\eq{cella}$$
where $\bal_j$ is a simple root of the algebra.
For Lie algebras, the inner product of two roots is defined by
$$\langle \bal_j,\bal_k\rangle
\equiv \balpha_j\cdot\balpha_k =\sum_{m=1}^{l} (\al_j)_m (\al_k)_m
\,\, .
\eq{defdotprod}$$
(As we show, a more general definition applies for the inner product
of roots in a KM algebra. See section 2.1.a.)
Cartan matrices are defined by four properties:

\n\hi {(a)} $A_{jj}= 2$ for $j= 1,2,\dots ,l$;

\n\hi {(b)} $A_{jk}= 0$, $-1$, $-2$, or $-3$ if $j\neq k$;

\n\hi {(c)} for $j\neq k$, $A_{jk}=0$ if and only if (iff) $A_{kj}=0$;

\n\hi {(d)} det$\,\bmit A$ and all proper principal minors of $\bmit A$
are positive.\footnote{A {\it principal minor of} $\bmit A$ is the
determinant of a {\it principle submatrix of} $\bmit A$, which is a
submatrix consisting of elements $A_{jk}$ in which $j$ and $k$ both vary
over the same subset of indices. These quantities are {\it proper} if the
subset of indices is a proper subset of the set of indices.}

\no Classification of all $l\times l$-dimensional
matrices with these properties
completely classifies Lie algebras of rank-$l$.

In the late 1960's Ka\v c and Moody discovered that some of these
properties could be relaxed to produce a new, enlarged set of algebras,
with the primary difference being that the new algebras were
infinite-dimensional. By infinite-dimensional, we mean that there is
an infinite
number of roots (equivalently an infinite number of generators)
of the algebra.
Their generalized\footnote{There is additionally a slight
change of notation. For Lie algebras, $I^{\kml}$ should be altered to
$I^{\lal}= \{ 1,2,\dots , l\equiv d_{\bmit A}\}$.}
Cartan matrix, ${\bmit A}^{\rm KM}$,
is defined as $d_{\bA}\times d_{\bA}$-dimensional with the
properties that

\n\hi {(a')} $A_{jj}=2$ for $j\in I^{\kml} = \{0,1,\dots ,d_{\bA}-1\}$;

\n\hi {(b')} for $j\neq k$ $(j,k\in I)$, $A_{jk}$ is  a non-positive integer;

\n\hi {(c')} for $j\neq k$ $(j,k\in I)$, $A_{jk}= 0$ iff $A_{kj}= 0$.

\no One modification is that property (d) has
been lifted.
No longer must the determinant or all proper minors
of the matrix be positive.
det$\, \bA \leq 0$ is now
allowed, with the rank-$l$ of the matrix, $\kml$,
determined by the largest square submatrix of
$\bA$ with non-zero determinant.
Thus, $l= d_{\bA}$
only when det$\, \bA\neq 0$.
Otherwise $l<d_{\bA}$.
Second, non-diagonal elements $A_{jk}< -3$, for $j\neq k$,
are permitted.

The basic ideas and terminology for roots and root subspaces for a complex
\KM algebra, $\kml$, are very similar to those for a semi-simple complex Lie
algebra.
The commutative subalgebra, $\cal H$, of $\kml$ is referred to as the
Cartan subalgebra (CSA) of $\kml$, and the set of elements $E^{\bal}$
of $\kml$
possessing the property that
$$\lbrack {\bmit h}, E^{\bal} \rbrack = \langle {\bal}, {\bmit h} \rangle
E^{\bal}
\,\, ;\quad\quad {\rm\,\, for~all~} h\in {\cal H}\,\, ,
\eq{comrelrv}$$
form the root subspace $\kml_{\bal}$ corresponding to
the root $\bal$. The set of roots $\bal_i$, for $i\in I^{\kml}$,  are the
simple
roots upon which the generalized Cartan matrix is based. A generic root,
$\bal$, has the form
$$\bal =  \sum_{i\in I} c^{\bal}_j \bal_i, \eq{defgenroot}$$
where the set of $c^{\bal}_j$ are either all non-negative integers
or all non-positive integers.

A distinctive difference between Lie algebras and \KMAS is whereas
the dimension, $n_{\lal}$, of the CSA of Lie algebras
is equal to the rank, $l$, of the Cartan matrix, this relation does not hold
for \KM algebras. Rather, for \KM algebras\mpr{cornwell89}
the dimension of the generalized CSA is
$$n_{\kml}= 2d_{\bA} - l\,\, .\eq{dimkmcsa}$$
Only when $l= d_{\bA}$ does $n_{\kml}= l$.

For any \KM algebra, the CSA $\cal H$ can be divided into two parts,
$\cal H'$ and $\cal H''$:
$\cal H'$ being a
$d_{\bA}$-dimensional algebra with
$ \{ H^i ,\, i\in I^{\kml}\}$  as its basis; and
$\cal H''$ simply defined to be the
$(d_{\bA} - l)$-dimensional complimentary subspace of
$\cal H'$ in $\cal H$.
The $ H^i$ are
the generators giving the first $d_{\bA}$ components,
$$\alpha_j (H^i) \equiv \langle {\bal}_j , H^i \rangle \,\, ,
\eq{lacompsr}$$
of the simple roots, $\balpha_j$.
$\cal H''$  is non-trivial only when det $\bA = 0$.
Within $\cal H'$ is a subset, $\cal C$,
that forms the center of the \KM algebra.
The elements ${\bmit h}\in {\cal C}$ commute with all the members of $\kml$.
That is, if ${\bmit h}\in {\cal C}$ then
$$\langle {\bal}_j, {\bmit h}\rangle =0 {\rm ~for~all~}
j\in I^{\kml}.\eq{defcenter}$$
That $\cal C$ is $(d_{\bA} - l)$-dimensional is shown by elementary matrix
theory. The proof is short and is as follows:
Any element of ${\bmit h}\in\cal H'$ has the form
$${\bmit h}= \sum_{j\in I} \mu_j H^j \,\, .\eq{hincenter}$$
If ${\bmit h}$ is also in $\cal C$, then
$$ \sum_{j\in I} \mu_j {\bal}\cdot H^j = 0\,\, .\eq{deqns}$$
The $\cal H'$ can always be rotated into the Chevalley basis\mpr{cornwell89}
where the set of
eqs.~(\puteqn{deqns})
becomes the matrix eq.
$$ {\bA} {\bmit \mu} =0\,\, .\eq{mateqnmu}$$
($\bmit \mu$ is a column vector with entries $\mu_j$.)
Since $\bA$ has rank $l$, elementary matrix theory shows that there are
$d_{\bA} - l$ linearly independent solutions to $\bmit \mu$.
The basis of the $d_{\bA}$-dimensional subspace
${\cal H}^-\equiv {\cal H} - {\cal C}$
(which includes $\cal H''$)
can be formed from those elements ${{\bmit h}^k_+}\in {\cal H}$,
where $\langle {\bal}_j, {\bmit h}^k_+ \rangle = \delta_{jk}$
and $j,k\in I^{\kml}$.
Thus, no non-trivial element
${\bmit h}''= \sum_{k\in I}\lambda_k h^k_+\in{\cal H''}\in {\cal H_+}$
can be in the $\cal C$, since
$$ \langle {\bal}_j , {\bmit h}'' \rangle = \lambda_j\,\, . \eq{noch}$$

\vhalf
{\hc{2.1.a}{\sl Categories of Ka\v c-Moody Algebras}}
\vhalf

Matrices satisfying properties (a'--c') defining a generalized Cartan
matrix can be divided
into three categories, each corresponding to a unique
class of \KM algebras. The following three theorems define these
classes
\vskip 2mm
\n {\bf Theorem} 2.1: A complex \KM algebra, $\kml$, is ``finite''
(equivalently, it is a Lie algebra), iff
all the principle minors of the corresponding
generalized Cartan matrix, $\bA$, are positive.

\n This constraint on the principle minors is equivalent to
demanding that:

\n\hi {(F.1)} det$\, \bA \neq 0$;

\n\hi {(F.2)} there exists a
vector
${\bmit u}> 0$
of dim $d_{\bA}$ such that
$\bA {\bmit u}>0$;\footnote{${\bmit u}> 0$ is defined to mean
$u_j>0$ for all $j\in I$. Similar definitions apply when
``$<$'', ``$\geq$'', or ``$\leq$'' appear in vector relations.} and

\n\hi {(F.3)} $\bA {\bmit v}\geq 0$ implies ${\bmit v}\geq 0$.

\n Properties (F.1-3) imply that the associated algebra does not contain
any {\it imaginary} roots, {\it i.e.}, roots $\alpha$ such that
$\langle \alpha , \alpha \rangle \leq 0$,\mpr{cornwell89}
which corresponds to reimposing constraints (b) and (d).
Hence, these Cartan matrices define finite Lie algebras.

\vskip 2mm
\n {\bf Theorem} 2.2:  A complex \KM algebra, $\kml$, is ``affine''
iff its generalized Cartan matrix, $\bA$, satisfies
det$\bA =0$ and all the proper minors of $\bA$
are positive.

\n An equivalent definition of this class is to
require that the matrix is such that:

\n\hi {(A.1)} det$\, \bA =0$ but $l= d_{\bA } -1$;

\n\hi {(A.2)} there exists a vector ${\bmit u}>0$ such that
$\bA {\bmit u} = 0$; and

\n\hi {(A.3)} $\bA {\bmit v}\geq 0$ implies
             $\bA {\bmit v}= 0$.

\no With these properties, this class of \KMAS
must contain imaginary roots. The term ``affine'' is derived
from the special characteristics of its generalized Weyl
group.\mpr{cornwell89} Each complex affine \KM algebra,
$\kml^{\rm aff}$,
can be constructed from an associated complex simple Lie algebra,
$\lal$. The properties that $l= d_{\bA } -1$  along with det${\, \bA}=0$,
place a severe constraint on the one
additional simple root$\, \equiv {\bal}_0$.
In terms of its $l$-dimension projection onto the Lie algebra subspace,
which we denote by ${\bal}_0^{\cal L}$, the constraint is
$${\bal}_0^{\cal L}= -\sum_{j=1}^{d_{\bA} -1}{\bal}_j \equiv {-{\bmit\psi}}
\,\, ,\eq{alzerodef}$$
where ${\bal}_j$ are the simple roots of the Lie algebra, $\lal$, and
${\bmit \psi}$ is its highest root.
The affine algebras are the class of \KMAS
upon which the spacetime gauge groups of string theory are based;
therefore, affine algebras are discussed in greater detail in the following
(sub)sections.

The last class of \KM algebras, called ``indefinite'' algebras is most simply
defined by those generalized Cartan matrices that satisfy neither of the
conditions of Theorems 2.1 or 2.2. Indefinite matrices have the following
properties.

\n\hi {(I.1)} there exists a ${\bmit u}>0$ such that
$\bA {\bmit u}<0$; and

\n\hi {(I.2)} $\bA {\bmit v}\geq0$ and ${\bmit v}\geq 0$ imply that
${\bmit v}=0$.

\no As (I.1-2) indicate, indefinite algebras also have imaginary roots.

For a specific $d_{\bA }$ there are only a finite number
of possible generalized Cartan matrices in the finite or affine classes.
In the finite case,
where $l=d_{\bA }$, these matrices
correspond to the standard simple Lie algebras, which are denoted by
$A_l\,$, $B_l\,$, $C_l\,$, $D_l\,$, $E_{6,7,8}\,$,
$F_4\,$, and $G_2\,$. In the affine case, where
$l= d_{\bA} -1$, there is an untwisted generalization
of each Cartan matrix associated with a Lie algebra of rank-$l$.
Common notation for the affine algebras is to add
a superscript of $(1)$ to the Lie algebra symbol.
For example, the untwisted affine version of $A_l$
is denoted by $A_l^{(1)}$.
Additionally, for $A_l\,$, $D_l\,$, and $E_{6}$ there is
a ``twisted'' generalization denoted by superscripts $(2)$.
There is a also a
second twisted affinization of  $D_4$, denoted by $D_4^{(3)}$. The
twisted algebras result from
either a $\Z_2$ or a $\Z_3$ rotation and projection
of the roots of the untwisted
affine algebra,\mpr{cornwell89}
and are non-simply-laced affinizations
of simply-laced Lie algebras.

The third type of \KM algebra,
the ``indefinite'' class, is appropriately named because
there is an infinite number of matrices
that meet neither ``finite'' nor ``affine'' requirements
for each value of $d_{\bA}$.
All of these matrices
correspond to non-isomorphic algebras not grounded in generalizations of
the Lie algebras.  Very few (if any) applications for indefinite
\KMAS have been found.
In particular,  they appear to play no part in string theory.

To illustrate the differences of the three classes, consider the simplest
non-trivial generalized Cartan matrix possible, the $2\times 2$-dimensional
$${\bA}= \left(\matrix{2&-r\cr
                   -s&2\cr}\right)\,\, ,$$
where $r$ and $s$ are positive integers.
Now let us classify all of the possible \KM algebras in each class
associated with specific values for $r$ and $s$.

\vskip 2mm
\item{1.} Finite (Lie) algebras: det$\, {\bA}> 0$ so $rs<4$. There are only
three possibilities for non-equivalent algebras,
\item\item{a.} $r=1$, $s=1$ corresponding to $A_2$;
\item\item{b.} $r=1$, $s=2$ corresponding to $B_2$; and
\item\item{c.} $r=1$, $s=3$ corresponding to $G_2$.

\vskip 2mm
\item{2.} Affine algebras: det$\, {\bA} =0 $ so $rs=4$. There are
only two inequivalent possibilities,
\item\item{a.} $r=1$, $s=4$ corresponding to $A^{(2)}_2$; and
\item\item{b.} $r=2$, $s=2$ corresponding to $A^{(1)}_1$.

\vskip 2mm
\item{3.} Indefinite algebras: det$\, {\bA}<0 $ so $rs>4$. There is
an infinite number of choices for $r$ and $s$ resulting in non-isomorphic
algebras.

\vskip 2mm
Since the finite class of \KM algebras is simply composed of
Lie algebras
and the indefinite class, although the largest, seems to have
little application to physics, we
cease our study of them with this example
of classification of $2\times 2$-dimensional generalized Cartan matrices.
We now focus in greater detail on the affine
algebras and their role in string theory.

\vhalf
{\hc{2.1.b}{\sl Affine Algebras}}\vhalf

Having discussed the three classes of \KM algebras, we focus here in detail
on affine \KMAS. We
generalized Cartan-Weyl basis.
Recall from eq.~(\puteqn{dimkmcsa})
that the CSA, $\cal H$, of a \KM algebra, $\kml$, has dimension
$$n_{\kml}= 2d_{\bA} - l\,\, ,$$
where $d_{\bA}$ is the number of simple roots and $l$ is the rank of the
associated generalized Cartan matrix.
$\cal H$ can be divided into two parts,
$\cal H'$ of dimension $d_{\bA}$, and its compliment $\cal H''$ of
dimension $d_{\bA} -l$.
Within $\cal H'$ is the $(d_{\bA} -l)$-dimensional
center, $\cal C$, of $\kml$.
Applying this to the affine class of \KM algebras,
shows that:
\item{(1)} $\cal H$ has dimension $l+2$.
\item{(2)} $\cal H'$ is $(l+1)$-dimensional with
only one generator in the center.
\item{(3)} $\cal H''$ is one-dimensional.

The $l$ generators, denoted by
$H^p$ for $p\in I^{\lal}=\{ 1,\, 2,\, \dots \, l\}$,
 in $\cal H$ but not in $\cal C$
form the CSA of the Lie algebra $\lal\subset\kml$.
Thus, affine CSA's contain two additional generators of
$\cal H$ not present in
the Lie subalgebra. The single generator of the
center is known as the level, $K$, of the algebra,
and the generator of $\cal H''$ is called the scaling element, $d$.
We can express generic roots of the \KMA in the
form\footnote{$\alpha_j({\bmit h})\equiv \langle {\bal}_j, {\bmit h}\rangle$.}
$$ {\bal}_j = \left( {\bal}_j^l, \alpha_j(K), \alpha_j(d) \right)\,\, ;
\quad j\in I^{\kml}\,\, ,
\eq{defkmsimcur}$$
where ${\bal}_j^l$ forms the $l$-dimensional subvector that is
associated solely with the Lie algebra $\lal$.
In this notation, the simple roots can be taken as
$${\bal}_p=\left({\bal}_p^l,0,0\right) {\rm ~~for~~} p\in I^{\lal}
\quad\quad {\rm and} \quad\quad
{\bal}_0 = \left( -{\bmit \psi}, 0, \alpha_0(d) \right) \,\, .\eq{simroots}$$
(Since $K$  forms the center of the algebra,
$\alpha_j (K) = 0$ by (\puteqn{comrelrv}).)

Based on eq.~(\puteqn{mateqnmu}), to
a given affine Cartan matrix ${\bA}$
is associated a single linearly independent $d_{\bA}$-dimensional
vector ${\bmit \mu}>0$ such that ${\bA}{\bmit \mu}=0$.
This vector is related to $\alpha_0(d)$ by
$$ {\bmit \delta} = {\sum_{j=0}^{l} \mu_j {\bal}_j}
= \left( {\bmit \delta}^{l}= 0, 0, \delta(d)= \alpha_0(d)\right)
\,\, .\eq{deltadefal}$$
In other words, $\delta (H^p) = \delta(K) = 0$.
$d$ can be defined so that $\delta(d)\equiv\alpha_0(d)=1$.
${\bmit \delta}$ is an actual root of the theory, as are all integer
multiples, $m{\bmit \delta}$; $m\in \Z$.
Thus, a general root has the form $\bal= (\bal^l,0,m)$.
For $\bal^0=0$ we denote the associated operator by
$H^0_m$; otherwise we denote the operator by $E^{\bal}_m$.

Consistency of the algebra\mpr{cornwell89} forces ${\bmit \delta}$ to be
a ``null root'' (imaginary root as previously defined)
with the property that
$$\langle {\bmit \delta}, {\bmit \delta} \rangle =
  \langle {\bmit \delta}, {\bal}_j \rangle= 0
\,\, ,\eq{imagdelta}$$
(this is more clearly seen in the Chevalley basis\mpr{cornwell89})
and determines the generalization  of (\puteqn{defdotprod})
for two generic roots, $\bal$ and $\bbe$ of an affine theory:
$$\langle {\bal} , {\bmit \beta}\rangle
=   {\bal}^l\cdot {\bmit \beta}^l + \alpha(K)\beta(d)
                              + \alpha(d)\beta(K)
\,\, .\eq{newdefdotprod}$$
Thus, only the Lie algebra components
contribute to the inner product of any two simple roots,
 $$\langle {\bal}_i, {\bal}_j \rangle=
{\bal}_i^l \cdot {\bal}_j^l\,\, .\eq{inprodsimcur}$$
Using this generalized definition of an inner product,
the Weyl reflection of a weight,
${\bmit \lambda}= ({\bmit \lambda}^l, k, n)$,
about a root, $\bal= (\bal^l,0,m)$, is
\subon
$$\eqalignno{
w_{\bal}({\bmit\lambda})
&= {\bmit\lambda} - {\bal}\langle {\bmit\lambda}, {\bal} \rangle
          & \eqnlabel{weyldef-a}\cr
&= \left(
{\bmit \lambda}^l
-2\lbrack
   {\bmit \lambda}^l\cdot\bal^l + k m
\rbrack
{\bal^l\over {\bal\cdot\bal}}, k, n
- 2\lbrack
    {\bmit \lambda}^l\cdot\bal^l + k m\rbrack
     {m\over {\bal\cdot\bal}}
\right)\,\, .
&\eqnlabel{weyldef-b}}$$
This reflection can be split into two parts, a series of $m$ translations by
$$t_{\bal^l}({\bmit\lambda}) = \left(
    {\bmit\lambda}^l + k {{\bal}^l\over {\bal\cdot\bal}}, k,
n + {1\over 2k}\left\{ {\bmit \lambda}^l \cdot{\bmit\lambda}^l
                     -  ({\bmit\lambda}^l
                          + 2k {\bal^l\over {\bal\cdot\bal}})^2  \right\}
                             \right)
 \,\, .\eq{weyldef-c}$$
followed by a Weyl reflection about ${\bal}^l$.
The affine Weyl rotation is the product of these
transformations,
$$ w_{\bal}({\bmit\lambda}) =
w_{\bal^l}\left( t^m_{\bal^l}({\bmit\lambda}) \right)
\,\, .\eq{weyldef-d}$$
\suboff
We conclude this general discussion of affine \KMAS with
a listing of the algebra, itself.
Adding the two additional generators, $K$ and $d$,
to the CSA of a Lie algebra,
$\{ H^p\equiv H^p_0\, ;\, \, p \in I^{\lal}=\{1,\, 2,\, \dots,\, l\} \}$,
forms the affine CSA and
enlarges the Lie algebra,
$\lal$,\footnote{Eqs.~(2.1.21) correspond to an {\it untwisted} affine
KM algebra. The {\it twisted} algebras involve a
{\tenBbb Z}$_{q=2 {\rm ~or~} 3}$
rotation by an outer automorphism of the untwisted
KM algebra, which creates a {\tenBbb Z}$_q$ projection on the roots.
The roots, $\bal = (\bal^{\lal}, 0 , m \}$, may be classified
by their
eigenvalues, $\exp \{ 2\pi i p/q \}$ with
$p\in \{ 0, 1,\, 2,\, \dots\, , q-1\}$,
under this rotation.
The related projection requires
\hbox{$m \pmod{q} = p$. The surviving $m=0$ roots}
are isomorphic to the simple roots of a non-simply-laced Lie subalgebra,
$\lal^{(q)}$, of $\lal$.}
from:
\subon
$$
\eqalignno{
\lbrack H^p, H^q \rbrack
    &= 0
                         & \eqnlabel{eqcr-a}\cr
\lbrack  H^p, E^{\bal} \rbrack
    & =  \bal(H^i) E^{\bal}
                         & \eqnlabel{eqcr-b}\cr
\lbrack E^{\bal}, E^{\bbe}\rbrack
    & =
{\cases {\epsilon(\bal,\bbe) E^{\bal+\bbe}
           & if $\bal,\bbe$ is a root\cr
         {2\over \bal^2} \bal \cdot {\bf H}\, ,
           & if $\bal+\bbe = 0$\cr
         0
           & otherwise\cr}}
                         & \eqnlabel{eqcr-c}\cr
\cr}
$$
(with $p, q \in I^{\lal}$), to the full affine algebra
\suboff
\subon
$$
\eqalignno{
\lbrack H^i_m, H^j_n \rbrack
    &= Km\delta^{ij}\delta_{m,-n}
        & \eqnlabel{eqcrkm-a}\cr
\lbrack  H^p_m, E^{\bal}_n \rbrack
    & =  \bal(H^i), E^{\bal}_{m+n}
                         & \eqnlabel{eqcrkm-b}\cr
&\cr
\lbrack E^{\bal}_m, E^{\bbe}_n\rbrack
    & =
{\cases {\epsilon(\bal,\bbe) E^{\bal+\bbe}_{m+n}
           & if $\bal+\bbe$ is a root\cr
         {2\over \bal^2}
          \lbrack \bal \cdot {\bf H}_{m+n} + Km\delta_{m,-n}\rbrack\, ,
           & if $\bal+\bbe = 0$\cr
         0
           & otherwise\cr}}
                         & \eqnlabel{eqcrkm-c}\cr
           &\cr
\lbrack K,  T^a_n\rbrack
    &= 0 &\eqnlabel{eqcrk-d}\cr
\lbrack d, T^a_n\rbrack
    &= n T^a_n &\eqnlabel{eqcrk-e}\cr}
$$
where $i,j\in I= \{0,\, 1,\, \dots ,\, l\}$; $m,\, n\in \Z$,
and $T^a_n$ is any element of the algebra.
\suboff
The operators of the algebra have the hermiticity properties
$$ {H^i_m}^{\dagger}= H^i_{-m}\, ,\quad
   {E^{\bal}_m}^{\dagger}= E^{-\bal}_{-m}\, ,\quad
   K^{\dagger}= K\, , {\rm~~and~~} d^{\dagger}=d\,\, .
   \eq{hermitprop}$$
\vhalf
{\hb{2.2}{\bfs Application to String Theory}}
\sectionnumstyle{arabic}\sectionnum=2\equationnum=0
\vhalf

Now we consider the specific role of affine \KMAS in string theory,
where they provide the world sheet realization of spacetime
gauge theories.
Present in string models are sets of $(h,\bar h)=(1,0)$
conformal fields, $J^a(z)$, called currents, which satisfy the OPE of a \KM
generator,
$$J^a(z)J^b(w) = {\tilde K^{ab}\over (z-w)^2}
                + {i f^{abc}\over (z-w)} J^c +
({\rm non-singular~terms})\,\, .\eq{kmope}$$
$f^{abc}$ are the structure constants for the Lie algebra, $\lal$,
with normalization
$$f^{abc}f^{dbc}= C_A\delta^{ad}= {\bmit \psi}^2\dc \delta^{ad}
\,\, .\eq{defstrcon}$$
$C_A$ is the quadratic Casimir of the adjoint representation of $\lal$,
$\dc$ is the dual Coxeter number, and ${\bmit \psi}$
denotes the highest root.
For each simple factor of the algebra, a basis can be chosen such that
$\tilde K^{ab}= \tilde K\delta^{ab}$.
$K\equiv 2\tilde K/{\bmit \psi}^2$
is defined as the level of a simple factor of
the \KMA (as discussed in 2.1.b).
Commonly in string theory the normalization of ${\bmit \psi}^2=2$ is used,
which results in $K=\tilde K\in\Z$ and $\tilde h=C_A/2$.
Also recall that
\subon
$$\eqalignno{
C_A &= r^{-1}_{\cal L}\sum^{{\rm dim}\, \lal}_{a=1} {\bal}^2_{a}
&\eqnlabel{cadef-a}\cr
& = {1\over r_{\lal}}\left( n_L + ({S\over
L})^2 n_S \right){\bmit \psi}^2
&\eqnlabel{cadef-b}\cr
& \rightarrow \left( {{\rm dim}\, \lal\over r_{\lal}} -1 \right)
{\bmit \psi}^2
\quad {\rm for~simply-laced~algebras}
&\eqnlabel{cadef-c}\cr}$$
where $r_{\lal}$ is the
rank of the algebra and ${\bal}_{a}$ is a simple
root.
$n_S$ and $n_L$ are the number of short and long roots, respectively,
and $S$ and $L$ are the lengths.
\suboff

The presence of an underlying \KMA is alternatively seen from the
related commutation relations of the modes of the
currents,\footnote{Any field, $\phi$, with conformal dimension,
$h_{\phi}$,
can be written in terms of the normal modes, $\phi_n$,
in a Laurent expansion,
$$\phi(z) = \sum_{n= -\infty}^{\infty} z^{-n-h_{\phi}}\phi_n\, ,
{\rm ~~where~~}
\phi_n = \oint {{\rm d}z\over (2\pi i z)} z^{n+h_{\phi}} \phi (z)\,\, .$$}
$$J^a(z)= \sum_{n\in Z} J^a_n z^{-n-1}\,\, ,
{\rm ~where~} J^a_n= \oint {{\rm d}z\over (2\pi i z)} z^{n+1} J^a(z)\, .
\eq{jmodcur}$$
The commutation relations have the form
$$ \lbrack J^a_m,J^b_n \rbrack = i f^{abc} J^c_{m+n} +
\tilde K m\delta^{ab}\delta_{m,-n}\,\, , \eq{kmcommrel}$$
where $m,n\in \Z$.
As was discussed previously, these commutators define the untwisted affine
\KM algebra, $\kml$, associated with a compact (semi)-simple Lie
algebra, $\lal$.
The horizontal Lie subalgebra, $\lal$, is formed from the algebra
of the zero modes, $J^a_0$, for which the level does not appear.
The full (infinite) set of $J^a_n$'s provides the affinization of the
finite dimensional subalgebra of $J^a_0$'s.

In a heterotic string model these currents appear in the vertex operator
for a spacetime gauge boson,\mpr{lewellen} {\it e.g.},
$$V^a= \zeta_{\mu} \psi^{\mu}(\bar z) J^a(z)\exp\{ip\cdot X\}\,\, ;
\quad p^{\mu}p_{\mu}= p^{\mu}\zeta_{\mu}=0\,\, .\eq{vertexop}$$
$X^{\mu}$ is the spacetime string coordinate and  $\psi^{\mu}$ is
a left-moving Ramond-Neveu-Schwarz fermion.  Thus,
spacetime gauge fields
imply the existence of a \KMA on the world sheet.
In other words,
there is an extension to the standard Virasoro algebra that includes
the affine \KM currents. In OPE language, the extended Virasoro--\KMA
takes the form
\subon
$$\eqalignno{
T(z) T(w)
&= {c/2\over (z-w)^4} + {2\over (z-w)^2}T(w) +
{1\over (z-w)}\partial T(w) + \dots
{\hbox to .5cm{\hfill}}
&\eqnlabel{tjope-a}\cr
T(z)J^a(w)
&= {J^a(w)\over (z-w)^2} + {\partial J^a(w)\over (z-w)} + \dots
&\eqnlabel{tjope-b}\cr
J^a(z)J^b(w)
&= {\tilde K^{ab}\over (z-w)^2}
+ {i f^{abc}\over (z-w)} J^c + \dots \,\, .
&\eqnlabel{tjope-c}}
$$
\suboff
Equivalently, the algebra can be expressed in terms of commutation
relations of normal modes:
\subon
$$\eqalignno{
\lbrack L_m, L_n \rbrack
&= (m-n) L_{m+n}
+ {c\over 12} (m^3 -m)\delta_{m,-n}
&\eqnlabel{tjcr-a}\cr
\lbrack L_m , J^a_n \rbrack
&= -n J^a_{m+n}
&\eqnlabel{tjcr-b}\cr
\lbrack J^a_m , J^b_n \rbrack
&= i f^{abc} J^c_{m+n} +
\tilde K m\delta^{ab}\delta_{m,-n}\,\, .
&\eqnlabel{tjcr-c}}$$
\suboff

In eq.~\pe{tjope-a}
the contribution to the Virasoro central charge (conformal anom-
aly)
from the \KM algebra is
$$c_{\kml} = {\Kt {\rm dim}\, \lal \over \Kt + C_A/2}
      = {K {\rm dim}\,  \lal \over K + \dc}\,\, .\eq{cccakm}$$
The
energy-momentum tensor, itself, may be written in terms of
the \KM currents:\mpr{ginsparg89}
$$ T(z)= {1\over \beta} \sum_{a=1}^{{\rm dim}\, \lal}
: J^a(z) J^a(z): = {1\over \beta}\left( \lim_{z\rightarrow w}
\sum_{a=1}^{{\rm dim}\, \lal} J^a(z) J^a(w)
- {\tilde K {\rm dim}\, \lal\over (z-w)^2}
\right)\,\, . \eq{emtensm}$$
$\beta\equiv 2\tilde K + C_A= 2(K+\dc)$ is a constant fixed either by the
requirement that $T(z)$ satisfy (\puteqn{tjope-a})
or, equivalently,
that the $J^a(z)$'s transform as dimension $(1,\, 0)$ primary fields.
In terms of the mode expansion for $T(z)$, (\puteqn{emtensm}) translates into
\subon
$$\eqalignno{
L_n
&= \oint {{\rm d} z\over (2\pi i z)} z^{n+2} T(z)
&\eqnlabel{ln-a}\cr
&= {1\over \beta} \sum_{m= -\infty}^{\infty} : J^a_{m+n} J^a_{-m}:
\,\, .
&\eqnlabel{ln-b}}$$
\suboff

All states in the
theory necessarily fall into representations of the
Virasoro--\KM
algebra.\mpr{lewellen}
Each representation (Verma module),
$\lbrack \phi_{(r)} \rbrack$,
is composed of a primary field $\phi_{(r)}$
(actually, a multiplet of fields $\phi_{(r)}^{\lambda})$,
and all of its ``descendent'' fields.
The descendent fields are the set of fields formed by acting
on a primary field with all possible products of
the raising operators $L_{-m}$ and $J^a_{-n}$ for $m,\, n\in\Z^+$,
$$\left\{ \prod_{i=1}^{\infty}(L_{-i})^{A_i}
  \prod_{a=1}^{{\rm dim}\, \lal} (J^a_{-i})^{B^a_i}\vert \phi_{(r)}
\rangle\right\} \,\, ,\eq{vermamod}$$
where $A_i\, ,\,\, B^a_i\in \{ 0, \Z^+ \}$.
$\phi_{(r)}$ transforms as a
highest weight representation $(r)$ of $\lal$, as indicated by
the leading term in the OPE of $\phi_{(r)}$ with the current $J^a(z)$,
\subon
$$\eqalignno{
J^a(z) \phi_{(r)}(w)
&= {(T^a)_{(r)}^{(r')}\over (z-w)}
\phi_{(r)}\,\, ,
&\eqnlabel{opejphi-a}\cr
&= {(t^a_{(r)})^{\lambda'\lambda}\over (z-w)}
\phi_{(r)}^{\lambda}\,\, .
&\eqnlabel{opejphi-b}}$$
\suboff
$t^a_{(r)}$ are representation matrices for $\lal$ in the representation
$(r)$.
These primary fields create states, called highest weight states, defined by
$$\vert \phi_{(r)}\rangle \equiv \phi_{(r)}(0) \vert {\rm vacuum}\rangle\,\, ,
\eq{hws}$$
that are representations of the zero-mode (Virasoro-Lie)
algebra,\footnote{$-L_0$ can be identified with the scaling
element, $d$, of the KM algebra.}
\subon
$$
   L_0 \vert \phi_{(r)}\rangle = h_{(r)}\vert \phi_{(r)}\rangle
\,\, , \quad\quad{\rm and} \quad\quad
J^a_0 \vert \phi_{(r)}\rangle
         = {(T^a)}_{(r)}^{(r')}\vert \phi_{(r')}\rangle
\,\, \,\, {\rm for~} n= 0\,\, ,
\eq{reptrans-a}$$
and
$$
 L_n \vert \phi_{(r)}\rangle  =
J^a_n \vert \phi_{(r)}\rangle = 0 \,\, \,\, {\rm for~} n>0\,\, .
\eq{reptrans-b}
$$

{}From (\puteqn{reptrans-a}), the general form for the
the conformal dimension, $h_{(r)}$, of the primary field,
$\phi_{(r)}$, is
\subon
$$\eqalignno{
h_{(r)} &=
 {C_{(r)}/2\over \tilde K + C_A/2}
& \eqnlabel{cdpf-a}\cr
&= {C_{(r)}/\psi^2\over K + \dc} \,\, ,
& \eqnlabel{cdpf-b}}$$
\suboff
where
$$C_{(r)} \equiv l_{(r)} { {\rm dim}\, \lal \over {\rm dim}\, (r)}$$
is the quadratic Casimir of the representation $(r)$,
with ${\rm tr}\, t^a_{(r)} t^b_{(r)}= l_{(r)} \delta^{ab}$.
The dimensions the descendent fields are $h_{(r)} + \Z^+$. Specifically,
$$h= h_{(r)} + \sum_{i=1}^{\infty} \left( i A_i +
\sum_{a=1}^{{\rm dim}\, \lal} i B^a_i\right)\eq{hdf}$$
for the field
$${ \prod_{i=1}^{\infty}(L_{-i})^{A_i}
  \prod_{a=1}^{{\rm dim}\, \lal} (J^a_{-i})^{B^a_i}\vert \phi_{(r)}
\rangle} \,\, .\eq{dfdef2}$$

An issue we wish to stress is that
only states in representations $(r)$ satisfying,
$$K\geq \sum_{i=1}^{r_{\lal}} n_i m_i\,\, ,\eq{unitaryreq}$$
may appear in sensible string models.
$n_i$ are the Dynkin labels of the highest weight of the
representation $(r)$ and $m_i$ are the related co-marks\footnote{See
Figure A.1 of Appendix A for listings of the co-marks for each
of the compact simple Lie algebras.}
group associated with the Lie algebra, $\lal$.\mpr{lust89}
Eq.~(\puteqn{unitaryreq})
is the condition for unitarity of a representation.
Within any \KM algebra, the subset
$$\{ J^{-\bmit\alpha}_1, J^{\bmit\alpha}_{-1},
K- {\bmit\alpha}\cdot {\bmit H}\}, $$
where ${\bmit\alpha}$ is a root in $\lal$ and ${\bmit H}$ is the vector of
currents in the Cartan subalgebra, forms an $SU(2)$ subalgebra.
If $\lambda$ is the weight of a component, $\phi_{(r)}^{\lambda}$,
of the multiplet $\phi_{(r)}$, then
$${\eqalign{
 0\leq \langle\phi_{(r)}^{\lambda}\vert J^{-\bmit\alpha}_1
    J^{\bmit\alpha}_{-1}\vert\phi_{(r)}^{\lambda}\rangle
& = \langle\phi_{(r)}^{\lambda}\vert \lbrack J^{-\bmit\alpha}_1,
    J^{\bmit\alpha}_{-1}\rbrack \vert\phi_{(r)}^{\lambda}\rangle \cr
& = \langle\phi_{(r)}^{\lambda}\vert (K- {\bmit\alpha}\cdot {\bmit H})
                            \vert\phi_{(r)}^{\lambda}\rangle \cr
 & = (K-{\bmit\alpha}\cdot{\bmit\lambda})\langle\phi_{(r)}^{\lambda}
\vert\phi_{(r)}^{\lambda}\rangle \,\, .}}\eq{unitproof}$$
Hence, $(K-{\bmit\alpha}\cdot{\bmit\lambda})$ must be positive for all
roots ${\bmit\alpha}$ and all weights ${\bmit\lambda}$ in the
representation $(r)$. Thus,
\subon
$$\eqalignno{
K & \geq {\bmit\psi}\cdot {\bmit\Lambda} & \eqnlabel{unitproof2-a}\cr
&  = \sum_{i=1}^{r_{\lal} } n_i m_i\,\, , & \eqnlabel{unitproof2-b}}$$
\suboff
where ${\bmit\psi}$ is the highest root of $\lal$ and ${\bmit\Lambda}$ is the
highest weight in the $(r)$ representation.
This is the first major constraint placed on
highest weight states of Lie algebras and the associated primary fields
that can appear in consistent string models.

One consequence of this, as we mentioned in chapter 1, is that
string models based on level-1 \KM algebras cannot have spacetime scalars
in the adjoint representation. Naively, there would appear a way of
escaping this. Since the \KM currents transform in the adjoint
representation we might use them to form spacetime scalars.
Unfortunately, this cannot be done, at least for models with
{\it chiral} fermions.\mpr{dixon87,dreiner89a,lewellen}
The basic argument is as follows:
The vertex operator, $V_{\rm scalar}^a$ for a spacetime scalar in a
level-1 adjoint representation would necessarily have the form
$$ V_{\rm scalar}^a= O(\bar z) J^a(z)\,\, , \eq{vertscalad}$$
where $J^a$ is one of the \KM currents. Masslessness of the state requires
that the anti-holomorphic operator, $O(\bar z)$, have
conformal dimension $\bar h_O=\half$ and behave both
in its OPE's and under GSO
projections like an additional RNS fermion. Hence, the spacetime spinor
degrees of freedom would fall into representations of the five-dimensional
Lorentz group, $SO(4,1)$. Decomposition into $SO(3,1)$ spinors always gives
non-chiral pairs. Thus, adjoint scalars and chiral fermions are mutually
exclusive. Further,
$N=1$ SUSY, at least for models based on free field construction,
also disallows these adjoint scalars.\mpr{lewellen,dreiner89b}
We have assumed in these arguments that the currents are not
primary fields of the full Virasoro-\KM algebra;
comparison of
eq.~(\puteqn{tjope-c}) with (\puteqn{opejphi-a})
proves this is, indeed, a valid assumption.

A second constraint on states is a bit more trivial.
Since the gauge groups come from the
bosonic sector of the heterotic string, the total contribution to the
conformal anomaly from the gauge groups cannot exceed 22, {\it i.e.,}
$$ c_{\rm KM} = \sum_i { K_i{\rm dim}\, \lal_i\over K_i + \dc_i}\leq 22\,\, ,
\eq{maxckma}$$
where the sum is over the different factors in the algebra
and every $U(1)_K$
contributes 1 to the sum. This condition gives an upper bound to the
levels for a GUT.\mpr{font90,ellis90}
For example, if the gauge group is $SO(10)$, the maximum
level is seven, for $E_6$ it is four, while it is  $55$ for $SU(5)$.

In terms of the massless representations of the Lie algebras that appear,
there is one additional constraint that is stronger
than either of
the first two. This constraint is on the conformal
dimension of a primary field, \pe{cdpf-a}.
Since the intercept for the bosonic sector of a heterotic string is one,
a potentially massless state in an $(r)$ representation cannot have
$h_{(r)}$ greater than one. That is,
$$h_{(r)}= {C_{(r)}/{\bmit \psi}^2\over K + \dc}\leq 1\,\, .\eq{maxcdone}$$
\hfill\vfill\eject

\pagenum=23

\chapternumstyle{blank}\subsectionnumstyle{blank}
\n {\bf Chapter 3: Modular Invariant Partition Functions}\vskip .8cm
{\hb{3.1}{\bfs Review of Characters, Partition Functions, and
Modular Invariance}}\vhalf
\chapternumstyle{arabic}\chapternum=3
\sectionnumstyle{arabic}\sectionnum=1\equationnum=0

Recently, studies of classical string solutions
have provided impetus for further research into
two-dimensional conformal field theories.
In particular, considerable effort has been spent in classifying
modular invariant partition functions (MIPF's) of these
theories.  In any string model, there is
corresponding to each (chiral)
Verma module representation, $\lbrack \phi (z)\rbrack$,
of the Virasoro algebra
(or an extension of it such as a super-Virasoro or Virasoro-\KM algebra)
a character (a.k.a. partition function), $\chi_{\lbrack \phi \rbrack}$.
The character is a trace over the Verma module on a cylinder,
$$\chi_{\lbrack \phi \rbrack}=
{\rm Tr}_{\lbrack \phi \rbrack} q^{L_{0_{\rm cyl}}}
= q^{-c/24} {\rm Tr}_{\lbrack \phi \rbrack}
q^{L_{0_{\rm plane}}}\,\, , \eq{defchar}$$
where $q= \exp(2\pi i\tau)$, $\tau= \tau_1 + i\tau_2$ ,
with $\tau_1\, $, $\tau_2\in\R$,
 and the trace containing the conformal anomaly
factor is defined on the complex plane.\footnote{The factor of $q^{-c/24}$
results from the stress-energy tensor, $T(z)$ not transforming
homogeneously under a conformal transformation, but picking up a quantity
equal to $c/12$ times the Schwartzian, $S(z,w)$. That is, under
$w\ra z= e^w$,
$$T_{\rm cyl}(w)= \left({ \partial z\over \partial w}\right)^2 T(z)
+ {{c\over12}}\, S(z(w),w)\,\, ,$$
where
$S(z(w),w)\equiv {\partial z\partial^3 z - (3/2)(\partial^2 z)^2\over
(\partial z)^2} = -\half\,\, .$ Thus, the $L_0$ defined on the cylinder is
not equivalent to the $L_0$ defined on the complex plane, rather
$L_{0_{\rm cyl}}= L_{0_{\rm plane}} - c/24$.  If the only purpose for
partition functions were to count the number of states at each level, the
anomaly term could be effectively discarded. However, this term is very
important with regard to modular invariance.}
If we expand this character in terms of powers of $q$,
$$\chi_{\lbrack \phi \rbrack} = q^{-c/24}\sum_{i=0}^{\infty}
n_i q^{h_{\phi} +i}\,\, ,\eq{charexpq}$$
the integer coefficient, $n_i$ counts how many (descendent) fields the
Verma module contains at the $i^{th}$ energy level.
The one-loop partition function of the string model can be expressed in
terms of bi-linears of the characters of the Verma modules,
$$\eqalignno{Z(\tau,\bar\tau)
&= \sum_{a,b} N_{ab}\chi_a(\tau)\bar\chi_b(\bar\tau)
&\eqnlabel{pfintro1}\cr
&= \sum_{a,b} N_{ab}{\rm Tr}\e^{2\pi i\tau_1 P}\e^{- 2\pi \tau_2 H}\,\, .
&\eqnlabel{pfintro2}\cr}$$
$H= L_0 + \bar L_0$ and  $P= L_0 - \bar L_0$
are the Hamiltonian and momentum operators of the
theory\footnote{Thus, from the statistical mechanics perspective,
$\tau_2$ can be viewed as either a Euclidean time
that propagates the fields over the one-loop world sheet cylinder
or as the inverse of a temperature.
Analogously, $P$ can be interpreted
as a momentum operator that twists an end of
the world sheet cylinder by $\tau_1$
before both ends meet to form a torus.}
and $N_{ab}\in\Z$ corresponds to the number of times that the primary field
associated with $\chi_a\bar\chi_b$ appears in the theory.

\begin{ignore}Further,
As with the characters from which it is formed, when the partition function
is expanded in terms of powers of $q$ and $\bar q$,
the coefficient of $q^r$ corresponds
to the number of states in the theory at energy (mass) level-$r$.
\end{ignore}

The term ``modular invariant partition function''
is  understood, as above,
to generally mean the MIPF for a genus-1 world sheet
(a torus). In conformal field theory,
a torus is characterized by a single complex parameter,
the $\tau$ of the above equations.
Geometrically, $\tau$
may be defined by making the following identifications on the
complex plane:\mpr{lust89}
$$ z\approx z + n + m\tau\,\, , \quad\quad n,m\in\Z\,\, ,
\quad\quad \tau\in\C\,\, . \eq{deftorus1}$$
The more general definition
$$ z\approx z + n\lambda_1 + m\lambda_2\,\, , \quad\quad n,m\in\Z\,\, ,
\quad\quad \lambda_1,\lambda_2\in\C\,\, , \eq{deftorus2}$$
leads to conformally equivalent tori under rescaling and rotation of
$\lambda_1$ and $\lambda_2$ by the conformal transformation
$z\rightarrow \alpha z$. Hence, only their ratio,
$\tau\equiv {\lambda_1\over \lambda_2}$, is a conformal invariant. Therefore
$\lambda_1$ is set to one. Also, the freedom to interchange
$\lambda_1$ and $\lambda_2$ allows us to impose Im$\,\tau >0$. Thus, tori
are characterized by complex $\tau$ in the upper-half plane.
(See Figure 3.1.)

This is not the whole story though. It is not quite true that $\tau$ is
a conformal invariant that cannot be changed by rescalings and
diffeomorphisms. There are global diffeomorphisms, not smoothly
connected to the
identity, that leave the torus invariant, but change the parameter $\tau$.
They correspond to cutting the torus along either cycle $a$ or $b$,
twisting
one of the ends by a multiple of $2\pi$, and then gluing
the ends back together. (See Figure 3.2.) Such
operations are know as Dehn twists and generate all global diffeomorphisms
of the torus.  A Dehn twist around the $a$ cycle, transforms $\tau$ into
$\tau +1$. (The related transformations of $\lambda_1$ and $\lambda_2$ are
$\lambda_1\rightarrow\lambda_1$ and $\lambda_2\rightarrow\lambda_1 +
\lambda_2$.)
This transformation is commonly denoted as ``$T$'',
\subon
$$ \hbox to 1cm{\n $T$:\hfill}\tau\rightarrow\tau +1 \,\, .\eq{deftranst-a}$$
The  twist
around the $b$ cycle corresponds (after rotation and rescaling to bring
$\lambda_1$ to one) to $\tau\rightarrow {\tau\over\tau+1}$
and can be expressed in terms of $T$ and another transformation,
``$S$'', defined by
$$ \hbox to 1cm{\n $S$:\hfill}\tau\rightarrow\-{ 1\over \tau}\,\, .
\eq{deftransst-b}$$
\suboff
Specifically, $TST: \tau\rightarrow{\tau\over\tau+1}$.

$S$ and $T$ are the generators of the symmetry group of
$PSL(2,\Z)=SL(2,\Z)/\Z_2$, called the modular group of the torus.
General modular transformations take the form
$$PSL(2,\Z): \tau\rightarrow \tau^{'} = {a\tau + b\over c\tau + d}\,\, ,
\eq{psltrans}$$
with $a,b,c,d\in \Z$ and  $ad-bc=1$. (The $Z_2$ projection equates
$(a,b,c,d)$ with $(-a,-b,-c,-d)$ since both correspond to the same
transformation of $\tau$.)
Thus, the true moduli space of conformally inequivalent tori is  the
upper-half plane modded out by the modular group.
This region is called the fundamental domain, ${\cal F}$, of $\tau$ .
The range for the fundamental domain
is normally chosen to be
$${\cal F} = \left\{ -\half\leq {\rm Re}\tau\leq 0,
\vert\tau\vert^2\geq 1\cup 0<{\rm Re}\tau<\half, \vert\tau\vert^2>1\right\}
\,\, .\eq{funddomain}$$
A value of $\tau$ outside of the fundamental domain corresponds to a
torus that is conformally equivalent to another produced by a $\tau$ in the
fundamental domain.
Any value of $\tau$ in the complex plane
outside of the fundamental domain can be transformed,
by a specific element of $PSL(2,\Z)$,
to the inside.

For a consistent string model,
physical quantities, such as amplitudes, must be invariant under
transformations of $\tau$ that produce conformally equivalent tori.
That is, physical quantities must be ``modular invariant''.
This implies the necessity of a modular invariant
partition function, because
the one-loop vacuum-to-vacuum
amplitude, $A$, of a theory
is the integral of the  partition function, $Z(\tau,\bar\tau)$
over the fundamental domain, ${\cal F}$,
$$A= \int\limits_{\cal F} {d\tau d\bar\tau\over({\rm Im}\tau)^2}
Z(\tau,\bar\tau)\,\, .\eq{amplitude}$$

Thus, consistency of a string theory requires that the one-loop partition
function be invariant under both $S$ and $T$ transformations.
Please note that although
invariance of the one-loop partition function under $S$ and $T$ is
necessary for a consistent model, it is not
sufficient.\mpr{alvarez86,kawai87a,antoniadis87, antoniadis88}
Multi-loop partition
functions must also be invariant under generalized
\n modular transformations. Multi-loop
invariance holds if, in addition to invariance at one-loop,
there is invariance under a symmetry that mixes the cycles of
neighboring tori of Riemann surfaces of genus $g>1$ world sheets.
This mixing is generally referred to as a $U$
transformation.
\hfill\vfill\eject

\hbox to 10cm{\hfill}
\vskip 3.5cm
\centertext{Fig.~3.1 Two conformally inequivalent tori}
\vskip 7.5cm
\centertext{Fig.~3.2 Lattice representation of a two-dimensional torus\\
defined by complex number $\tau$}
\hfill\vfill
\centertext{Fig.~3.3 Lattice representation of a two-dimensional torus\\
defined by complex numbers $\lambda_1$ and $\lambda_2$}
\eject

\hbox to 10cm{\hfill}
\vskip 3.5cm
\centertext{Fig.~3.4 The two independent cycles on the torus}
\vskip 7.7cm
\centertext{Fig.~3.5 Transformation of $\tau$ from Dehn twist around the
$a$ cycle}
\hfill\vfill
\centertext{Fig.~3.6 Transformation of $\tau$ from Dehn twist around the
$b$ cycle}
\eject

\hbox to 10cm{\hfill}
\vskip 17.3cm
\centertext{Fig.~3.7 Fundamental domain $\cal F$ in moduli space\\
and its images under $S$ and $T$}
\hfill\vfill\eject

{\hb{3.2}{\bfs Complications for Models Based on General \KM Algebras}}
\sectionnumstyle{arabic}\sectionnum=2\equationnum=0
\vhalf

Complete classification of modular invariant one-loop partition functions
exists only for some of the simplest
conformal field theories, in particular the minimal discrete series with
$c<1$, and the
models based on $SU(2)_K$ Ka\v c-Moody algebras.\mpr{capelli87}
These MIPFs are formed from bilinears of
characters, $\chi_l^{(K)}$, of $SU(2)_K$,
which we label by twice the spin,
$l=2s$, of the corresponding $SU(2)$ representation
($l = 0$ to $K$).\footnote{The values of $l$
correspond to the dimensions of the highest weight representations
(primary fields) meeting unitary conditions
for an $SU(2)_K$ algebra. Throughout this chapter generic highest weight
representations of an  $SU(2)_K$ algebra are denoted by $\Phi_l$ or,
where there will be no confusion,
simply by $l$.  When discussing holomorphic and
anti-holomorphic primary fields, generically $l$ will represent the former
and $\bar l$ the latter.}
These MIPF's
were constructed and found to be in one-to-one correspondence
with the simply-laced Lie algebras:
$$\eqalign{\noindent Z(A_{K+1}) &= \sum_{l = 0}^{K} |\chi_\l|^2\,;\,\, K \geq
1\cr
Z(D_{{K\over 2} + 2}) &= \cases{\sum_{l_{EVEN}=0}^{{K\over 2} -2}
|\chi_l + \chi_{K-l}|^2 + 2|\chi_{{K\over 2}}|^2;\,\, K\in
4 \Z^+\cr
\sum_{l_{EVEN}=1}^{K} |\chi_l|^2 + |\chi_{{K\over
2}}|^2\cr
\, \, \, \, + \sum_{l_{ODD}=1}^{{K\over 2}-2} (\chi_l
\chi_{K-l} + c.c.); \,\, K\in 4 \Z^+ +2\cr}\cr
Z(E_6) &= |\chi_0 + \chi_6|^2 + |\chi_3 + \chi_7|^2 + |\chi_4 +
\chi_{10}|^2 \,; \,\, K = 10\cr
Z(E_7) &= |\chi_0 + \chi_{16}|^2 + |\chi_4 + \chi_{12}|^2 +
|\chi_6 + \chi_{10}|^2 + |\chi_8|^2\cr
       &\,\,\,\,\,\,\,\,\,\, + [(\chi_2 + \chi_{14})\chi_8^* + c.c.]\,;\,\,
K = 16\cr
Z(E_8) &= |\chi_0 + \chi_{10} + \chi_{18} + \chi_{28}|^2 +
|\chi_6 + \chi_{12} + \chi_{16} + \chi_{22}|^2\,; \,\, K =
27\cr}\eqno\eqnlabel{c1}$$
The $D_{{K\over 2}+2}$ partition function is formed from twisting of the
$A_{K+1}$ partition function by the simple current $J_S = (0, K)$ for
$K\in 4\Z^+$ or by the non-simple current
$J_{NS} = (1,K-1)$\footnote[\phantom{4}]{Fields with both holomorphic and
anti-holomorphic components are denoted by either
$(\Phi _l,\bar {\Phi } _{\bar l})$ or $(l,\bar l)$.  In either case
the first element is holomorphic and the second is
anti-holomorphic.  The product of two such fields, resulting from
tensoring $SU(2)_{K_A}$ and $SU(2)_{K_B}$ algebras, is
denoted by $(l_A,\bar l_A;m_B,\bar m_B)$, where $l_A$ and $\bar l_A$ are
holomorphic and antiholomorphic primary fields for the $SU(2)_{K_A}$
algebra. $m_B$ and $\bar m_B$ are to be interpreted
similarly for the $SU(2)_{K_B}$.}
for $K\in 4\Z^+ +2$.\footnote{We follow the standard definition for simple
currents.\mpr{schellekens89c}
  A simple current $J$ is a primary field which when fused with
any other primary field (including itself) $\Phi_l$ of the K-M algebra produces
only one primary
field as a product state:
$$ J\otimes \Phi_l = \Phi_{l'}$$
A non-simple current $J'$, when fused with at least one other primary field
(possibly itself), produces more than one primary field:
$$ J'\otimes \Phi_l = \sum_{l'}\Phi_{l'}$$}

  The exceptional invariants of $E_6$ and
$E_8$ originate via conformal embeddings\mpr{bouwknegt87}
of $A_1 \subset C_2$ and $A_1
\subset G_2$
respectively.  $Z(E_7)$ can be derived by the more involved process of first
conformally embedding  ${SU(2)\over \Z_2} \otimes {SU(3)\over \Z_3}$ in $E_8$,
and then gauging away the $SU(3)\over \Z_3$ contribution.
The reason for the correspondence between $SU(2)_K$ modular invariants and
simply-laced Lie groups is not fully understood.  General arguments have shown
that for any simply-laced Lie group a modular invariant solution can be
constructed for affine ${SU}(2)$ at a specific level.\mpr{kaku91}
But we are not aware of a
complete explanation
as to why these are one-to-one.
Expressing these partition functions in the general form
$$ Z = \sum_{l,\bar l} N_{l,\bar l}\chi_l\bar\chi_{\bar l}\,\, ,
\eq{aorbpartfn} $$
it was realized that: (1) for each MIPF the values of $N_{l,\bar l}$ for
$l=\bar l$ coincide with the exponents of the associated simply-laced Lie
algebra. These exponents give the degree (minus one) of a system of
independent generators of the ring of invariant polynomials in these
algebras; (2) the level $K$ at which a specific modular invariant exists
obeys the rule $K+2=\kappa$, where $\kappa$ is the Coxeter number of the Lie
algebra. Classification of MIPFs for tensor products of $SU(2)_{K_i}$ may
shed more light on the underlying significance of this.

For tensor products of
other theories no procedures have been developed that give all of the
possible
modular invariants, but a few simple algorithms exist for modifying a known
modular invariant to produce another one, in particular the orbifold
construction\mpr{dixon85} and the related
operation of twisting by a simple current.\mpr{schellekens89}
In this chapter,
we make some proposals aimed at the general problem of
classifying all possible modular invariants for conformal field theories
constructed by tensoring together models whose modular invariants are already
known. By a tensor product of two theories, say $A$ and $B$, we mean a theory
whose chiral algebra includes the chiral algebras of both the $A$ and $B$
theories. As a consequence, the central charge of the combined theory will be
the sum of those for the individual factors, the chiral blocks that make up
amplitudes will be constructed from the products of the individual chiral
blocks, and the characters will be
products of the individual characters. Thus the partition function
of the $A\otimes B$ theory is restricted to
the form
$$Z^{AB}=\sum_{l,m,\bar{l},\bar{m}}
N^{AB}_{lm\bar{l}\bar{m}}\chi^A_l\chi^B_m\bar{\chi}^A_{\bar{l}}
\bar{\chi}^B_{\bar{m}}\,\, .\eqno\eqnlabel{zab}$$

The approach taken here derives rules by iteration in the
number of terms in the tensor products,
{\it i.e.}, we consider the conditions placed on higher
order tensor products by the requirements of modular invariance of lower
order tensor products.  We also discuss the degrees of freedom in MIPFs
that remain after these conditions have been applied to the higher order
terms.
For an application of this process, we concentrate
on the specific case of tensor products of two
$SU(2)_{K_i}$ K-M algebras
and their MIPFs.
This is investigated for two reasons: ~(1) for insight
into the density of MIPFs derived by simple currents compared to the
total space of
MIPFs;\footnote{Knowledge of the density of simple current MIPFs will play a
significant role in understanding the total space of MIPFs.  In the last
few years A.N. Schellekens {\it et al.}\mpr{schellekens89} have made
significant progress
towards complete classification of simple current modular
invariants (SCMI's)
for rational conformal field theories (RCFTs). These classifications
appear amenable for
generalization to SCMI's for tensor products of
RCFTs.  Therefore, understanding of the density of
SCMI's compared to the total space of MIPFs is very constructive,
for this will reveal the size of the space of solutions that cannot be found
through Schellekens' approach.}
and (2) as a first step towards developing a
systematic set of rules for constructing MIPFs out of tensor products of
characters for general K-M algebras and minimal models.\mpr{warner90}
The latter issue was first discussed in Ref.~\pr{lewellen}.

Completion of this set of rules generalizes the work
in \pr{kawai87a}, \pr{antoniadis87} and \pr{antoniadis88}, wherein the
process for creating consistent
({\it i.e.}, modular invariant) models from tensor products of Ising
models (the free fermion
approach) is derived.  These papers reveal how an infinite set of
consistent free fermion models can be constructed, with the majority based on
left-right (L-R) asymmetric
modular invariants.  Ref.~\pr{lewellen} suggests that the majority  of
consistent models formed from tensor products of K-M algebras and
minimal models may likewise be L-R asymmetric.  As
with the free fermion models, the L-R asymmetric cases may comprise the
larger, and perhaps more interesting, class of models.

The combined tensor product theory is {\it not} restricted to be simply the
product of the
individual theories; the operators in the combined theory need not
be diagonal ({\it i.e.}, left-right symmetric), and in general the fusion rules
for
the operator products will be modified. The latter point is the chief
complication in the general problem. The allowed tensor product theories built
from free bosons or fermions have been successfully categorized, because the
possible fusion rules in these theories are almost trivial; likewise
twisting a theory by a simple current gives unambiguously a new theory, because
the new fusion rules are unambiguous.
The  difficulty of this
procedure in general (compared to that for the free fermion or boson models)
becomes clear from the transformation properties of the
characters under  $S$ and $T$ transformations,
the generators of the modular group $PSL(2,\Z)$.

For an Ising (free fermion) model, there are three non-zero characters. Each of
these transforms
under  $S$ or $T$ into another one of the three characters
(possibly times a phase):
$$	\eqalign{T:\qquad &\chi\pmatrix{A\cr A\cr} \rightarrow e^{i\pi/24}
\chi\pmatrix{A\cr P\cr}\cr
\qquad &\chi \pmatrix{A\cr P\cr} \rightarrow e^{-i\pi/24} \chi \pmatrix{A\cr
A\cr}\cr
\qquad &\chi\pmatrix{P\cr A\cr} \rightarrow e^{i\pi/12} \chi\pmatrix{P\cr
A\cr}\cr
\qquad & \cr
S:\qquad &\chi \pmatrix{A\cr A\cr} \rightarrow \chi \pmatrix{A\cr A\cr}\cr
\qquad & \chi \pmatrix{P\cr A\cr} \rightarrow \chi \pmatrix{A\cr P\cr}\cr
\qquad & \chi \pmatrix{A\cr P\cr} \rightarrow \chi \pmatrix{P\cr A\cr}\cr}
\eqno\eqnlabel{c2}$$
where,
$$
\chi \pmatrix{A\cr A\cr} = \chi_0 + \chi_{1/2}\, , \,\,
\chi \pmatrix{A\cr P\cr} = \chi_0 - \chi_{1/2}\, , \,\,
 {\rm ~and~}\,\,
\chi \pmatrix{P\cr A\cr} = \sqrt{2} \chi_{1/16}\, ,
\eq{chisttrans}$$
Here $\chi_i, i = 0, 1/16, 1/2$ are the characters of the primary
fields of conformal dimension $h = 0$,
$1/16$, $1/2$ in the $c = 1/2$ critical Ising model.
$P(A)$ denotes (anti-)periodic
boundary conditions around one of the two non-contractible
loops of the world sheet torus.\mpr{kaku91}  In this case, the $S$ and $T$
transformations act on the
characters in the manner of generic simple currents denoted by
$J_S$ or $J_T$,
respectively, twisting the corresponding primary states $\Phi_i$.
In other words, the outcome of the
transformation or fusion is, respectively, a single character or primary
field:

{\settabs 7\columns
\+ &&  $S:\quad\chi_i \rightarrow \chi_{j}$
   &$\,\,$ ;& $T:\quad\chi_i \rightarrow \chi_{k}$\cr
\+ &&  $J_S \otimes \Phi_i = \Phi_{j}$
   &$\,\,$ ;& $J_T \otimes \Phi_i = \Phi_{k}\,\, .$\cr}

However, in the generic case for a K-M algebra or minimal model, $S$
transforms a character $\chi_l$
in the manner of a
non-simple current, $J_{NS}$, acting on a generic primary field $\Phi_l$.
The outcome of the transformation or fusion is in general is not a single
term, but a sum of terms:
$$\eqalign {S: ~\chi_l &\Rightarrow \sum_{l'} S(l,l') \chi_{l'}\cr
J_{NS} \otimes \Phi_l &= \sum_{l'} N_{J_{NS}l}^{l'}\Phi_{l'}}\,\, .
\eqno\eqnlabel{c4}$$
(As shown by E. Verlinde\mpr{verlinde88}, the (positive integer) coefficients
$N_{J_{NS}l}^{l'}$ are related to the matrix elements of the $S$
transformation matrix:
$$ N_{J_{NS}l}^{l'}= \sum_{n} {S(J_{NS},n)~S(l,n)~S(l',n)\over S(0,n)}
\,\, ,\eq{verlindeneq}$$
where $0$ denotes the vacuum state or identity field.) The complicated
transformations of tensor products of generic K-M characters $\chi_l$
make
complete classification of associated MIPFs much more difficult than in the
free fermion or boson case.

    In section 3.3 we consider the extent to which the integer coefficients
$N^{AB}_{lm\bar{l}\bar{m}}$ in the partition function of the tensor product
theory are constrained once we know all of the allowed possibilities for the
corresponding coefficients $N^A_{l\bar{l}}$ and $N^B_{m\bar{m}}$ in the factor
theories. In section 3.4 we investigate the more general problem of combining
theories whose holomorphic and anti-holomorphic degrees of freedom
need not possess the same chiral algebras. That is, we consider
partition functions of the form,
$Z^{AB}=\sum_{l,\bar{m}}N_{l\bar{m}}\chi^A_l\bar{\chi}^B_{\bar{m}}$.
In the following
sections we are interested ultimately
in classifying consistent conformal field theories, not just modular invariant
combinations of characters. Accordingly, we invoke consistency
conditions for amplitudes on the plane when they constrain
the states that can appear in the partition function.

\sectionnumstyle{blank}\vhalf
{\hb{3.3}{\bfs Constraints on Tensor Product Modular Invariants}}\vhalf
\sectionnumstyle{arabic}\sectionnum=3\equationnum=0

   In order for the tensor product partition function (\puteqn{zab}) to be
invariant
under the generators of modular transformations, $T$ and $S$,  we must have,
$$\eqalign{T\quad{\rm invariance:}\quad &h_l+h_m=h_{\bar{l}}+
h_{\bar{m}}\pmod{1}\quad {\rm if}\quad
N^{AB}_{lm\bar{l}\bar{m}}\ne 0\cr
S\quad{\rm invariance:}\quad &N^{AB}_{lm\bar{l}\bar{m}}=\sum_{l^\prime,
m^\prime,\bar{l}^\prime,\bar{m}^\prime} N^{AB}_{l^\prime m^\prime
\bar{l^\prime}\bar{m^\prime}}S^A_{ll^\prime}S^B_{mm^\prime}
\bar{S}^A_{\bar{l}\bar{l}^\prime}\bar{S}^B_{\bar{m}\bar{m}^\prime}
\,\, ,\cr}\eqno\eqnlabel{minv}$$
where $h_l$ denotes the conformal dimension of the primary field represented
by the label $l$, {\it etc.}
   We assume that the solutions to the corresponding equations for the
factor theories are known. That is, we are given all possibilities
(labeled by $i$)
for non-negative integer coefficients $N^{A,i}_{l\bar{l}}$ such that,
$$\eqalign{&h_l=h_{\bar{l}}
\pmod{1}\quad {\rm if}\quad N^{A,i}_{l\bar{l}}\ne 0\cr
{\rm and}\quad\quad&N^{A,i}_{l\bar{l}}=\sum_{l^\prime,
\bar{l}^\prime} N^{A,i}_{l^\prime  \bar{l^\prime}}S^A_{ll^\prime}
\bar{S}^A_{\bar{l}\bar{l}^\prime}\,\, ,\cr}\eqno\eqnlabel{aminv}$$
and similarly for $N^{B,j}_{m\bar{m}}$.  We can get relations between the
integer coefficients in equations (\puteqn{minv}) and (\puteqn{aminv})
by multiplying (\puteqn{minv}) by $N^{A,i}_{l\bar{l}}$ and summing
over $l$ and $\bar{l}$,
$$\eqalign{\sum_{l,\bar{l}}N^{A,i}_{l\bar{l}}N^{AB}_{lm\bar{l}\bar{m}}
&=\sum_{l,\bar{l},l^\prime,
m^\prime,\bar{l}^\prime,\bar{m}^\prime}N^{A,i}_{l\bar{l}}
 N^{AB}_{l^\prime m^\prime
\bar{l^\prime}\bar{m^\prime}}S^A_{ll^\prime}S^B_{mm^\prime}
\bar{S}^A_{\bar{l}\bar{l}^\prime}\bar{S}^B_{\bar{m}\bar{m}^\prime}\cr
&=\sum_{l^\prime,
m^\prime,\bar{l}^\prime,\bar{m}^\prime}N^{A,i}_{l^\prime\bar{l^\prime}}
 N^{AB}_{l^\prime m^\prime
\bar{l^\prime}\bar{m^\prime}}S^B_{mm^\prime}
\bar{S}^B_{\bar{m}\bar{m}^\prime}
\,\, ,\cr  }\eqno\eqnlabel{amom}$$
where we have used (\puteqn{aminv}) and the symmetry of $S$ to
simplify the right-hand side. The resulting
equation is precisely of the form (\puteqn{aminv}) for the $B$ theory,
therefore we must have,
\subon
$$\sum_{l,\bar{l}}N^{A,i}_{l\bar{l}}N^{AB}_{lm\bar{l}\bar{m}}
  =\sum_jn^{A,i}_j N^{B,j}_{m\bar{m}}\,\, ,\eqno\eqnlabel{constr-a}$$
where $n^{A,i}_j$ are integers.
This constrains some combinations of coefficients in the $AB$ theory to be
linear combinations (with integer coefficients) of the  allowed
coefficients in the $B$ theory, which are presumed known. There is an analogous
constraint arising from taking the appropriate traces over the $B$ theory
indices in (\puteqn{minv}),
$$\sum_{m,\bar{m}}N^{B,j}_{m\bar{m}}N^{AB}_{lm\bar{l}\bar{m}}
  =\sum_j n^{B,j}_i N^{A,i}_{l\bar{l}}\,\, ,\eqno\eqnlabel{constr-b}$$
and a further constraint arising from taking
appropriate
traces over both sets of indices in either possible order,
$$\eqalignno{
\sum_{l,\bar{l}\atop m,\bar m}
N^{B,j}_{m\bar{m}}N^{A,i}_{l\bar{l}}
N^{AB}_{lm\bar{l}\bar{m}}
&  = \sum_{m,\bar m}N^{B,j}_{m,\bar m}
\left( \sum_j n^{A,i}_j N^{B,j}_{m\bar{m}}\right)\,\, ,
& \eqnlabel{constr-c}\cr
&  = \sum_{l,\bar l}N^{A,i}_{l,\bar l}
\left( \sum_i n^{B,j}_i N^{A,i}_{l\bar{l}}\right)\,\, .
& \eqnlabel{constr-d}}$$
\suboff
Note that the number
of constraint equations increases as the factor theories become more complex
(in the sense of having more possible modular invariants), and also as the
tensor products theories have more factors.

    These equations constrain part of the
operator content of the tensor product theories, which we wish to classify.
Often, this information, together with some simple consistency requirements
for
conveniently chosen amplitudes on the plane, serves to completely determine
the
allowed possibilities for the tensor product modular invariants. For
concreteness, we illustrate with a simple example.

\vhalf
{\hc{3.3.a} {\sl Example: $SU(2)_{K_A}\otimes SU(2)_{K_B}$ Tensor Product
Theories.}
\vhalf

  $SU(2)_{K}$ has $K+1$ unitary primary fields, which we label by
twice the spin, $l=2s$, of the corresponding SU(2) representation.
Their conformal dimensions are $h_l= {{l(l+2)}\over 4(K+2)}$.
The matrix $S$,
implementing the modular transformation $\tau\rightarrow-1/\tau$
on the Ka\v c-Moody characters, is
$$S^K_{ll^\prime}=\left(2\over K+2\right)^{1/2}\sin\left(\pi (l+1)(l^\prime +1)
\over K+2\right)\,\, .\eqno\eqnlabel{ssu}$$
The fusion rules, which we will make use of
momentarily, are
$$\phi_l\times\phi_{l^\prime}=\sum^{{\rm min}(l+l^\prime,2K-l-l^\prime)}_{
{m=\vert l-l^\prime\vert \atop m-\vert l-l^\prime\vert\,\,{\rm even}}}
\phi_m \,\, .\eqno\eqnlabel{fra}$$

   For simplicity we only consider the tensor product
theories with holomorphic and anti-holomorphic chiral algebras
SU(2)$_{K_A}\otimes$SU(2)$_{K_B}$ for both $K_A$ and $K_B$ odd. Then the only
possible modular invariants for the factor theories are the diagonal ones,
$N_{l\bar{l}}=\delta_{l\bar{l}}$.\footnote{For our purposes we need to
consider all sets of
non-negative integer coefficients $N_{l\bar{l}}$ that give rise to $S$ and
$T$ invariant partition functions, but not necessarily only ones
with a unique vacuum state ($N_{00}=1$).
Relaxing this condition in the SU(2) case
does not expand the space of possible solutions, aside from a trivial
multiplicative constant.} Applying the constraint equations
(\puteqn{constr-a}-d) gives the conditions,
$$\eqalign{\sum_{l=\bar{l}}
N^{AB}_{lm\bar{l}\bar{m}}&=a\delta_{m\bar{m}}\,\, ;\quad a\in \Z^+\cr
 \sum_{m=\bar{m}} N^{AB}_{lm\bar{l}\bar{m}}&=b\delta_{l\bar{l}}\,\, ;\quad\quad
b\in \Z^+\cr
\sum_{l=\bar{l}}\sum_{m=\bar{m}}
N^{AB}_{lm\bar{l}\bar{m}}&=a(K_B+1)=b(K_A+1)\,\, .
\cr}\eqno\eqnlabel{nabcon}$$

    If we label the primary operators in the tensor product theory by the
corresponding $l$ values of the factor theories, {\it e.g.},
$(l,m\vert\bar{l},\bar{m})$, then the integer $a$ is equal, in particular, to
the number of primary operators in the theory of the form $(j,0\vert j,0)$.
These are pure $A$ theory operators and so must form a closed operator
subalgebra of the $A$ theory. Similarly, $b$ must equal the dimension of some
closed operator algebra in the $B$ theory. This is useful because we know
(from studying the consistency of amplitudes on the plane) all
consistent closed operator sub-algebras of
SU(2) Ka\v c-Moody theories.\mpr{christe87}  For $K$
odd these sub-algebras (labeling them by their dimensions, $d$) are
$$      \eqalign{d = 1: &~~\{\Phi_0\} {\rm \, \, (the~identity)}\cr
d = 2: &~~\{\Phi_0,\, \,  \Phi_{K}\}\cr
d = {K + 1\over 2}: &~~\{\Phi_l;~~0\leq {\rm ~even~}l\le K\}{\rm
{}~(the~allowed~integer~spin~representations)}\cr
d = K + 1: &~~\{\Phi_l;~0\leq l\leq K\}\cr}\,\, .\eqno\eqnlabel{clsa}$$
Thus, in the tensor product theory we know all of the possibilities for
operators of the form $(j,0\vert j,0)$ or $(0,j\vert 0,j)$. Given
(\puteqn{nabcon})  and
the uniqueness of the vacuum state $(0,0\vert 0,0)$ in the tensor product
theory, the multiplicities of the operators in the closed sub-algebras must be
as given in (\puteqn{clsa}).

   We can now write down all of the possibilities for $a,b,K_A$ and $K_B$
that are consistent with (\puteqn{nabcon})  and (\puteqn{clsa}),
and consider each type of tensor product modular invariant individually:

\vhalf
\noindent (1) $a = K_A +1, ~~b = K_B+1$ :  \hskip 1.5em
Here we have $N^{AB}_{l\bar l m\bar m} =
\delta_{l\bar l} \delta_{m\bar m} + M^{AB}_{l \bar l m \bar m}$,
with $M^{AB}_{l \bar l m \bar m}$  traceless with respect to both $l, \bar l$
and $m, \bar m$. It is easy to see that $M^{AB}$ must in fact vanish, leaving
us with the simple uncorrelated tensor product of the SU(2)$_{K_A}$ and
SU(2)$_{K_B}$ diagonal modular invariants. Were this not the case, then
$M^{AB}$ by itself would give rise to a modular invariant which did not include
the term containing the identity operator. But this is not possible since
(from (\puteqn{ssu})  and quite generally
in a unitary theory) $S_{0l}>0$ for all $l$.

\vhalf
\noindent (2) $a = {K_A + 1\over 2}, ~~b = {K_B +
1\over2}$ :  \hskip 1.5em
In this case the operators diagonal in either of the factor theories comprise
the set $\{(l,m\vert l,m)\}$ with $l$ and $m$ both odd or both even. This set
contains the operators $(1,K_B\vert 1,K_B)$ and $(K_A,1\vert K_A, 1)$.
The non-diagonal operators in the theory, $(i,j\vert m,l)$ must have a
consistent operator product with these two operators, in particular at least
some of the operators appearing in the naive fusion with them (using the rules
(\puteqn{fra}) must have integer spins ($h-\bar{h}\in \Z$).
This restricts
the non-diagonal operators $(i,j\vert m,l),~i\ne m,~j\ne l$, to those
satisfying $i+m=K_A$ and $j+l=K_B$. For these operators, in turn, to have
integer spin we have either:  $i-j$ even and $K_A+K_B=0 \pmod{4}$; or
$i-j$ odd and $K_A-K_B=0 \pmod{4}$. Taking all such operators,
the former case gives the modular invariant obtained from the
simple tensor product invariant of case (1) by twisting by the simple current
$(K_A,K_B\vert 0,0)$; the latter is obtained by twisting by $(K_A,0\vert
0,K_B)$. An extension of the argument given in (1) using the fact that
$S^{K_A}_{iK_A}S^{K_B}_{jK_B} >0$ for all $i-j$ even, shows that these are the
only possibilities in this category.

\vhalf
\noindent (3) $a=1,~b=2,~K_B=2K_A+1$ or
  $a=2,~b=1,~K_A=2K_B+1$:  \hskip 1.5em
Take $a=1$, $b=2$ so $K_B=2K_A+1$. The model must include
the states $(0,0\vert 0,0)$ and $(0,K_B\vert 0,K_B)$ but no other states of the
form $(0,l\vert 0,l)$, $(i,0\vert i,0)$ or $(j,K_B\vert j,K_B)$.
There must also be two states of
the form $(K_A,j\vert K_A,j)$. Demanding that the fusion products of these
states with themselves are consistent with the above restriction requires
$j=0$, or $K_B$, but then the states themselves are inconsistent with the
restriction. Thus there are no possible consistent theories within this
category.

\vhalf
\noindent (4) $a=2,~b={K_B+1\over 2},~K_A=3$ or
 $a={K_A+1\over 2},~b=2,~K_B=3$: \hskip 1.5em
This case differs from case (2) with $K_A=3$ and/or $K_B=3$,
in that the $d=2$ closed subalgebra of the SU(2)$_3$ theory consists of
$\{\Phi_0,\Phi_3\}$  instead of $\{\Phi_0,\Phi_2\}$ as in (2). If $a=2$,
$b={K_B+1\over 2}$ and $K_A=3$, then the operators diagonal in either factor
theory comprise the set $\{ (0,l\vert 0,l), (3,l\vert 3,l)~l$ even; $(1,j\vert
1,j), (2,j\vert 2,j)~j$ odd$\}$. There must be additional non-diagonal
operators, $(i,j\vert l,m)$ $i\ne l$, $j\ne m$, if there are to be any modular
invariants in this category. If \hbox {$(i,j\vert l,m)$} appears then
\hbox {$(3-i,j\vert 3-l,m)$}
appears also. For both operators to have integer spin, $i$ and $l$ must be both
even or both odd. Thus there must be operators of the form $(0,j\vert 2,m)$
or $(1,p\vert 3,l)$. Fusing these
with the operators
\hbox {$(1,K_B\vert 1,K_B)$}
from
the diagonal part of the theory produces the operators
\hbox {$(1,K_B-j\vert 1,K_B-m)$}
and/or
\hbox {$(1,K_B-j\vert 3,K_B-m)$} and \hbox {$(0,K_B-p\vert 2,K_B-l)$}
and/or
\hbox {$(2,K_B-p\vert 2,K_B-l)$,}
respectively.
It is easy to see that if the former
fields have integer spin then none of the possible fusion products do. Thus,
there can be no consistent theories in this category.

\vhalf
\noindent (5) $ K_A=K_B\equiv K $, $a= b= 1$ or $a = b = 2$ : \hskip 1.5em
The situation
becomes more complicated for $K_A = K_B \equiv K$.  For these cases we have
additional trace equations,
$$\eqalign{\sum_{l=\bar{m}} N^{AB}_{lm\bar{l}\bar{m}}\, &=a^\prime
\delta_{m\bar{l}}\,\, ;\quad\quad
a^\prime\in \Z^+\cr
 \sum_{m=\bar{l}} N^{AB}_{lm\bar{l}\bar{m}}\, &=b^\prime
\delta_{l\bar{m}}\,\, ;\quad\quad b^\prime\in \Z^+\,\,
.\cr}\eqno\eqnlabel{morcon}$$
If the values of $a^\prime$ and $b^\prime$ correspond to any of cases
(1)---(4),
then the invariants are precisely as given above, with the factor
theories permuted. Thus we only need to consider the cases:
(5a) $a=b=a^\prime=b^\prime=1$, (5b) $a=b=a^\prime=b^\prime=2$,
and (5c) $a=b=1,~
a^\prime=b^\prime=2$. Case (5b) is most quickly disposed of. The operators in
the theory include the closed subalgebra $\{(0,0\vert 0,0),(0,K\vert 0,K)\}$,
$\{(0,0\vert 0,0),(K,0\vert K,0)\}$, $\{(0,0\vert 0,0),(K,0\vert 0,K)\},$ and
$\{(0,0\vert 0,0),(0,K\vert K,0)\}$. For the operator algebra with these
together to be closed the chiral fields $(K,K\vert 0,0)$ and $(0,0\vert K,K)$
must appear, but for K odd these do not have integer conformal dimension.
Therefore this case is ruled out.

    Cases (5a) and (5c) can also be ruled out as follows. In both cases there
must be fields $(1,j\vert 1,j)$ and $(p,1\vert p,1)$ with a single choice for
$j$ and $p$ in each case. Consider the four-point correlation function on the
plane $\langle (1,j\vert 1,j)(1,j\vert 1,j)(p,1\vert p,1)(p,1\vert p,1)
\rangle$. In one channel the only possible intermediate state primary fields
that can appear, consistent with the restrictions of cases (5a) or (5c), are
$(0,0\vert 0,0)$ and $(2,2\vert 2,2)$. In the cross channels only a subset of
the states of the form $(p\pm 1,j\pm 1\vert p\pm 1,j\pm 1)$ can appear as
intermediates. We know from the four-point amplitudes $\langle (1\vert 1)
(1\vert 1) (j\vert j) (j\vert j)\rangle$ and $\langle (1\vert 1)
(1\vert 1) (p\vert p) (p\vert p)\rangle$ in the factor theories that the chiral
blocks making up the amplitudes have two-dimensional monodromy, so the
blocks appearing in the tensor product theory must have four-dimensional
monodromy.
There is no way, then, to assemble the two chiral blocks corresponding to the
allowed intermediate primaries $(0,0\vert 0,0)$ and $(2,2\vert 2,2)$ in such a
way that the four-point function in the tensor product theory can be monodromy
invariant ({\it i.e.}, single-valued).

    To summarize: We have used the constraints
(\puteqn{constr-a}-\puteqn{constr-d}) and the
consistency of
conveniently chosen fusion rules and amplitudes to find the only consistent
tensor product theories of the type $SU(2)_{K_A}\otimes SU(2)_{K_B}$, with
$K_A,K_B$ odd. These turn out to be the simple (uncorrelated) product of the
diagonal invariants of the factor theories and all theories obtained from
them by twisting by the allowed simple current fields that can be built
from the identity and fields labeled $K_A$ and $K_B$.

\sectionnumstyle{blank}\vhalf
{\hb{3.4}{\bfs Left-Right Asymmetric Modular Invariants}}\vhalf
\sectionnumstyle{arabic}\sectionnum=4\equationnum=0

    So far we have considered tensor product conformal field theories that
are diagonal in the sense that for each holomorphic conformal field theory
factor there is a corresponding anti-holomorphic conformal field theory factor
with an isomorphic chiral algebra.
While these are the relevant theories to consider
for statistical mechanics applications, it is natural in the construction of
heterotic string theories to consider conformal field theories that are
inherently left-right asymmetric as well. For these, the methods discussed
above
do not apply. Nonetheless, we can exploit known properties of left-right
symmetric conformal field theories to construct modular invariants even for
inherently asymmetric theories by using the following result: Given two
consistent diagonal rational conformal field theories ({\it a priori}
with different
chiral algebras) with modular invariant partition functions
$Z^A=\sum\chi^A_i\bar{\chi}^A_i$ and $Z^B=\sum\chi^B_i\bar{\chi}^B_i$,
the left-right asymmetric partition function given by
$Z^{A\bar{B}}=\sum\chi^A_i\bar{\chi}^B_i$ will be modular invariant
if and only if: (1) the conformal dimensions
agree modulo 1, or more precisely
$h^A_i-c^A/24=h^B_i-c^B/24\pmod{1}$; and (2) the
fusion rules of the two theories coincide, $\phi^A_i\times \phi^A_j=\sum_k
N_{ij}^k\phi^A_k$ and $\phi^B_i\times \phi^B_j=\sum_k N_{ij}^k\phi^B_k$.

   Condition (1) is obviously necessary and sufficient for $Z^{AB}$ to be
$T$ invariant. Condition (2) is almost immediate given Verlinde's
results.\mpr{verlinde88} $Z^{AB}$ is invariant under the $S$
transformation only if the
$S$ matrices implementing the modular transformations on the characters of the
$A$ and $B$ theories coincide, $S^A_{ij}=S^B_{ij}$. As Verlinde showed, the
fusion rule coefficients determine the $S$ matrix,\footnote{To be precise,
Verlinde showed that the eigenvalues, $\lambda_i^{(j)}$, of the matrices
$(N_i)_l^{\ k}$ satisfy
$\lambda_i^{(j)}=S_{ij}/S_{0j}$ but there could be an ambiguity in the
 choice of superscript $(j)$ labeling each member of the set
of eigenvalues of $(N_i)_l^{\ k}$. We believe in the present case that this
ambiguity is fixed given $T$ and the requirement $(ST)^3=1$, but have no
proof.} so condition (2) is required
for $S^A=S^B$. In employing this relation it is crucial to define the primary
fields with respect to the full $A$ and $B$ chiral algebras.

     As a simple example, consider the theories $A= SO(31)$ level-one,
and $B= (E_8)$ level-two, both with central charge $c=31/2$.
The consistent diagonal theories have partition functions,
$$Z^A=
\chi_0\bar{\chi}_0+ \chi_{{1\over 2}}\bar{\chi}_{{1\over 2}}+
\chi_{{31\over 16}}\bar{\chi}_{{31\over 16}}\eq{so31pf}$$
and
$$Z^B=
\chi_0\bar{\chi}_0+ \chi_{{3\over 2}}\bar{\chi}_{{3\over 2}}+
\chi_{{15\over 16}}\bar{\chi}_{{15\over 16}}\,\, ,\eq{e8pf}$$
where the characters are labeled
by the conformal dimensions of the associated primary fields.
The fusion rules in
both theories are analogous to those in the Ising model.
The asymmetric partition function,
$$Z^{AB}=
\chi^A_0\bar{\chi}^B_0+ \chi^A_{{1\over 2}}\bar{\chi}^B_{{3\over 2}}+
\chi^A_{{31\over 16}}\bar{\chi}^B_{{15\over 16}}\,\, ,\eq{asymppf}$$
satisfies conditions (1)
and (2), and so is itself a modular invariant. This can
also be constructed by choosing appropriate boundary conditions for a
collection of 31 free, real fermions.

    A more interesting example, which cannot be constructed from free bosons or
fermions or by twisting a known invariant by a simple current, is the
following. For the $A$ theory we take the simple tensor product of the
diagonal
theories for G$_2$ level-one and $SU(3)$ level-two;
for the $B$ theory the simple
tensor product of the diagonal theories for F$_4$ level-one and the three
state
Potts model. The central charges coincide: $c^A=14/5+16/5=6$;
$c^B=26/5+4/5=6$.
The primary fields appearing in each theory are, for G$_2$ level-one
the identity and 7 ($h={2\over 5}$); for $SU(3)$ level-two the identity, 3 and
$\bar{3}$ ($h={4\over 15}$), 6 and ${\bar 6}$ ($h={2\over 3}$), and 8
($h={3\over 5}$); for F$_4$ level-one the identity and 26
($h={3\over 5}$); and for
the Potts model, the primaries, labeled by their conformal dimensions, are 0,
${2\over 5}$, ${2\over 3}$, ${\bar {2\over 3}}$, ${1\over 15}$, and
${\bar {1\over 15}}$.

To economically list the fusion rules for these theories we can simply list
the non-vanishing three-point amplitudes (where we represent the field by its
conformal dimension). Besides the obvious ones involving the identity operator
($\langle \phi\bar{\phi} 0\rangle$) these are: for G$_2$ level-one
$\langle {2\over 5}, {2\over 5}, {2\over 5}\rangle$; for $SU(3)$ level-two,
$\langle
{4\over 15},{4\over 15},{4\over 15}\rangle$, $\langle {4\over 15},{4\over 15},
\bar{{2\over 3}}\rangle$, $\langle
{4\over 15},\bar{{4\over 15}},{3\over 5}\rangle$, $\langle {4\over 15},{3\over
 5},{2\over 3}\rangle$, $\langle
{2\over 3},{2\over 3},{2\over 3}\rangle$, $\langle {3\over 5},{3\over 5},
{3\over 5}\rangle$, and the conjugates of
these; for F$_4$ level-one, $\langle {3\over 5},{3\over 5},{3\over 5}\rangle$;
 and for the three
state Potts model, $\langle
{1\over 15},{1\over 15},{1\over 15}\rangle$, $\langle {1\over 15},{1\over 15},
\bar{{2\over 3}}\rangle$, $\langle
{1\over 15},\bar{{1\over 15}},{2\over 5}\rangle$, $\langle {1\over 15},{2\over
 5},{2\over 3}\rangle$, $\langle
{2\over 3},{2\over 3},{2\over 3}\rangle$, $\langle {2\over 5},{2\over 5},
{2\over 5}\rangle$, and the conjugates of
these. All of the non-zero fusion rule coefficients, $N_{ijk}$, for these four
theories are equal to one.

Given the obvious similarities of the fusion rules and conformal dimensions
of these theories, it is not difficult to verify that the asymmetric partition
function given by,
$${\eqalign{Z^{AA^{\prime}BB^{\prime}}=
&\chi^A_{0}\chi^{A^{\prime}}_{0}\bar{\chi}^B_{0}
\bar{\chi}^{B^{\prime}}_{0}+\chi^A_{0}\chi^{A^{\prime}}_{{3\over 5}}
\bar{\chi}^B_{{3\over 5}}
\bar{\chi}^{B^{\prime}}_{0}+\chi^A_{0}\chi^{A^{\prime}}_{{2\over 3}}
\bar{\chi}^B_{0}
\bar{\chi}^{B^{\prime}}_{{2\over 3}}+\chi^A_{0}\chi^{A^{\prime}}_{\bar{{2\over
 3}}}
\bar{\chi}^B_{0}\bar{\chi}^{B^{\prime}}_{\bar{{2\over 3}}}+\cr &\chi^A_{0}
\chi^{A^{\prime}}_{{4\over 15}}\bar{\chi}^B_{{3\over 5}}
\bar{\chi}^{B^{\prime}}_{\bar{{2\over 3}}}+\chi^A_{0}\chi^{A^{\prime}}_{
\bar{{4\over 15}}}
\bar{\chi}^B_{{3\over 5}}\bar{\chi}^{B^{\prime}}_{{2\over 3}}+\chi^A_{{2\over
 5}}
\chi^{A^{\prime}}_{0}\bar{\chi}^B_{0}\bar{\chi}^{B^{\prime}}_{{2\over 5}}+
\chi^A_{{2\over 5}}\chi^{A^{\prime}}_{{3\over 5}}\bar{\chi}^B_{{3\over 5}}
\bar{\chi}^{B^{\prime}}_{{2\over 5}}+\cr &\chi^A_{{2\over 5}}
\chi^{A^{\prime}}_{{2\over 3}}
\bar{\chi}^B_{0}\bar{\chi}^{B^{\prime}}_{\bar{{1\over 15}}}+\chi^A_{{2\over 5}}
\chi^{A^{\prime}}_{\bar{{2\over 3}}}\bar{\chi}^B_{0}\bar{\chi}^{B^{\prime}}_{
{1\over 15}}+
\chi^A_{{2\over 5}}\chi^{A^{\prime}}_{{4\over 15}}\bar{\chi}^B_{{3\over 5}}
\bar{\chi}^{B^{\prime}}_{{1\over 15}}+\chi^A_{{2\over 5}}\chi^{A^{\prime}}_{
\bar{{4\over 15}}}
\bar{\chi}^B_{{3\over 5}}\bar{\chi}^{B^{\prime}}_{\bar{{1\over 15}}}\,\,
,\cr}}\eq{gsfp1}
 $$
satisfies the two conditions for modular invariance.
Here $A$, $A^{\prime}$, $B$, and $B^{\prime}$ denote the G$_2$, $SU(3)$,
F$_4$,
and Potts theories, respectively.

An alternative sewing of the operators in these four conformal
field theories gives rise to the diagonal E$_6$ level-one modular invariant
$${\eqalign{Z^{{\rm E}_6}=
&\chi^A_{0}\chi^{A^{\prime}}_{0}\bar{\chi}^B_{0}
\bar{\chi}^{B^{\prime}}_{0}+\chi^A_{{2\over 5}}\chi^{A^{\prime}}_{{3\over 5}}
\bar{\chi}^B_{0}
\bar{\chi}^{B^{\prime}}_{0}+\chi^A_{0}\chi^{A^{\prime}}_{0}\bar{\chi}^B_{
{3\over 5}}
\bar{\chi}^{B^{\prime}}_{{2\over 5}}+\chi^A_{{2\over 5}}\chi^{A^{\prime}}_{
\bar{{3\over 5}}}
\bar{\chi}^B_{{3\over 5}}\bar{\chi}^{B^{\prime}}_{{2\over 5}}+\cr &\chi^A_{0}
\chi^{A^{\prime}}_{{2\over 3}}\bar{\chi}^B_{0}
\bar{\chi}^{B^{\prime}}_{\bar{{2\over 3}}}+\chi^A_{0}\chi^{A^{\prime}}_{\bar{
{2\over 3}}}
\bar{\chi}^B_{0}\bar{\chi}^{B^{\prime}}_{{2\over 3}}+\chi^A_{0}
\chi^{A^{\prime}}_{{2\over 3}}\bar{\chi^B_{{3\over 5}}}\bar{\chi}^{
B^{\prime}}_{{1\over 15}}+
\chi^A_{0}\chi^{A^{\prime}}_{\bar{{2\over 3}}}\bar{\chi}^B_{{3\over 5}}
\bar{\chi}^{B^{\prime}}_{\bar{{1\over 15}}}+\cr &\chi^A_{{2\over 5}}\chi^{A^{
\prime}}_{{4\over 15}}
\bar{\chi}^B_{{3\over 5}}\bar{\chi}^{B^{\prime}}_{\bar{{1\over 15}}}+\chi^A_{
{2\over 5}}
\chi^{A^{\prime}}_{\bar{{4\over 15}}}\bar{\chi}^B_{{3\over 5}}\bar{\chi}^{B^{
\prime}}_{{1\over 15}}+
\chi^A_{{2\over 5}}\chi^{A^{\prime}}_{{4\over 15}}\bar{\chi}^B_{0}
\bar{\chi}^{B^{\prime}}_{{2\over 3}}+\chi^A_{{2\over 5}}\chi^{A^{\prime}}_{
\bar{{4\over 15}}}
\bar{\chi}^B_{0}\bar{\chi}^{B^{\prime}}_{\bar{{2\over 3}}}\,\, .\cr}}
\eq{gsfp2} $$

It is natural to suppose that the asymmetric modular invariant,
\pe{gsfp1}, can be
obtained from the symmetric one, \pe{gsfp2},
by twisting by the appropriate field or
fields.
This intuition is correct, but the twisting is not by a simple current
operator, and correspondingly there is no definite algorithm for achieving
it. In the symmetric theory the chiral algebra is enlarged (to
E$_6\otimes$E$_6$). Twisting by a simple current\mpr{schellekens89c}
cannot reduce the chiral algebra, and here gives back the same
theory. There is, however, a candidate field that is primary under the {\it
smaller} chiral algebra of the asymmetric theory, and which has simple fusion
rules when defined with respect to this algebra, namely the field $(0,{2\over
3}\vert 0,{2\over 3})$. Twisting $Z^{E_6}$ by this
operator, that is throwing
out those operators which when fused with
$(0,{2\over 3}\vert 0,{2\over3})$ give
$T$ noninvariant states while adding those $T$ invariant operators which
result from fusing, gives only a subset of the characters in the
asymmetric theory. To
get the full set we must add the operators formed by fusing $({2\over
5},{3\over 5}\vert {3\over 5},{2\over 5})$ with itself under the now modified
fusion rules of the new theory ({\it i.e.,} those preceding eq.~\pe{gsfp1}),
(which {\it a priori} is an ambiguous procedure).
Similarly, twisting the asymmetric invariant by any combinations of simple
currents in that theory gives back the same invariant.  In order to obtain
$Z^{E_6}$ we have to twist
by the non-simple current $({2\over 5},{3\over 5}\vert
0,0)$, with suitably modified fusion rules, which again is an ambiguous
procedure.
\hfill\vfill\eject

{\hb{3.5}{\bfs Concluding Comments on MIPFs}}
\vhalf

The techniques introduced in sections 3.3 and 3.4
make the classification of modular
invariants for tensor product theories built from a small number of factors
feasible. A complete classification of the invariants for
$SU(2)_{K_A}\otimes SU(2)_{K_B}$ theories, that is the straightforward
extension of the results of section 3.2 to even $K$, may, in particular, prove
interesting if there is some generalization of the ADE classification found for
the single theories. Nonetheless, a complete classification for tensor product
theories built with many factors is not likely to be found,
given the enormous number of
possibilities. For the purposes of string model building a procedure for
constructing any new class of invariants,
such as the one known for free field
constructions, would be progress. Perhaps a generalization of the twisting
procedure to operators with nontrivial (or altered) fusion rules, as suggested
by the example in section 3.3, could produce one. In this regard, the results
of \pr{warner90, roberts92},
(which have been extensively exploited recently by Gannon\mpr{gannon92})
are an intriguing step though not totally satisfactory. In these works,
new tensor product modular invariants
are obtained by shifting the momentum
lattice of a free boson theory, but at the cost of
sacrificing positivity of the coefficients in the partition function.

   Finally we must stress that the condition of (one-loop)
modular invariance alone is
insufficient to guarantee a consistent conformal field theory; for
constructions not based on free fields we must still check that there is a
consistent operator algebra.
\hfill\vfill\eject

\pagenum=46

\chapternumstyle{blank}\sectionnumstyle{blank}
\n {\bf Chapter 4: Fractional Superstrings}\vskip .8cm
{\hb{4.1}{\bfs Introduction to Fractional Superstrings}}\vhalf
\chapternumstyle{arabic}\chapternum=4
\sectionnumstyle{arabic}\sectionnum=1
\equationnum=0

In the last few years, several generalizations of standard (supersymmetric)
string
theory have been proposed.\mpr{schwarz87,schellekens89,kaku91,pope92}
One of
them\mpr{dienes92b,argyres91b,argyres91d,argyres91a,argyres91c}
uses the (fractional spin)
parafermions introduced from the perspective of
2-D conformal field
theory (CFT) by Zamolodchikov and Fateev\mpr{zamol87} in 1985
and further developed by Gepner\mpr{gepner87} and
Qiu.\footnote{This is not to be confused with
the original definition of ``parafermions.''
The term ``parafermion'' was introduced by H. S. Green in
1953.\mpr{green53}  Green's parafermions are defined as
spin-1/2 particles that do not obey standard
anticommutation rules, but instead follow more general trilinear
relations.\mpr{ardalan74,mansouri87,antoniadis86}}
In a series of papers, possible
new string theories with local parafermionic world-sheet
currents (of fractional conformal spin) giving critical dimensions
\hbox{$D=6$, $4$, $3, {\rm ~and~}2$} have been
proposed.\mpr{dienes92b,argyres91b,argyres91d,argyres91a,argyres91c}

At the heart of these new ``fractional superstrings'' are $\Z_K$
parafermion conformal field theories (PCFT's) with central charge
$c= {2(K-1)\over K+2}$. (Equivalently, these are $SU(2)_K/U(1)$ conformal
field theories.)
The (integer) level-$K$
PCFT contains a set of unitary primary fields $\phi^j_m$, where
$0\leq j$, $\vert m\vert\leq K/2$; $j,\, m\in \Z/2$, and
$j-m = 0 \pmod{1}$.
These fields have the identifications
$$\phi^j_m = \phi^j_{m+K} = \phi^{{K\over 2}-j}_{m-{K\over 2}}
\,\, .\eqno\eqnlabel{phidents}$$
In the range $\vert m\vert\leq j$, the conformal dimension is
$h(\phi^j_m) = {j(j+1)\over K+2} - {m^2\over K}$.
At a given level
the fusion rules are
$$\phi^{j_1}_{m_1}\otimes\phi^{j_2}_{m_2} = \sum^r_{j=\vert j_1 - j_2\vert}
\phi^j_{m_1+m_2}\, ,\eqno\eqnlabel{fusion}$$
where $r\equiv {\rm min}\, (j_1 + j_2, K-j_1-j_2)$.
This CFT contains a subset of primary fields,
$$\{\phi_i\,\equiv\phi^0_i\equiv\phi^{K/2}_{-K/2 + i}\,\, ;\,\, 0\leq i
\leq
K-1\}
\eqno\eqnlabel{subset}$$
$(\phi^{\dag}_i\equiv\phi_{K-i})$
which, under fusion, form a closed subalgebra  possessing
a $\Z_K$ Abelian symmetry:
$$\phi_i\otimes\phi_j=\phi_{(i+j)}\pmod{K}\,.\eqno\eqnlabel{Zk}$$
The conformal dimensions, $h(\phi_i)$, of the fields in this subgroup
have the form
$$h(\phi_i)= {i(K-i)\over K}\,\, . \eq{condimzkcsa}$$
It has been proposed that string models based on tensor products of a
level-$K$ PCFT are
generalizations of the Type II $D=10$
superstring.\mpr{dienes92b,argyres91b,argyres91d,argyres91a,argyres91c}
In these potential models,
the standard $c={1\over 2}$ fermionic superpartner of the
holomorphic  world sheet
scalar, $X(z)$, is replaced by the ``energy operator,''
$\epsilon\equiv\phi^1_0$,
of the {\tenBbb Z}$_K$ PCFT.\footnote{Note that $\epsilon$ is not in
the {\tenBbb Z}$_K$  Abelian subgroup, and thus is not a
{\tenBbb Z}$_K$ parafermion, except
for the degenerate $K=2$ superstring case where $\phi^1_0\equiv\phi^0_1$.}
 $\epsilon$ has conformal dimension (spin)
$2\over K+2$, which is ``fractional''
({\it i.e.}, neither integral nor half-integral),
for $K\ne 2$.
This accounts for to the name of these models.
Each $\epsilon-X$ pair has a total conformal anomaly (or central charge)
$c={3K\over K+2}$.

The naive  generalization of the (holomorphic)
supercurrent (SC) of the standard superstring,
$J_{\rm SC}(z)= \psi (z)\cdot\partial X(z)$
(where $\psi$ is a real world sheet fermion),
to $J_{\rm FSC}=\phi^1_0(z)\cdot\partial X(z)$
proves to be inadequate.\mpr{argyres91c}
Instead, the proposed ``fractional supercurrent'' (FSC) is
$$J_{\rm FSC}(z)= \phi^1_0(z)\cdot \partial_z X(z) + \desc\,\, .
\eqno\eqnlabel{current}$$
$\desc$ (which vanishes for $K=2$ since $\phi^1_0=\psi$ at $K=2$)
is the first
descendent field of $\phi^1_0$.
$J_{\rm FSC}(z)$ is the generator of a local ``fractional'' world sheet
supersymmetry between $\epsilon(z)$  and $X(z)$,
extending the Virasoro algebra of the stress-energy
tensor $T(z)$.  This local current of spin
\hbox{$h(J_{\rm FSC})= 1 + {2\over K+2}$}
has fractional powers of $1\over (z-w)$ in
the OPE with itself, implying a non-local world-sheet interaction and,
hence, producing cuts on the world sheet.
The corresponding chiral ``fractional superconformal
algebra''\mpr{argyres91c} is,
\subon
$$T(z)T(w)={{1\over 2}c\over (z-w)^4}+{2T(w)\over
(z-w)^2}+...\eqno\eqnlabel{FSCalgebra-a}$$
$$T(z)J_{\rm FSC}(w)={h J_{\rm FSC}(w)\over (z-w)^2}+
{\partial J_{\rm FSC}(w)\over
(z-w)}+...\eqno\eqnlabel{FSCalgebra-b}$$
$$J_{\rm FSC}(z)J_{\rm FSC}(w)={1\over (z-w)^{2h}}+{{2h\over c}T(w)\over
(z-w)^{2h-2}}+{\lambda_K(c_0)J_{\rm FSC}(w)\over (z-w)^{h}}+{{1\over
2}\lambda_K(c_0)\partial J_{\rm FSC}(w)\over
(z-w)^{h-1}}+...\eqno\eqnlabel{FSCalgebra-c}$$
\suboff
where $c=Dc_0$. $D$ is the critical dimension,
$c_0= {3K\over K+2}$ is the central charge for one dimension,
and $\lambda_K$ is a constant.\mpr{argyres91f}

The relationship between critical dimension, $D$,
and the level, $K$, of the paraferm-
ion CFT may be shown to be
$$ D= 2 + {16\over K}\, , \eqno\eqnlabel{dimeqn}$$
for $K=2,\, 4,\, 8,\, 16,\, {\rm and~} \infty$.
In \pr{dienes92b,argyres91b,argyres91d,argyres91a,argyres91c}
the relationship \pe{dimeqn} is
derived by requiring a massless spin-1 particle in the open string
spectrum, produced by $\phi^1_0(z)^{\mu}$
(where $\mu$ is the spacetime index) operating on the vacuum.

\vhalf
{\hc{4.1.a}{\sl Parafermion Characters}}
\vhalf

Before we present the computer generated
fractional superstring partition functions of
refs.~\pr{dienes92b, argyres91b,argyres91d,argyres91e} and follow with
a new derivation of these these partition functions, as a prerequisit
we wish to
discuss the characters $Z(\phi^j_m)$ for the Verma
modules, $[\phi^j_m]$,\footnote{From here on, we do not distinguish between
the primary field $\phi^j_m$ and its complete Verma module $[\phi^j_m]$.
Thus, $\phi^j_m$ can represent either, depending on the context.}
for $j, \vert m\vert< K/2$.
Each verma module contains
a single (holomorphic) parafermionic primary field
$\phi^j_m(z)$ and its parafermion descendents.
\begin{ignore}
$$L_{-1}^a L_{-2}^b L_{-3}^c L_{-4}^d \cdots
  J_{-1}^x J_{-2}^y J_{-3}^z J_{-4}^w \cdots (\phi^j_m)$$,
where $L_{-i}$ and $J_$
\end{ignore}
The form of the characters is
$$\eqalignno{
Z(\phi^j_m)&= q^{-c/24} {\rm tr}\, q^{L_0}\cr
           &= \eta(\tau)c^{2j}_{2m}(\tau)\, ,&\eqnlabel{2a}}$$
where $q= e^{2\pi\tau}$
and $\eta$ is the Dedekind eta-function,
$$ \eta(\tau) = q^{1/24}\prod^{\infty}_{n=1}(1-q^n)\,\, .
\eqno\eqnlabel{defeta}$$
$c^{2j}_{2m}(\tau)$ is a
string function\mpr{kac80} defined by
\subequationnumstyle{alphabetic}
$$\eqalignno{
c^{2j}_{2m}(\tau)
    &= {1\over \eta^3(\tau)}\sum_{x,y} {\rm sign}(x)
       q^{x^2(K+2) -y^2K} &\eqnlabel{cfn-a}\cr
&= q^{h^j_m + {1\over 4(K+2)}}{1\over \eta^3}
\sum^{\infty}_{r,s=0} (-1)^{r+s}
q^{r(r+1)/2 + s(s+1)/2 + rs(K+1)}\times\cr
&\quad \left\{ q^{r(j+m) + s(j-m)} - q^{K+1-2j+r(K+1-j-m)+s(K+1-j+m)}\right\}
    &\eqnlabel{cfn-b}\cr
    &= q^{h^j_m - {c(SU(2)_K)\over 24}}(1 + \cdots)& \eqnlabel{cfn-c}}$$
\subequationnumstyle{blank}
where in \pe{cfn-a} the conditions
\item{1.} $-\vert x\vert<y<\vert x\vert\, ,$
\item{2.} either $x={2j+1\over 2(K+2)} \pmod{1}$
         or     $(\half - x)= {2j+1\over 2(K+2)} \pmod{1}\, ;$ and
\item{3.} either $y= {m\over K} \pmod{1}$
         or     $(\half + y) = {m\over K} \pmod{1}$

\no must be met simultaneously.
($h^j_m \equiv h(\phi^j_m)$ and $c(SU(2)_K)= {3K\over K+2}$.)
These
string functions obey the same equivalences as their associated primary
fields $\phi^j_m$:
\subequationnumstyle{alphabetic}
$$c^{2j}_{2m}= c^{2j}_{2m+2K} = c^{K-2j}_{2m-K}\, .\eqno\eqnlabel{cid-a}$$
Additionally, since $\phi^j_{m}=(\phi^j_{-m})^{\dagger}$,
$$c^{2j}_{2m}= c^{2j}_{-2m}\, .\eq{cid-b}$$
\subequationnumstyle{blank}

Since the $K=2$ theory
is the standard Type II superstring
theory,\footnote{The $K=2$ parafermion model is a
$c={1\over2}$ CFT that
corresponds to a critical Ising (free fermion) model.}
expressing its partition function
in terms of string functions rather than
theta-functions
can be accomplished simply using the following set of identities:
$$ {K=2: \cases{2\eta^2(c^1_1)^2 = \vartheta_2/\eta\, ;\cr
               \eta^2(c^0_0 + c^2_0)^2 = \vartheta_3/\eta\, ;\cr
               \eta^2(c^0_0 - c^2_0)^2 = \vartheta_4/\eta\, .}}\eq{fpequiv}$$

For each spacetime dimension in these theories, a term in the
partition function of the form (\puteqn{2a})
is tensored with the partition function $Z\left(X\right)$
for an uncompactified chiral boson $X(z)$. Since
$$ Z\left(X\right)\propto {1\over\eta(\tau)}\,\, ,
\eqno\eqnlabel{partstdim}$$
the $\eta(\tau)$ factors cancel out in
$Z(\phi^j_m)\times Z(X)$. Similar cancellation of
$\bar\eta(\bar\tau)$ occurs in the antiholomorphic sector.
In the
following partition functions, we suppress the trivial factor of
$({\rm Im}\, \tau)^{-8/K}$ contributed together by the $D-2$ holomorphic and
anti-holomorphic world sheet boson partition functions.

The purpose of this chapter is to examine a number of issues relating
to these models:
In section 4.2 we derive the
partition functions of the $D=6$, $4$, and $3$ theories (corresponding to
$K=4$, $8$ and $16$ respectively), using the factorization method of Gepner
and Qiu,\mpr{gepner87}
as well as
demonstrating a new approach to obtaining the superstring partition function.
In section 4.3 we consider other necessary elements of string
theory. In particular, we propose a generalization of the GSO
projection that applies to the fractional superstring and we address
the question of whether similar theories at different Ka\v c-Moody levels
can be constructed.
Additionally, a comparison with the superstring is made and
we attempt to elucidate its features in the current, more general
context.
\vhalf
{\hb{4.2}{\bfs Fractional Superstring Partition Functions}}
\vhalf
\sectionnum=2\equationnum=0

Computerized searches demonstrated that for each (and only those) $K$
listed above, there is a unique one-loop partition function
(written in light-cone gauge) that is (1) modular invariant,
(2) contains a term, $(c^0_0)^{D-3}(c^2_0)$,
which is the character for a
massless spacetime spin-2 particle generated by an untwisted
non-chiral
$\phi^1_0(z){\bar{\phi}}^1_0(\overline z)$ field acting on the vacuum,
and (3) has no characters for
tachyonic states.\mpr{dienes92b, argyres91b,argyres91d,argyres91e}
Partition functions with these properties were found to exist only in
10, 6, 4, and 3 dimensions and were presented as:
{\settabs 8 \columns
\+ \cr
\+ $D=10$ & $(K=2)$: &&$Z =  \vert A_2\vert^2$, ~where\cr}
$$\eqalignno{ A_2 &= 8(\cc0 0)^7(\cc2 0)+56(\cc0 0)^5(\cc2 0)^3+56(\cc0
0)^3(\cc2
0)^5+8(\cc0 0)(\cc2 0)^7-8(\cc1 1)^8\cr &= {1\over
2}\eta^{-12}(\theta^4_3-\theta^4_4-\theta^4_2) &\eqnlabel{ipart2}\cr}$$
{\settabs 8 \columns
\subon
\+ $D=6$ & $(K=4)$: &&$Z= \vert A_4\vert^2 + 3\vert B_4\vert^2$, ~where\cr}
$$\eqalignno {A_4 &= 4(\cc0 0+\cc4 0)^3(\cc2 0)-4(\cc2 0)^4-4(\cc2 2)^4+32(\cc2
2)(\cc4 2)^3 &\eqnlabel{ipart4-a}\cr B_4 &= 8(\cc0 0+\cc4 0)(\cc2 0)(\cc4
2)^2+4(\cc0 0+\cc4 0)^2(\cc2 2)(\cc4 2)-4(\cc2 0)^2(\cc2
2)^2&\eqnlabel{ipart4-b}\cr}$$
{\settabs 8 \columns
\suboff
\subon
\+ $D=4$ & $(K=8)$:
&&$Z= \vert A_8\vert^2 + \vert B_8\vert^2 + 2\vert C_8\vert^2$, ~where\cr}
$$\eqalignno { A_8 & = 2(\cc0 0+\cc8 0)(\cc2 0+\cc6 0)-2(\cc4 0)^2-2(\cc4
4)^2+8(\cc6 4\cc8 4)&\eqnlabel{ipart8-a}\cr B_8 &= 4(\cc0 0+\cc8 0)(\cc6
4)+4(\cc2
0+\cc6 0)(\cc8 4)-4(\cc4 0\cc4 4)&\eqnlabel{ipart8-b}\cr
C_8 &= 4(\cc2 2+\cc6 2)(\cc8 2+\cc8 6)-4(\cc4 2)^2&\eqnlabel{ipart8-c}\cr}$$
{\settabs 8 \columns
\suboff
\subon
\+ $D=3$ & $(K=16)$:
&&$ Z = \vert A_{16}\vert^2 + \vert C_{16}\vert^2 $, ~where\cr}
$$\eqalignno{A_{16} &= \cc2 0+c^{14}_0-\cc8 0-\cc8 8+2\cc{14}
8&\eqnlabel{ipart16-a}\cr
C_{16} &= 2\cc2 4+2\cc{14} 4-2\cc8 4\,\, .&\eqnlabel{ipart16-b}\cr}$$
\suboff

These closed-string partition functions
$Z({\rm level-} K)$
all have the
general form $$ Z({\rm level-} K) = \vert A_K\vert^2 + \vert B_K\vert^2 + \vert
C_K\vert^2\,\, .
\eqno\eqnlabel{pf1}$$
The $D=10$ partition function, in string function format, was
obtained by the authors of refs.~\pr{dienes92b, argyres91b, argyres91d}
as a check of their program, both by computer generation and by the
$K=2$ string functions/Jacobi $\theta$-functions equivalences.

In the above partition functions,
the characters for the massless graviton (spin-2 particle) and
gravitino (spin-${3\over 2}$) are terms in the
$A_K$--sector, $\vert A_K\vert^2$.
The $D<10$ fractional superstrings have a new feature not
present in the standard $D=10$ superstrings.
This is the existence of the
massive $B_K$-- and $C_K$--sectors.  These additional
sectors were originally derived in the computer program of the authors of
refs. \hbox{\pr{dienes92b,argyres91b,argyres91d} by applying $S$
transformations}
to the $A_K$--sector and then demanding modular invariance of the
theory.

An obvious question with respect to these partition functions is
how to interpret the relationship between the spacetime spin of the
physical
states and the subscripts of the corresponding characters in the partition
functions.
The solution is not immediately transparent for general $K$.
$K=2$ is, of course, the exception.
Based on the aforementioned identites \pe{fpequiv},
we see that terms with all $n_i\equiv 2m_i=0$
correspond to spacetime bosons, while those with all $n_i\equiv 2m_i= K$
correspond to spacetime fermions.  This rule also seems to be followed
by terms in $A_K$ for all $K$.
In the $B_K$-- and $C_K$--sectors, interpretation is much less clear.
There have been two suggested meanings for
the terms in $B_K$ and $C_K$. The first hypothesis is that
these terms
correspond to massive spacetime anyons, specifically
spin-${1\over 4}$ particles for $B_K$, and
massive spin-${1\over 8}$ particles for $C_K$.
The second alternative is that
the $B_K$--sector particles are spacetime fermions and bosons, but
with one (for $K=8$) or two (for $K=4$) spatial dimensions
compactified.\mpr{dienes92b, argyres91b,argyres91d,argyres91e}
Along this line of thought, the $C_K$--sector particles are still
interpreted to be spacetime anyons,
but with spin-${1\over4}$ rather than spin-${1\over8}$.

Until present, the general concensus has been that spacetime
anyons can presumably exist only in three or less uncompactified dimensions.
This would seem to contradict suggestions that the $D=4$ or $D=6$ models
may contain spacetime anyons unless at least one or two dimensions,
respectively, are compactified.
Based on other, independent, reasons
we will suggest this compactification does automatically occur in
fractional superstring models.
Further, we will also show that there are possibly no physical states in
the $C_K$--sector for $K=8$.

At each level of $K$, the contribution of each sector is separately zero.
This is
consistent with spacetime SUSY and suggests cancellation
between bosonic and fermionic terms at each mass level.
This leads to the following
identities:\mpr{dienes92b}
$$A_2=A_4=B_4=A_8=B_8=C_8=A_{16}=C_{16}=0\,\, .\eqno\eqnlabel{partident}$$

In this section, we will introduce
a new method for generating these partition functions that reveals
(1) new aspects of
the relationship between
the $B_K$-- and $C_K$--sectors and the $A_K$--sector,
and (2) the evidense for
spacetime supersymmetry in all sectors.
(Specifically, these type II models should have $N=2$ spacetime SUSY, with the
holomorphic and antiholomorphic sectors each effectively contributing an
$N=1$ SUSY. Hence, heterotic fractional superstrings would only possess $N=1$
SUSY.)
We will
demonstrate that cancellation suggestive of spacetime SUSY results from
the action of a simple twist current used in the
derivation of these partition functions.  Only by this twisting can
cancellation between bosonic and fermionic terms occur at each mass level
in the $A_K$-- and $B_K$--sectors.  The same twisting
results in a
``self-cancellation'' of terms in the $C_K$--sector, and does, indeed,
suggest the anyonic spin-${1\over4}$
interpretation of the $C_K$--sector states.

\vhalf
{\hc{4.2.a}{\sl New Derivation of the Partition Functions}}\vhalf

We find the computer generated partition functions listed above
not to be in the most suggestive form.
By using the string function equivalences,
(\puteqn{cid-a}-b),
the partition functions for the level-$K$ fractional superstrings in
refs.~\pr{dienes92b,argyres91b,argyres91d,argyres91e} with critical spacetime
dimensions
\hbox {$D= 2 + {16\over K}= 10, 6,$} {$\,4,\, {\rm~and~}3$}
can be rewritten (in light-cone gauge) in the form below.
\begin{ignore}
Effectively, the rewritten partition functions
correspond to first replacing a single character, $c^l_n$, by the symmetrized
sum of characters, $\half (c^l_n + c^{K-l}_n)$ and then forming subsectors
in each mod-squared term, based on the values of the subscripts of the
characters.
\end{ignore}

{\settabs 8 \columns
\+ \cr
\+ $D=10$ & $(K=2)$: &&$Z =  \vert A_2\vert^2$, ~where\cr}
$$\eqalignno{ A_2 &= {1\over 2}\left\{ (c^0_0 + c^2_0)^8 - (c^0_0 -
c^2_0)^8 \right\}_{\rm boson} - 8(c^1_1)^8_{\rm fermion}\cr &= 8\left\{
(c^0_0)^7
c^2_0 + 7(c^0_0)^5(c^2_0)^3 +7(c^0_0)^3(c^2_0)^5 +
c^0_0(c^2_0)^7\right\}_{\rm boson}  -
8(c^1_1)^8_{\rm fermion} &\eqnlabel{part2}}$$
{\settabs 8 \columns
\subon
\+ \cr
\+ $D=6$ & $(K=4)$: &&$Z= \vert A_4\vert^2 + 3\vert B_4\vert^2$, ~where\cr}
$$\eqalignno {A_4 &= {\rm\hskip .37 truecm}4\left\{(c^0_0 + c^4_0)^3
(c^2_0) - (c^2_0)^4\right\}\cr &\quad + 4\left\{(c^0_2 + c^4_2)^3 (c^2_2) -
(c^2_2)^4\right\}&\eqnlabel{part4-a}\cr B_4 &= {\rm\hskip
.37truecm}4\left\{(c^0_0 +
c^4_0)(c^0_2+c^4_2)^2(c^2_0) - (c^2_0)^2(c^2_2)^2\right\}\cr &\quad +
4\left\{(c^0_2 + c^4_2)(c^0_0+c^4_0)^2(c^2_2) -
(c^2_2)^2(c^2_0)^2\right\}&\eqnlabel{part4-b}\cr}$$
{\settabs 8 \columns
\suboff
\hfill\vfill\eject
\subon
\+ $D=4$ & $(K=8)$:
&&$Z= \vert A_8\vert^2 + \vert B_8\vert^2 + 2\vert C_8\vert^2$, ~where\cr}
$$\eqalignno { A_8 & = {\rm\hskip .37 truecm}2\left\{(c^0_0 + c^8_0)(c^2_0
+ c^6_0) - (c^4_0)^2\right\}\cr &\quad +2\left\{(c^0_4+ c^8_4)(c^2_4+c^6_4)
- (c^4_4)^2\right\}&\eqnlabel{part8-a}\cr B_8 &= {\rm\hskip .37
truecm}2\left\{(c^0_0+c^8_0)(c^2_4+c^6_4)-(c^4_0c^4_4)\right\}\cr &\quad +
2\left\{(c^0_4+c^8_4)(c^2_0+c^6_0)-(c^4_4c^4_0)\right\}&\eqnlabel{part8-b}\cr
C_8 &= {\rm
\hskip .37truecm}2\left\{(c^0_2 + c^8_2)(c^2_2 + c^6_2) -
(c^4_2)^2\right\}\cr &\quad +2\left\{(c^0_2 + c^8_2)(c^2_2 + c^6_2) -
(c^4_2)^2\right\}&\eqnlabel{part8-c}\cr}$$ {\settabs 8 \columns
\suboff
\subon
\+ $D=3$ & $(K=16)$:
&&$ Z = \vert A_{16}\vert^2 + \vert C_{16}\vert^2 $, ~where\cr}
$$\eqalignno{A_{16} &= {\rm\hskip .3 truecm}\left\{(c^2_0 + c^{14}_0) -
c^8_0\right\}\cr &\quad +\left\{(c^2_8 + c^{14}_8) - c^8_8\right\}
&\eqnlabel{part16-a}\cr
C_{16} &= {\rm\hskip .3 truecm}\left\{(c^2_4 + c^{14}_4) -
c^8_4\right\}\cr &\quad +\left\{(c^2_4 + c^{14}_4) -
c^8_4\right\}\,\, .&\eqnlabel{part16-b}\cr}$$
\suboff

The factorization method of Gepner and Qiu\mpr{gepner87} for
string function partition functions allows us to rederive
the fractional superstring partition functions in this new form
systematically.  Using this approach we can express a
general parafermion partition function (with the level
of the string functions henceforth suppressed),
\subequationnumstyle{alphabetic}
$$ Z= \vert \eta\vert^2 \sum N_{l,n,\bar l,\bar n} c^l_n \bar c^{\bar
l}_{\bar n}\, ,\eqno\eqnlabel{partfn2-a}$$
in the form
$$Z= \vert\eta\vert^2\sum
{1\over 2}L_{l,\bar l}M_{n,\bar n}c^l_n \bar c^{\bar l}_{\bar n}\,
,\eqno\eqnlabel{partfn2-b}$$
\subequationnumstyle{blank}
(with $c^{l=2j}_{n=2m}=0$ unless $l-n\in 2\Z$ since $\phi^j_m=0$ for
$j-m\not\in\Z$).  As a result of the factorization,
$$N_{l,n,\bar l,\bar n} = {1\over 2}L_{l,\bar l}\, M_{n,\bar n}\,\, ,
\eqno\eqnlabel{Nfactorization}$$
we can construct all modular invariant
partition functions (MIPF's) for parafermions from a tensor product
of modular invariant solutions for the $(l,\bar l)$ and
$(n,\bar n)$ indices separately. This results from the
definition of level-$K$ string functions, $c^l_n$, in terms of the
$SU(2)_K$ characters $\chi_l$ and the Jacobi theta-function,
$\theta_{n,K}$:\footnote{The associated relationship between the level-$K$
$SU(2)$ primary fields $\Phi^j$ and the parafermionic $\phi^j_m$ is
$$\Phi^j= \sum_{m=-j}^j \phi^j_m\,  :\exp\left\{ {i{m\over \sqrt{K}}
\varphi}\right\}:$$
where $\varphi$ is the $U(1)$ boson field of the $SU(2)$ theory.}
$$\chi_l(\tau)
= \sum^K_{n= -K+1} c^l_n(\tau)\theta_{n,K}(\tau)\,
,\eqno\eqnlabel{partfn4}$$ where the theta-function is defined by
$$\theta_{n,K}(\tau) = \sum_{p\in\Z + {n\over 2K}}\e^{2\pi i K p^2\tau}\,
,\eqno\eqnlabel{thetafn}$$ and $\chi_l$ is the character for the spin-${l\over
2}$ representation of $SU(2)_K$, $$\chi_l(\tau) =
{\theta_{l+1,K+2}(\tau)-\theta_{-l-1,K+2}(\tau)\over
\theta_{1,2}(\tau)-\theta_{-1,2}(\tau)}
\, .\eqno\eqnlabel{chifn}$$
This factorization is seen in the transformation properties of
$c^l_n$ under the modular group generators $S$ and
$T$,
\subequationnumstyle{alphabetic}
$$\eqalignno{
S: c^l_n &\rightarrow {1\over \sqrt{-i\tau K(K+2)}}
\sum_{l'=0}^{K}\sum_{n'= -K+1 \atop l'-n`\in 2Z}^K \exp\left\{{i\pi n
n'\over K}\right\} \sin\left\{{\pi (l+1)(l'+1)\over K+2}\right\} c^{l'}_{n'}
{\rm ~~~~~~~~~}
&\eqnlabel{ctrans-a}\cr
T: c^l_n &\rightarrow \exp
\left\{ 2\pi i
\left(
{l(l+2)\over 4(K+2)} - {n^2\over 4K} - {K\over 8(K+2)}\right)
\right\} c^l_n\,\, .
&\eqnlabel{ctrans-b}\cr}$$
\subequationnumstyle{blank}
Thus, (\puteqn{partfn2-b}) is modular invariant if and only if the
$SU(2)$ affine partition function
$$ W(\tau,\overline\tau)=
\sum_{l,\bar l= 0}^K L_{l,\bar l}\chi_l(\tau)\bar\chi_{\bar
l}(\bar\tau)\eqno\eqnlabel{partfn3}$$
and the $U(1)$ partition function
$$ V(\tau,\overline\tau)= {1\over\vert\eta(\tau)\vert^2}
\sum_{n,\bar n= -K+1}^K M_{n,\bar n}\theta_{n,K}(\tau,\overline\tau)
\bar\theta_{\bar n,K}(\tau,\overline\tau)
\eqno\eqnlabel{partfn10}$$ are simultaneously modular invariant.
That is, $N_{l,n,\bar l,\bar n}= {1\over 2}L_{l,\bar l}M_{n,\bar n}$
corresponds to a
MIPF (\puteqn{partfn2-a}) if and only if $L_{l,\bar l}$ and $M_{n,\bar n}$
correspond to MIPF's of the forms (\puteqn{partfn3}) and
(\puteqn{partfn10}), respectively.

This factorization is also possible for parafermion tensor product theories,
with matrices $\bmit L$ and $\bmit M$ generalized to tensors.  Any
tensor $\bmit M$ corresponding to a MIPF for $p$ factors of $U(1)$
CFT's can be written as a tensor product of $p$ independent matrix $\bmit M$
solutions to (\puteqn{partfn10}) twisted by simple
currents $\cal J$.\mpr{cleaver}
This approach greatly simplifies the derivation of
the fractional superstring partition functions,
while simultaneously suggesting much about the meaning
of the different sectors, the origin of spacetime supersymmetry, and
related ``projection'' terms. We now proceed with the
independent derivations of $\bmit L$ and $\bmit M$ for the PCFT's.

\vhalf
{\hc{4.2.b}{\sl Affine Factor and ``W'' Partition Function}}\vhalf

In the $A_K$--sectors defined by
eqs.~(\puteqn{part4-a}, \puteqn{part8-a}, \puteqn{part16-a})
the terms inside the first (upper) set of brackets
carry ``$n\equiv 2m=0$'' subscripts
and can be shown, as our prior discussion suggested,
to correspond to spacetime bosons;
while the terms inside the second (lower) set carry ``$n\equiv 2m= K/2$''
and correspond to spacetime fermions.
(See eqs.~(4.2.36a-b).)
Expressing the $A_K$--sector in this form
makes a one--to--one correspondence
between bosonic and fermionic states in the
$A_K$--sector manifest. If we remove the subscripts on the string functions
in the bosonic and fermionic subsectors
(which is analogous to replacing $c^l_n$ with
$\chi_l$), we find the subsectors become equivalent.
In fact, under this
operation of removing the ``$n$'' subscripts and replacing
each string function by its corresponding affine character
(a process we denote by $\buildrel {\rm
affine}\over\Longrightarrow$), all sectors become the same up to an
integer coefficient:
\subon
{\settabs 8 \columns
\+ \cr
\+ $D=6$ &$(K=4)$:\cr} $$A_4,B_4 \hbox to 1cm{\hfill}{\buildrel {\rm
affine}\over\Longrightarrow}\hbox to 1cm{\hfill}
A_4^{\rm aff}\equiv (\chi_0+\chi_K)^3\chi_{K/2}-(\chi_{K/2})^4
\eqno\eqnlabel{affine-a}$$
\hfill\vfill\eject
{\settabs 8 \columns
\+ $D=4$ &$(K=8)$:\cr} $$A_8,B_8,C_8
\hbox to 1cm{\hfill}{\buildrel {\rm affine}\over\Longrightarrow}
\hbox to 1cm{\hfill}
A_8^{\rm aff}\equiv (\chi_0+\chi_K)(\chi_2+\chi_{K-2})
- (\chi_{K/2})^2 \eqno\eqnlabel{affine-b}$$
{\settabs 8 \columns\+
$D=3$
&$(K=16)$:\cr} $$A_{16},C_{16}
\hbox to 1cm{\hfill}{\buildrel {\rm affine}\over\Longrightarrow}
\hbox to 1cm{\hfill}
A_{16}^{\rm aff}\equiv (\chi_2 + \chi_{K-2}) -\chi_{K/2}\,\, .
\eqno\eqnlabel{affine-c}$$
\suboff

\noindent
We see that the $B$-- and $C$--sectors are not arbitrary additions,
necessitated only by modular invariance, but rather are naturally
related to the physically motivated $A$--sectors.
Thus, the affine factor in each
parafermion partition function is:
$$ Z_{{\rm affine}}
(K) = \vert A_K^{\rm aff}\vert^2\,\, ,\eqno\eqnlabel{affine2}$$
where we see that
eqs.~(\puteqn{affine-a}-c) all have the
general form
$$ A_K^{\rm aff} \equiv (\chi_0
+\chi_K)^{D-3}(\chi_2+\chi_{K-2}) -
(\chi_{K/2})^{D-2}\,\, . \eqno\eqnlabel{affall}$$
\no
(Note that the modular
invariance of $W$ requires that $A_K^{\rm aff}$ transforms back into
itself under $S$.)

The class of partition functions (\puteqn{affine2}) is indeed modular
invariant and possesses special properties.
This is easiest to show for
$K=16$.  The $SU(2)_{16}$ MIPF's for $D=3$ are trivial to classify, since
at this level the A--D--E classification forms a complete basis set of modular
invariants, even for MIPF's containing terms with negative coefficients. The
only free parameters in $K=16$ affine partition functions $Z\left(
SU(2)_{16}\right)$ are integers $a$, $b$, and $c$, where $$Z\left(
SU(2)_{K=16}\right) =
a\times Z({\rm A}_{17})+b\times Z({\rm D}_{10})+c\times Z({\rm E}_7)\,\,
.\eqno\eqnlabel{affine3}$$

Demanding that neither a left- nor a right-moving tachyonic state be in
the Hilbert space of states in the $K=16$
fractional superstring when the intercept $v$, defined by $$L_0\vert
{\rm physical}\rangle = v\vert {\rm physical}\rangle\, ,
\eqno\eqnlabel{intercept}$$
is positive, removes these degrees of freedom and requires $a= -(b+c)=0$,
independent of the possible $U(1)$ partition
functions.
These specific values for $a$, $b$, and $c$ give us (\puteqn{affine2}) for
this level:
$$W\left( K=16 \right) = Z({\rm D}_{10})-Z({\rm E}_{7})
= \vert A^{\rm aff}_{16}\vert^2
\,\, .\eqno\eqnlabel{affine4}$$

Though not quite as straightforward a process,
we can also derive the affine partition functions $W(K)$ for the
remaining levels.  The affine factors in the
\hbox {$K=4 {\rm ~and~} 8$}
partition
functions involve twisting by a non-simple current.
(See footnote p. for the definition of non-simple current.)
These cases
correspond to theories that are the difference between a
${\bigotimes\atop {D-2\atop {\rm factors}}} {\rm D}_{{K\over 2} +2}$ tensor
product model
and a ${\bigotimes\atop{D-2\atop {\rm factors}}} {\rm D}_{{K\over 2} +2}$
tensor
product model twisted by the affine
current
$$J_{\rm non-simple}^{K,~{\rm affine}}=(\Phi^{K\over
4})^{D-2}\bar\Phi^1(\bar\Phi^0)^{D-3}\,\, .\eqno\eqnlabel{jkaff}$$
The equivalent parafermionic twist current is obvious,
$$J_{\rm non-simple}^{K,~{\rm parafermion}}=
(\phi^{K\over 4}_0)^{D-2}(\bar\phi^1_0)
(\bar\phi^0_0)^{D-3}\,\, .\eqno\eqnlabel{affine11}$$ (This derivation
applies to
the $K=16$ case
also.)\footnote{We
have left off the spacetime indices on most of the following currents
and fields.  We are working in light-cone gauge so only indices for
transverse modes are implied.
The $D-2$ transverse dimensions are assigned
indices in the range 1 to $D-2$ (and are generically represented by
lowercase Greek superscripts.) When spacetime indices
are suppressed,
the fields, and their corresponding characters in the partition
function,
acting along directions 1 to $D-2$
are ordered in equations from left to right, respectively, for both
the holomorphic and antiholomorphic sectors separately.
Often, we will be still more implicit
in our notation and will express $r$ identical factors of $\phi^j_m$ along
consecutive directions (when these directions are either all compactified or
uncompactified) as $(\phi^j_m)^r$. Thus, eq.~\pe{affine11} for $K=8$ means
$$J_{\rm non-simple}^{K=8,~{\rm parafermion}}\equiv
(\phi^{K/4}_0)^{\mu=1}(\phi^{K/4}_0)^{\nu=2}(\bar\phi^1_0)^{\bar\mu=1}
(\bar\phi^0_0)^{\bar\nu=2}\,\, .$$}
\hfill\vfill\eject
{\hc{4.2.c}{\sl Theta-Function Factor and the ``$V$'' Partition
Function}}\vhalf

We now consider the theta-function factors, $\bmit M$,
carrying the  $(n,\bar n)$-indices in the fractional superstring partition
functions.
Since all
$A_K$--, $B_K$--, $C_K$--sectors in the level-$K$ fractional superstring
partition
function (and even the boson and fermion subsectors separately in $A_K$)
contain the
same affine factor, it is clearly the choice of the theta-function
factor which determines the spacetime supersymmetry of the
fractional superstring theories. That is, spacetime spins of particles in the
Hilbert space of states depend upon the
${\bmit M}'s$ that are allowed in tensored versions of
eq.~(\puteqn{partfn10}).  In the case of matrix $\bmit M$, rather than a more
complicated tensor, invariance of (\puteqn{partfn10})
under $S$ requires that the components $M_{n\bar n}$
be related by
\subon
$$ M_{n',\bar n'} = {1\over 2K}\sum_{n,\bar n= -K+1}^{K}
M_{n,\bar n} \e^{i\pi n
n'/K}\e^{i\pi \bar n \bar n'/K}\,\, , \eqno\eqnlabel{m-a}$$
and $T$ invariance
demands that $$ {n^2 - \bar n^2\over 4K}\in\Z\,\, ,
{\rm ~~if~} M_{n,\bar n}\neq 0\,\,. \eqno\eqnlabel{m-b}$$
\suboff
At every level-$K$ there is a unique modular
invariant function corresponding to
each factorization\mpr{gepner87},
$\alpha\times\beta=K$, where $\alpha,\,\, \beta\in\Z$.
Denoting the matrix elements of ${\bmit M}^{\alpha,\beta}$ by
$M^{\alpha,\beta}_{n,\bar n}$,
they are given by\footnote{By eq.~(2.23), $M^{\alpha,\beta}_{n,\bar n}
= M^{\beta,\alpha}_{n,-\bar n}$.
Hence, ${\bmit M}^{\alpha,\beta}$ and ${\bmit M}^{\beta,\alpha}$
result in equivalent fractional superstring partition functions.
To avoid this redundancy, we choose $\alpha\leq\beta$.

Throughout this subsection
we will view the $n$ as representing, simultaneously,
the holomorphic $\theta_{n,K}$ characters for $U(1)$ theories
and, in some sense,
the holomorphic string functions, $c^0_{n}$, for parafermions.
($\bar n$ represents the antiholomorphic parallels.)
However, we do not intend to imply that the string functions
can actually be factored into $c^l_0\times c^0_n=c^l_n$.  Rather,
we mean to use this in eqs.~(2.31b, 2.33b, 2.35b) only as an
artificial construct for
developing a deeper understanding of the
function of the parafermion primary fields (Verma modules) $\phi^0_m$ in
these models.  In the case of the primary fields, $\phi^j_m$,
factorization is, indeed, valid:
$\phi^j_0\otimes\phi^0_m=\phi^j_m$ (for integer $j,\, m$).}
$$M^{\alpha,\beta}_{n,\bar n} = {1\over 2}
\sum_{x\in\Z_{2\beta}\atop y\in\Z_{2\alpha}}
\delta_{n,\alpha x +\beta y}\delta_{\bar n,\alpha x -\beta y}\,\, .
\eqno\eqnlabel{m-c}$$

Thus,
for $K=4$ the two distinct choices for the matrix ${\bmit M}^{\alpha,\beta}$
are ${\bmit M}^{1,4}$ and ${\bmit M}^{2,2}$; for
$K=8$, we have ${\bmit M}^{1,8}$ and ${\bmit M}^{2,4}$; and
for $K=16$, the three alternatives are ${\bmit M}^{1,16}$,
${\bmit M}^{2,8}$, and ${\bmit M}^{4,4}$.
${\bmit M}^{1,K}$ represents the level-$K$ diagonal, {\ie} $n=\bar n$,
partition function.
${\bmit M}^{\alpha, \beta={K\over\alpha}}$
corresponds to the
diagonal partition function twisted by a $\Z_{\alpha}$ symmetry.
(Twisting by $\Z_{\alpha}$ and $\Z_{K/\alpha}$ produce isomorphic
models.)
Simple
tensor products of these ${\bmit M}^{\alpha,\beta}$ matrices are insufficient
for
producing fractional superstrings with spacetime SUSY (and, thus, without
tachyons).
We have found that twisting by a special simple $U(1)$ current is required to
achieve this.
Of the potential choices for the $U(1)$ MIPF's,
$V({\rm level~} K)$, the
following are the only ones that produce numerically zero
fractional superstring partition functions:
{\settabs 8 \columns
\+\cr
\+ $D=6$ & $(K=4)$:\cr
\+ \cr}

The ${\bmit M}= {\bmit M}^{2,2}\otimes{\bmit M}^{2,2}
\otimes{\bmit M}^{2,2}\otimes{\bmit M}^{2,2}$ model twisted by the
simple $U(1)$
current\footnote{Recall that the parafermion primary fields $\phi^0_m$
have simple fusion rules,
$$\phi^0_m\otimes\phi^0_{m'}=\phi^0_{m+m' \pmod{K}}$$
and form a {\tenBbb Z}$_K$ closed subalgebra.  This fusion rule, likewise,
holds for
the $U(1)$ fields $:\exp\{i{m\over K}\varphi\}:$.  This isomorphism
makes it clear that any simple $U(1)$ current,
${\cal J}_K$, in this subsection
that contains only integer $m$ can be
expressed equivalently either in terms of these parafermion fields
or in terms of $U(1)$ fields.
(We specify integer $m$ since $\phi^0_{m}=0$ for half--integer $m$.)
In view of the following discussion, we define all of the
simple twist currents, ${\cal J}_K$, as composed of the former.
(Please note, to distinguish between simple $U(1)$ currents
and affine currents, the U(1) currents
appear in calligraphy style, as above.)}
$${\cal J}_4\equiv \phi_{K/4}^0\phi_{K/4}^0\phi_{K/4}^0\phi_{K/4}^0
\bar\phi_0^0\bar\phi_0^0\bar\phi_0^0\bar\phi_0^0 \eqno\eqnlabel{j4}$$
results in the following $U(1)$ partition functions:
\subon
$$\eqalignno{V\left(K=4\right) &= \hbox to
.25cm{\hfill}[(\t_{0,4}+\t_{4,4})^4(\bt_{0,4}
 + \bt_{4,4})^4 +
(\t_{2,4} +\t_{-2,4})^4(\bt_{2,4}+ \bt_{-2,4})^4\cr
&\hbox to 1em{\hfill}
+ (\t_{0,4} +\t_{4,4})^2(\t_{2,4}+\t_{-2,4})^2(\bt_{
0,4} +\bt_{4,4})^2(\bt_{2,4} + \bt_{-2,4})^2\cr
&\hbox to 1em{\hfill}
+ (\t_{2,4}+\t_{-2,4})^2(\t_{0,4}+\t_{4,4})^2(\bt_{2,4} +\bt_{-2,4})^2
(\bt_{0,4} + \bt_{4,4})^2]_{\rm untwisted}\cr
&&\eqnlabel{nn4-a}\cr
&\hbox to 1em{\hfill}
+ [(\t_{2,4}+\t_{-2,4})^4(\bt_{0,4}+\bt_{4,4})^4 + (\t_{4,4}+\t_{0,4})^4
(\bt_{2,4}+\bt_{-2,4})^4\cr
&\hbox to 1em{\hfill}
+ \hbox to
.15cm{\hfill}(\t_{2,4}+\t_{-2,4})^2(\t_{4,4}+\t_{0,4})^2(\bt_{0,4}+\bt_{4,4})^2
(\bt_{2,4}+\bt_{-2,4})^2\cr
&\hbox to 1em{\hfill}
+ \hbox to .15cm{\hfill}(\t_{4,4}+\t_{0,4})^2(\t_{2,4}+\t_{-2,4})^2
(\bt_{2,4} + \bt_{-2,4})^2(\bt_{0,4} +\bt_{4,4})^2]_{\rm twisted}\,\, .}$$

Writing this in parafermionic form, and then
using string function identities, followed by
regrouping according to $A_4$ and $B_4$ components,
results in
$$Z({\rm theta~ factor,~} K=4)=
\vert (c^0_0)^4 + (c^0_2)^4\vert^2_{_{(A_4)}}
+ \vert (c^0_0)^2(c^0_2)^2 + (c^0_2)^2(c^0_0)^2
\vert^2_{_{(B_4)}}\,\, .
\eqno\eqnlabel{nn4-b}$$
\suboff
{\settabs 8\columns
\+ \cr
\+ $D=4$ & $(K=8)$:\cr
\+\cr}

The ${\bmit M}= {\bmit M}^{2,4}\otimes{\bmit M}^{2,4}$ model twisted by the
simple $U(1)$ current $${\cal J}_8\equiv
\phi_{K/4}^0\phi_{K/4}^0\bar\phi_0^0\bar\phi_0^0\eqno\eqnlabel{j8}$$
results in
\subon
$$\eqalignno{V\left(K=8\right) &= \hbox to
.35cm{\hfill}[(\t_{0,8}+\t_{8,8})(\bt_{0,8}
+ \bt_{8,8}) +
(\t_{4,8}+\t_{-4,8})(\bt_{4,8}+\bt_{-4,8})]^2_{\rm untwisted}\cr
&\hbox to 1em{\hfill}
+ [(\t_{2,8}+\t_{-6,8})(\bt_{2,8}+\bt_{-6,8})
+(\t_{-2,8}+\t_{6,8})(\bt_{-2,8}+\bt_{6,8})]^2_{\rm untwisted}\cr
&&\eqnlabel{nn8-a}\cr
&\hbox to 1em{\hfill}
+ [(\t_{4,8}+\t_{-4,8})(\bt_{0,8} + \bt_{8,8})
+ (\t_{0,8}+\t_{8,8})(\bt_{4,8} + \bt_{-4,8})]^2_{\rm twisted}\cr
&\hbox to 1em{\hfill} +
[(\t_{6,8}+\t_{-2,8})(\bt_{2,8}+\bt_{-6,8})
+(\t_{2,8}+\t_{-6,8})(\bt_{-2,8}  +\bt_{6,8})]^2_{\rm twisted}\,\, .}$$
Hence,
$$ Z({\rm theta~factor,~} K=8)=  \vert (c^0_0)^2 +
(c^0_4)^2\vert^2_{_{(A_8)}} + \vert (c^0_0)(c^0_4)
+(c^0_4)(c^0_0)\vert^2_{_{(B_8)}} + 4\vert
(c^0_2)^2\vert^2_{_{(C_8)}}\,\, . \eqno\eqnlabel{nn8-b}$$
\suboff
\hfill\vfill\eject
{\settabs 8\columns
\+ \cr
\+ $D=3$ & $(K=16)$:\cr
\+\cr}

The ${\bmit M}= {\bmit M}^{4,4}$ model twisted by the simple $U(1)$ current
$${\cal J}_{16}\equiv \phi_{K/4}^0\bar\phi_0^0\eqno\eqnlabel{j16}$$
produces,
\subon
$$\eqalignno{ V\left(K=16\right) &= \hbox to .38cm{\hfill}\vert (\t_{0,16} +
\t_{16,16}) +
(\t_{8,16} +\t_{-8,16})\vert^2_{\rm untwisted}\cr
&\cr
&\hbox to 1em{\hfill}
+ \vert (\t_{4,16} + \t_{-4,16}) + (\t_{12,16} +\t_{-12,16})
\vert^2_{\rm untwisted}\,\, .
 &\eqnlabel{nn16-a}}$$
Thus,
$$Z({\rm theta~factor,~} K=16)=\vert c^0_0 + c^0_8\vert^2_{_{(A_{16})}} +
4\vert c^0_4\vert^2_{_{(C_{16})}}\,\, .
\eqno\eqnlabel{nn16-b}$$
\suboff
(In this case the twisting is trivial since ${\cal J}_{16}$ is in the initial
untwisted model.)

The partition function for the standard $D=10$ superstring can also be
factored into affine and theta-function parts:
{\settabs 8\columns
\+ \cr
\+ $D=10$ & $K=2$:\cr}
\subon
$$ A_2 {\buildrel {\rm affine}\over\Longrightarrow}
\sum_{i {\rm ~odd~}= 1}^7 {8\choose i}(\chi_0)^{i}(\chi_K)^{8-i} -
(\chi_{K/2})^8\,\, . \eqno\eqnlabel{affine10-a}$$
The accompanying $U(1)$ factor is
$$\eqalignno{
Z\left({\rm theta~factor},\, K=2\right)&=
\phantom{35\times}\vert (\vartheta_{0,2})^8 +
(\vartheta_{1,2})^8 + (\vartheta_{-1,2})^8 + (\vartheta_{2,2})^8\vert^2\cr
&\phantom{= } + 35\vert(\vartheta_{0,2} + \vartheta_{2,2})^4
(\vartheta_{1,2} + \vartheta_{-1,2})^4\vert^2\cr
&\phantom{= } + 35[(\vartheta_{1,2}+ \vartheta_{-1,2})^4
(\vartheta_{0,2}+\vartheta_{2,2})^4]
[(\bar\vartheta_{0,2}+\bar\vartheta_{2,2})^4
 (\bar\vartheta_{1,2}+\bar\vartheta_{-1,2})^4]\cr
& &\eqnlabel{affine10-b}}$$
\suboff
which\footnote{Note that the effective
$Z\left((n,\bar n),K=2\right)$ contributing to eq.~(\puteqn{part2})
reduces to just the first mod-squared term in eq.~\pe{affine10-b} since
$c^l_n\equiv 0$ for $l-n\neq 0 \pmod{2}$.}
originates from the
$${\bmit M}= {\bmit M}^{2,1}\otimes{\bmit M}^{2,1}\otimes{\bmit
M}^{2,1}\otimes{\bmit M}^{2,1}\otimes{\bmit M}^{2,1}\otimes{\bmit M}^{2,1}
\otimes{\bmit M}^{2,1}\otimes{\bmit M}^{2,1}$$
model twisted by the (simple) current
$${\cal J}^{\rm theta}_2\equiv ( :\exp\{i\varphi/2 \}:)^8\,\, .
\eqno\eqnlabel{j2}$$
The difference between the factorization for $K=2$
and those for $K>2$ is that
here we cannot define an actual parafermion twist current $(\phi_{K/4}^0)^8$
since $\phi^0_{K/4}=0$ for $K=2$.

All of the above simple $U(1)$ twist currents are of the general form
$${\cal J}_K =
(\phi^0_{K/4})^{D-2}(\bar\phi^0_0)^{D-2}\,\, {\rm for~}K>2\,\, .
\eqno\eqnlabel{gensc}$$
We believe this specific class of twist currents
is the key to spacetime
supersymmetry in the parafermion models.\footnote{$\bar {\cal J}_K=
(\phi^0_0)^{D-2}(\bar\phi^0_{K/4})^{D-2}$ is automatically generated as a
twisted state.} Without its twisting effect,
numerically zero fractional superstring MIPF's in three, four, and six
dimensions  cannot be formed and, thus, spacetime SUSY would be impossible.
This twisting also reveals much about the necessity of
non-$A_K$--sectors.  Terms from the twisted and untwisted
sectors of these models become equally mixed in the $\vert A_K\vert^2$,
$\vert B_K\vert^2$, and $\vert C_K\vert^2$ contributions to the level $K$
partition function.
Further, this twisting keeps the string functions with $n\not\equiv 0,\, K/2
\pmod{K}$ from mixing with those possessing $n\equiv 0,K/2 \pmod{K}$.  This is
especially significant since we believe the former string functions
in the $C_K$--sector
likely
correspond to spacetime fields of fractional spin-statistics ({\it i.e.,}
anyons)
and the latter in both $A_K$ and $B_K$ to spacetime bosons and
fermions. If mixing were allowed, normal spacetime SUSY would be
broken and replaced by a fractional supersymmetry, most-likely ruining
Lorentz invariance for $D>3$.

Since in the antiholomorphic sector ${\cal J}_K$ acts as the identity, we
will focus on its effect in the holomorphic sector.
In the $A_K$--sector the operator
$(\phi_{K/4}^0)^{D-2}$ transforms the bosonic (fermionic)
nonprojection
fields into the
fermionic (bosonic) projection fields and
vice-versa.\footnote{We use the same language as the authors of
refs.~\pr{argyres91e}.  Nonprojection refers to the
bosonic and fermionic fields in the $A^{\rm boson}_K$ and $A^{\rm fermion}_K$
subsectors, respectively, corresponding to string functions with positive
coefficients, whereas projection fields refer to those
corresponding to string functions with negative signs.  With this definition
comes an overall minus sign coefficient on $A_K^{\rm fermion}$, as shown in
eq.~(4.2.38a).  For example,
in (4.2.38b), the bosonic non-projection fields are
\hbox {$(\phi^0_0 + \phi^2_0)^3(\phi^1_0)$}
 and the bosonic projection is
$(\phi^1_0)^4$. Similarly, in (4.2.38c)
the fermionic non-projection
field is $(\phi^1_1)^4$ and the projections are $(\phi^0_1 +\phi^2_1)^3
(\phi^1_1)$.}
For example, consider
the effect of this twist current on the fields represented in
\subon
$$ A_4= A^{\rm boson}_4 - A^{\rm fermion}_4\,\, ,
\eqno\eqnlabel{abosferm-a}$$
where
$$\eqalignno{A^{\rm boson}_4 &\equiv 4\left\{ (c^0_0 + c^4_0)^3(c^2_0) -
                                          (c^2_0)^4\right\}
                         &\eqnlabel{abosferm-b}\cr
             A^{\rm fermion}_4 &\equiv 4\left\{
                             (c^2_2)^4 - (c^0_2 + c^4_2)^3(c^2_2)\right\}
\,\, .
                         &\eqnlabel{abosferm-c}\cr}$$
\suboff
Twisting by $(\phi^0_{K/4})^{D-2}$ transforms the related fields according to
\subon
$$\eqalignno{ (\phi^0_0 + \phi^2_0)^3(\phi^1_0) &\hbox to 1cm{\hfill}
{\buildrel {(\phi^0_{K/4})^{D-2}}\over\Longleftrightarrow}\hbox to 1cm{\hfill}
(\phi^2_1 + \phi^0_1)^3 (\phi^1_1)
&\eqnlabel{phitwist-a}\cr (\phi^1_0)^4 &\hbox to 1cm{\hfill}
{\buildrel {(\phi^0_{K/4})^{D-2}}\over\Longleftrightarrow}\hbox to 1cm{\hfill}
(\phi^1_1)^4\,\, .
&\eqnlabel{phitwist-b}\cr}$$
\suboff

Although the full meaning of the projection fields is not yet understood,
the authors of refs.~\pr{argyres91b} and \pr{argyres91e} argue that the
corresponding string functions
should be interpreted as ``internal'' projections, {\it i.e.},
cancellations of degrees of freedom in the fractional superstring models.
Relatedly, the authors show that when the $A_K$--sector is
written as $A^{\rm boson}_K - A^{\rm fermion}_K$, as defined above,
the $q$-expansions of
both $A^{\rm boson}_K$ and $A^{\rm fermion}_K$ are all positive.
Including the
fermionic projection terms results in the identity
\subon
$$\eta^{D-2} A_K^{\rm fermion} = (D-2)\left( {(\theta_2)^4\over
16\eta^4}\right)^{{D-2\over 8}}\,\, .
\eqno\eqnlabel{af-a}$$
Eq.~(\puteqn{af-a}) is the standard theta-function expression for $D-2$
worldsheet Ramond
Major-
ana-Weyl fermions. Further, $$\eta^{D-2} A_K^{\rm boson} = (D-2)\left(
{(\theta_3)^4 - (\theta_4)^4\over 16\eta^4}\right)^{{D-2\over 8}}\,\, .
\eqno\eqnlabel{af-b}$$
\suboff

Now consider the $B_K$--sectors. For $K=4$ and $8$, the operator
$(\phi^0_{K/4})^{D-2}$ transforms the primary fields corresponding to the
partition functions terms in the first set of brackets on the RHS of
eqs.~(\puteqn{part4-b}, \puteqn{part8-b})
into the fields represented by the partition functions terms in the
second set. For example, in the $K=4$ ($D=6$) case
\hfill\vfill\eject
\subon
$$\eqalignno{(\phi^0_0 + \phi^2_0)(\phi^1_0)(\phi^0_1 + \phi^2_1)^2
&\hbox to 1cm{\hfill}
{\buildrel {(\phi^0_{K/4})^{D-2}}\over\Longleftrightarrow}\hbox to 1cm{\hfill}
(\phi^0_1 + \phi^2_1)
(\phi^1_1)(\phi^2_0+\phi^0_0)^2 \hbox to .5cm{\hfill}& \eqnlabel{phib-a}\cr
(\phi^1_0)^2(\phi^1_1)^2 &\hbox to 1cm{\hfill}
{\buildrel {(\phi^0_{K/4})^{D-2}}\over\Longleftrightarrow}
\hbox to 1cm{\hfill} (\phi^1_1)^2(\phi^1_0)^2\,\, .&
\eqnlabel{phib-b}\cr}$$
\suboff
Making an analogy with what occurs in the $A_K$--sector, we suggest that
$(\phi^0_{K/4})^{D-2}$ transforms bosonic (fermionic)
nonprojection fields into fermionic (bosonic) projection fields and
vice-versa in the $B_K$--sector also.
Thus, use of the twist current ${\cal J}_K$ allows for bosonic and fermionic
interpretation of these
fields\footnote{Similar conclusions have been reached by K. Dienes and P.
Argyres for different reasons. They have, in fact, found theta-function
expressions for the $B_K^{\rm boson}$-- and $B_K^{\rm fermion}$--
subsectors.\mpr{dienes92a}}:
\subequationnumstyle{alphabetic}
$$B_4= B^{\rm boson}_4 - B^{\rm fermion}_4\,\, ,\eqno\eqnlabel{bbf-a}$$
where
$$\eqalignno{B^{\rm boson}_4 &\equiv
4\left\{(c^0_0 + c^4_0)(c^2_0)(c^0_2+c^4_2)^2 -
(c^2_0)^2(c^2_2)^2\right\}&\eqnlabel{bbf-b}\cr
B^{\rm fermion}_4 &\equiv
4\left\{(c^2_2)^2(c^2_0)^2 - (c^0_2+c^4_2)(c^2_2)(c^0_0
+c^4_0)^2\right\}\,\, .&\eqnlabel{bbf-c}\cr}$$
\subequationnumstyle{blank}
What appears as the projection term, $(c^2_0)^2(c^2_2)^2$, for the
proposed bosonic part acts as the nonprojection term for the fermionic half,
when the subscripts are reversed.  One interpretation is
this implies a
compactification of two transverse dimensions.\footnote{This was also
suggested in ref.~\pr{argyres91b} working from a different approach.} The
spin-statistics of the physical states of the $D=6$
model as observed in four-dimensional uncompactified spacetime would
be determined
by the (matching) $n$ subscripts of the first two string
functions\footnote{Using the subscripts $n'$ of last two string functions
to define spin-statistics in $D=4$ uncompactified spacetime
corresponds to interchanging the definitions of $B^{\rm boson}_4$ and
$B^{\rm fermion}_4$.}
(corresponding to the two uncompactified transverse dimensions) in each term
of four string functions, $c^{l_1}_n c^{l_2}_n c^{l_3}_{n'} c^{l_4}_{n'}$ .
The $B_8$ terms can
be interpreted similarly when one dimension is compactified.

However, the
$C_K$--sectors are harder to interpret. Under
$(\phi^0_{K/4})^{D-2}$ twisting, string functions with $K/4$ subscripts
are invariant, transforming back into themselves.  Thus, following
the pattern of $A_K$ and $B_K$ we would end up writing, for example,
$C_{16}$ as
\subequationnumstyle{alphabetic}
$$C_{16}= C_{16}^a - C_{16}^b \eqno\eqnlabel{cspin-a}$$ where,
$$\eqalignno{
C^a_{16} &\equiv (c^2_4 + c^{14}_4) - c^8_4
& \eqnlabel{cspin-b}\cr
C^b_{16} &\equiv c^8_4 -
(c^2_4 + c^{14}_4)\,\, . & \eqnlabel{cspin-c}\cr}$$
\subequationnumstyle{blank}

The transformations of the corresponding primary fields are not quite as
trivial, though.
$(\phi^1_2 + \phi^7_2)$ is transformed into its conjugate field
$(\phi^7_{-2} + \phi^1_{-2})$ and likewise $\phi^4_2$ into
$\phi^4_{-2}$,
suggesting that $C^a_{16}$ and $C^b_{16}$ are the partition functions
for conjugate fields. Remember, however, that $C_{16}=0$. Even though we may
interpret this sector as containing two conjugate spacetime fields, this
(trivially) means that the partition function for each is identically zero.
We refer to
this effect in the $C_K$--sector as ``self-cancellation.'' One
interpretation is that there are no states in the $C_K$--sector of the Hilbert
space that survive all of the internal projections. If this is correct,
a question may arise as to the consistency of the $K=8$ and
$16$ theories.  Alternatively, perhaps anyon statistics allow two
(interacting?)  fields of either identical fractional spacetime spins
$s_1=s_2={2m\over K}$, or spacetime spins related by $s_1={2m\over K}= 1- s_2$,
where in both cases $0<m<{K\over 2} \pmod{1}$,
to somehow cancel each other's contribution to the partition function.

Using the $\phi^j_m\equiv\phi^j_{m+K}\equiv\phi^{{K\over
2}-j}_{m-{K\over 2}}$ equivalences at level $K\in 4\Z$, a PCFT has $K/2$
distinct classes of integer $m$ values. If one associates these classes with
distinct spacetime spins (statistics) and assumes $m$ and $-m$ are also
in the same classes since
$(\phi^0_m)^{\dagger}= \phi^0_{-m}$, then the
number of spacetime spin classes reduces to ${K\over 4} +1$. Since $m=0$
$(m= {K\over 4})$ is
associated with spacetime bosons (fermions),
we suggest that general $m$ correspond to particles of
spacetime spin ${2\vert m \vert \over K}$,
${2m\over K} +\Z^+ $, or $ \Z^+ -{2m\over K} $.
If this is so, most likely
${\rm spin}(m)\in \{ {2m\over K},
\Z^+ + {2m\over K} \}$ for $0< m< K/4 \pmod{K/2}$
and ${\rm spin}(m)\in\Z^+ - {2\vert m\vert \over K}$
for $-K/4< m<0 \pmod{K/2}$. This is one of the few
spin assignment rules that maintains the equivalences of the fields $\phi^j_m$
under
$(j,\, m)\rightarrow({k\over 2}-j,\, m-{K\over 2})\rightarrow(j,\, m+K)$
transformations.  According
to this rule, the fields in the $C_K$--sectors have quarter spins
(statistics),
which agrees with prior claims.\mpr{dienes92b,argyres91b,argyres91d}

Also, we do not believe
products of primary fields in different $m$ classes in the $B_K$--sectors
correspond to definite spacetime spin states unless some dimensions are
compactified. Otherwise by our interpretation of $m$ values above, Lorentz
invariance in uncompactified spacetime would be lost.
In particular, Lorentz invariance requires that either all or none of the
transverse modes in uncompactified spacetime be fermionic spinors.
Further, $B$--sector particles apparently cannot
correspond to fractional spacetime spin particles for a
consistent theory. Thus, the $D=6\, (4)$ model must have two (one) of its
dimensions compactified.\footnote{This implies the $D=6,\, 4$ partition
functions are incomplete. Momentum (winding) factors
for the two compactified dimensions would have to be added (with modular
invariance maintained).}

The $B_8$--sector of the $D=4$ model
appears necessary for more reasons than just modular invariance of the
theory.
By the above spacetime
spin assignments, this model suggests massive
spin-quarter states anyons in the $C_K$--sectors,
which presumably cannot exist in $D>3$ uncompactified dimensions.
However, the $B_K$--sector, by forcing compactification to three dimensions
where anyons are allowed, would save the model, making it self-consistent.
Of course, anyons in the  $K=16$ theory with $D_{\rm crit}=3$ are physically
acceptable. (Indeed, no $B_K$--sector
is needed and none exists, which would otherwise reduce the theory to zero
transverse dimensions.) Thus,
$K=8$ and $K=16$ models are probably both allowed solutions for three
uncompactified spacetime dimensional models.
If this interpretation is correct then it is
the $B_K$--sector for $K=8$ which makes that theory self-consistent.

An alternative, less restrictive, assignment of spacetime spin
is possible. Another view is that the $m$ quantum number is not
fundamental for determining spacetime spin. Instead, the
transformation of states under $\phi^{j}_{K/4}$ can be considered to
be what divides the set of states into spacetime bosonic and fermionic
classes. With this interpretation, compactification in the $B_K$--sector
is no more necessary than in the $A_K$--sector. Unfortunately, it is not
{\it a priori} obvious, in this approach, which group of states is bosonic,
and which fermionic. In the $A_K$--sector, the assignment can also
be made phenomenologically.
In the $B_K$--sector, we have no such guide.
Of course, using the $m$ quantum number to determine spacetime spin does
not truly tell us which states have bosonic or fermionic statistics,
since the result depends on the arbitrary choice of which of the
two (one) transverse dimensions to compactify.

A final note of caution involves multiloop modular invariance.
One-loop modular invariance amounts
to invariance under $S$ and $T$ transformations. However modular
invariance at higher orders requires an additional
invariance under $U$ transformations: Dehn twists mixing cycles of
neighboring tori of $g>1$ Riemann
surfaces.\mpr{kawai87a,antoniadis87,antoniadis88}
We believe neither our new method of generating the one-loop
partitions, nor the original method of Argyres {\it et al.} firmly proves the
multi-loop modular invariance that is required for a truly consistent
theory.

\vhalf
{\hb{4.3}{\bfs Beyond the Partition Function: Additional Comments}}\vhalf
\sectionnum=3\equationnum=0

In the last section, we introduced a new derivation of the fractional
superstring partition functions. However,
this previous discussion
did not fully demonstrate the consistency of the
fractional superstrings. Further comparisons
to the $K=2$ superstring are of assistance for this.
Here in this section, we comment on such related aspects of
potential string theories.
We consider the analog of the GSO
projection and the uniqueness of the ``twist'' field $\phi^{K/2}_{K/2}$ for
producing spacetime fermions.
First, however we investigate bosonized representations of the fractional
superstrings and what better understanding of the models, this approach
might reveal.

\vhalf
\sectionnumstyle{blank}
{\hc{4.3.a}{\sl Bosonization of the $K=4$ Theory.}}
\sectionnumstyle{arabic}\sectionnum=3\equationnum=0
\vhalf

Several papers\markup{[\putref{li88}]}
have examined the issue of bosonization of $\Z_K$
parafermion CFTs. Since $0\leq c(K)\leq 2$
for these theories,
generically a $\Z_K$ model can be bosonized using two distinct free bosonic
fields, with one carrying a background charge. The chiral
energy-momentum tensor
for a free bosonic field $X$ with background charge $\alpha_0$ is
$$ T(z)= {1\over 2}[\partial_z X(z)]^2 - {\alpha_0\over 2}\partial^2_z
X(z)\,\, ,
\eqno\eqnlabel{emtensor}$$
which results in
$$ c(X) = 1 - 3(\alpha_0)^2\,.\eqno\eqnlabel{central}$$
For $2<K<\infty$,\footnote{The only primary field for $K=1$ is the vacuum
and, as discussed prior, the $K=2$ theory is the
$c={1\over 2}$ critical Ising (free fermion) model.}
only two $\Z_K$ theories (those at $K= 3,\, 4$) do not require two free
real bosonic fields in the bosonized version and only for $K= 4$ is a
background charge unnecessary since $c(K=4)=1$.  The bosonization process
for the
$\Z_4$ parafermion CFT is straightforward since $c=1$ CFTs have only
three classes
of solutions,
corresponding to a boson propagating on (1) a torus of radius
$R$,
 (2) a
$\Z_2$ orbifold of radius $R$, or (3) discrete orbifold spaces defined on
$SU(2)/\Gamma_i$, where $\Gamma_i$ are discrete subgroups of
$SU(2)$.\mpr{kiritsis88}

The $\Z_4$ parafermion CFT is identical\mpr{ginsparg88}
to the $\Z_2$ orbifold at radius
$R=\sqrt{6}/2$ (and $R=1/\sqrt{6}$ by duality).
The $\Z_4$ primary fields with their conformal dimensions and corresponding
partition functions for the Verma modules are listed in Table 4.1.
\vskip .4cm

\centertext{Table 4.1 $\Z_4$ Primary Fields}
$$\vbox{\settabs 3\columns
\+ {\underbar{\hbox to 3.7cm{\hfill Primary Fields\hfill}}}
& {\underbar{Conformal Dimension h}}
& {\underbar{Partition Fn.}}\cr
\+ \hbox to 3.7cm{\hfill $\phi^0_0\equiv\phi_0$\hfill}
&\hbox to 4.3cm{\hfill 0\hfill}
&\hbox to .9cm{\hfill}$\eta c^0_0$ \cr
\+ \hbox to 3.7cm{\hfill $\phi^2_{-1}=\phi^0_1\equiv\phi_1
= \phi^{\dag}_3$\hfill}
&\hbox to 4.3cm{\hfill $3\over 4$\hfill}
&\hbox to .9cm{\hfill}$\eta c^4_2$\cr
\+ \hbox to 3.7cm{\hfill $\phi^2_0=\phi^0_2\equiv\phi_2$\hfill}
&\hbox to 4.3cm{\hfill 1\hfill}
&\hbox to .9cm{\hfill}$\eta c^4_0$ \cr
\+ \hbox to 3.7cm{\hfill $\phi^2_{1}=\phi^0_3\equiv\phi_3
=\phi^{\dag}_1$\hfill}
&\hbox to 4.3cm{\hfill $3\over 4$\hfill}
&\hbox to .9cm{\hfill}$\eta c^4_2$ \cr
\+ \hbox to 3.7cm{\hfill $\phi^1_0\equiv\ep$\hfill}
&\hbox to 4.3cm{\hfill $1\over 3$\hfill}
&\hbox to .9cm{\hfill}$\eta c^2_0$ \cr
\+ \hbox to 3.7cm{\hfill $\phi^1_1=\phi^1_{-1}$\hfill}
&\hbox to 4.3cm{\hfill $1\over 12$\hfill}
&\hbox to .9cm{\hfill}$\eta c^2_2$ \cr
\+ \cr
\+ \hbox to 3.7cm{\hfill $\phi^{1/2}_{-1/2}$\hfill}
&\hbox to 4.3cm{\hfill $1\over 16$\hfill}
&\hbox to .9cm{\hfill}$\eta c^1_{-1}$ \cr
\+ \hbox to 3.7cm{\hfill $\phi^{1/2}_{1/2}$\hfill}
&\hbox to 4.3cm{\hfill $1\over 16$\hfill}
&\hbox to .9cm{\hfill}$\eta c^1_1$ \cr
\+ \hbox to 3.7cm{\hfill $\phi^{3/2}_{-1/2}$\hfill}
&\hbox to 4.3cm{\hfill $9\over 16$\hfill}
&\hbox to .9cm{\hfill}$\eta c^3_{-1}$ \cr
\+ \hbox to 3.7cm{\hfill $\phi^{3/2}_{1/2}$\hfill}
&\hbox to 4.3cm{\hfill $9\over 16$\hfill}
&\hbox to .9cm{\hfill}$\eta c^3_1$ \cr}$$
\vskip .2cm
An $S^1/\Z_2$ orbifold at radius $R$ has the
partition function
\subequationnumstyle{alphabetic}
$$\eqalignno {Z_{\rm orb}(R) &= {1\over 2}\{Z(R)
+ {\vert\eta\vert\over\vert\theta_2\vert}
+ {\vert\eta\vert\over\vert\theta_3\vert}
+ {\vert\eta\vert\over\vert\theta_4\vert}\}
& \eqnlabel{orb-a}\cr
&= {1\over 2}\{Z(R) + {\vert\theta_3\theta_4\vert\over
\vert\eta\vert^2} +   {\vert\theta_2\theta_4\vert\over \vert\eta\vert^2}
                    + {\vert\theta_2\theta_3\vert\over
\vert\eta\vert^2}\} & \eqnlabel{orb-b}\cr}$$
where $\theta_{i= 1{\rm ~to~} 4}$ are the classical Jacobi theta-functions.
$$ Z(R) = {1\over \eta\bar\eta}\sum^{\infty}_{m,n=\infty}
q^{\{{m\over 2R} + nR\}^2/2}\bar q^{\{{m\over 2R} - nR\}^2/2}
\eqno\eqnlabel{pf99}$$
\subequationnumstyle{blank}
is the partition function for a free scalar boson compactified on a circle
of radius $R$.
For $R={\sqrt{6}\over 2}$ the generalized momentum states
$p={m\over\sqrt{6}}+{n\sqrt{6}\over 2}$ can be categorized into four
classes
based on the value of ${p^2\over 2} \pmod{1}$. The classes are
${p^2\over 2}= 0,{1\over 12},{1\over 3}, {\rm ~and~} {3\over 4}\pmod{1}$.
$p= {m\over\sqrt{6}} + {n\sqrt{6}\over 2}$ and
$\bar p= {m\over\sqrt{6}} - {n\sqrt{6}\over 2}$ belong to the same
class.
That is,
$${1\over 2}(p^2 -\bar p^2) \equiv 0 {\rm ~mod~} 1\, ,\eqno\eqnlabel{pf2}$$
(as required by modular invariance or, equivalently, by level
matching.)\mpr{narain86,narain87}

The untwisted sector of the model corresponds to the first two terms on the
right-hand side of eq.~(\puteqn{pf99}) and the twisted sector, the remaining
two terms. The factor of ${1\over 2}$ is due to the GSO projection from
the $\Z_2$ orbifolding, requiring invariance of states under
$g: ~~X(z,\bar z)\rightarrow -X(z,\bar z)$.  In the untwisted sector this
invariance requires pairing together of momentum
states $\vert m,n\rangle$ and $\vert -m,-n\rangle$ and so projects out
half the original number of states in the untwisted sector.
The second term in (\puteqn{orb-a}) and (\puteqn{orb-b}) correspond to
states antiperiodic along the ``time'' loop and thus can only be states
built from (net even numbers) of $\alpha(z)$ and $\bar \alpha(\bar z)$'s acting
on $\left\vert m=n=0\right\rangle$.
The twisted sector states correspond to a total even number of
$\alpha_{r}$ and $\bar \alpha_{r}$, $r\in\Z+{1\over2}$, oscillations acting on
the
$\left\vert m=n=0\right\rangle$
twisted vacuum with $h=\bar h={1\over16}$. Thus the twisted states have
conformal dimensions of the form
$(h,\bar h)\in ({1\over16}+\Z,\, {1\over16}+\Z)$
or $({1\over16}+\Z+{{1\over2}},\, {1\over16}+\Z+{{1\over2}})$.

The first six primary fields of the $\Z_K$ PCFT listed in Table 4.2 have
representations in the untwisted sector of $Z$(orbifold, $R={\sqrt6\over2}$)
and the latter four have representations in the twisted
sector.\footnote {We note that independent of the choice of the affine factor
in the partition functions of section 4.2, the required ($n$,$\bar n$)
partition functions of (\puteqn{nn4-a}, \puteqn{nn8-a}, \puteqn{nn16-a})
effectively remove from the theory
the primary fields with half integer $j$, $m$. The only theory which uses
the twisted sector is the $K=2$ superstring. The significance of this
observation is under investigation.} From the classes of
${p^2\over2}=0$, ${1\over12}$, $1\over3$,
$3\over4$ $\pmod{1}$
states we find the following identities for string functions:
\subon
$$\eqalignno
{\vert\eta c^0_0\vert^2 + \vert\eta c^4_0\vert^2
&={1\over2}
\left\{
{1\over\vert\eta\vert^2}
\sum_{{p^2\over2}\equiv 0 {\rm~mod~} 12}
q^{({m\over2R}+nR)^2/2}
\bar q^{({m\over2R}-nR)^2/2} +
{\vert\theta_3\theta_4\vert\over\vert\eta\vert^2}
\right\}{\hbox to .75cm{\hfill}}
&\eqnlabel{defc-a}
\cr
\vert\eta c^2_2\vert^2
&={1\over2}{1\over\vert\eta\vert^2}
\sum_{{p^2\over2}\equiv 1 {\rm ~mod~} 12}
q^{({m\over2R}+nR)^2/2}
\bar q^{({m\over2R}-nR)^2/2}
&\eqnlabel{defc-b}
\cr
\vert\eta c^2_0\vert^2
&={1\over2}{1\over\vert\eta\vert^2}
\sum_{{p^2\over2}\equiv 4 {\rm ~mod~} 12}
q^{({m\over2R}+nR)^2/2}
\bar q^{({m\over2R}-nR)^2/2}
&\eqnlabel{defc-c}
\cr
\vert\eta c^4_{2}\vert^2 =
\vert\eta c^4_{-2}\vert^2
&={1\over4}{1\over\vert\eta\vert^2}
\sum_{{p^2\over2}\equiv 9 {\rm ~mod~} 12}
q^{({m\over2R}+nR)^2/2}
\bar q^{({m\over2R}-nR)^2/2}
&\eqnlabel{defc-d}
\cr
\vert\eta c^1_{1}\vert^2 +
\vert\eta c^3_{1}\vert^2
&=
\vert\eta c^1_{-1}\vert^2 +
\vert\eta c^3_{-1}\vert^2
={1\over4}
\left\{
{\vert\theta_2\theta_4\vert\over\vert\eta\vert^2} +
{\vert\theta_2\theta_3\vert\over\vert\eta\vert^2}
\right\}\,\, .
&\eqnlabel{defc-e}
\cr}$$
\suboff

Identities for the primary fields partition functions are
possible from this bosoni-
zation. Since only the $\phi^j_m$ with integer
$j$, $m$
are in the $\Z_4$ model, we will henceforth concentrate on the untwisted
sector of the $S/\Z_2$ model. We make the following identifications between
the primary fields of the two models:
\vskip .4cm

\centertext{Table 4.2 Primary Field Representation From Orbifold Bosonization}
$$\vbox{\settabs 3 \columns
\+{$\Z_4$ Primary Field}
&{\hbox to 1.2cm{\hfill} $S/\Z_2$ \hbox to 1.2cm{\hfill}}
&\hbox to .4cm{\hfill $h$\hfill}
\cr
\+ {\overline{\phantom{$\Z_4$ Primary Field}}}
&{\overline{\phantom{\hbox to 1.2cm{\hfill} $S/\Z_2$ \hbox to 1.2cm{\hfill}}}}
&{\overline{\phantom{\hbox to .4cm{\hfill $h$\hfill}}}}
\cr
\+\hbox to 3.1cm{\hfill $\phi_0(z)$\hfill}
&\hbox to 3.6cm{\hfill 1\hfill}
&\hbox to .4cm{\hfill 0\hfill}
\cr
\+\hbox to 3.1cm{\hfill $\phi_1(z)+\phi_{-1}(z)$\hfill}
&\hbox to 3.6cm{\hfill $\e^{i{3\over\sqrt6} X(z)}
+ \e^{-i{3\over\sqrt6} X(z)}$\hfill}
&\hbox to .4cm{\hfill $3\over4$\hfill}
\cr
\+\hbox to 3.1cm{\hfill $\phi_2(z)$\hfill}
&\hbox to 3.6cm{\hfill $i\partial X$\hfill}
&\hbox to .4cm{\hfill 1\hfill}
\cr
\+\hbox to 3.1cm{\hfill $\epsilon(z)$\hfill}
&\hbox to 3.6cm{\hfill $\e^{i{2\over\sqrt6} X(z)}
+ \e^{-i{2\over\sqrt6} X(z)}$\hfill}
&\hbox to .4cm{\hfill $1\over3$\hfill}
\cr
\+\hbox to 3.1cm{\hfill $\phi^1_1(z)$\hfill}
&\hbox to 3.6cm{\hfill $\e^{i{1\over\sqrt6} X(z)}
+ \e^{-i{1\over\sqrt6} X(z)}$\hfill}
&$1\over12$\cr}$$\vskip .2cm
\noindent ($\phi_1$ and $\phi_{-1}$ must be paired together since the
$S/\Z_2$ physical state is $\e^{+}+\e^{-}$, where $\epsilon^{+}\equiv
\e^{i{3\over\sqrt6} X}$, $\epsilon^{-}\equiv \e^{-i{3\over\sqrt6} X}$.)
Perhaps the first aspect that becomes apparent is how to represent the
fractional supercurrent, $J_{\rm FSC}$ as $J_{\rm FSC}^+ + J_{\rm FSC}^-$:
\subon
$$\eqalignno{
J_{{\rm FSC},\, {-{4\over3}}}
&=\epsilon\partial X + :\epsilon\epsilon:
&\eqnlabel{supcur-a}\cr
&=\epsilon^+\partial X + :\epsilon^+\epsilon^+:+\epsilon^-\partial X +
:\epsilon^-\epsilon^-:
&\eqnlabel{supcur-b}\cr
&= J^+_{{\rm FSC},\, {-{4\over3}}}
+  J^-_{{\rm FSC},\, {-{4\over3}}}
&\eqnlabel{supcur-c}\cr}$$
\suboff
with $\e^{\pm i{4\over\sqrt6}}$ the only candidates for
$:\epsilon^{\pm}\epsilon^{\pm}:\,$.\footnote{Subsequently,
[\putref{argyres91e}]
has shown that closure of the fractional current and energy-momentum OPEs
requires $:\epsilon^{\pm}\epsilon^{\pm}:= \e^{\pm i{4\over\sqrt6}}$
be the descendent term in $G^{\mp}$, respectively.}

Since the identities (\puteqn{defc-a}--\puteqn{defc-e}) involve
$\vert\eta c^{2j}_{2m}\vert^2$ rather
than just $\eta c^{2j}_{2m}$, they do not necessarily imply the exact
equivalence of the parafermion and orbifold models. However, more
fundamental identities for the string functions do exist.
Since none of the $\Z_4$ parafermion fields
connected with the twisted orbifold sector appear in the $
K=4$ FSC model, we can
look just at a left-moving (holomorphic) boson compactified
on a circle
with $R={\sqrt6}$, but not $\Z_2$ twisted.
\subon
$$Z(z,R={\sqrt6})=
{1\over\eta}
\sum_{m=-\infty}^{\infty}
q^{[{m\over R}]^2/2}\,\, .\eqno\eqnlabel{zpart-a}$$
If we change summation using $m=6n+i$, $i=0 {\rm ~to~}5$,
then the partition
function can be split into\footnote{Note that $m=i \pmod{6}$ terms
are equivalent to $m=-i \pmod{6}$ terms, so if we include a factor of
two, we need only sum over $i=0 {\rm ~to~} 3$.}
$$Z(z,R={\sqrt6})=
{1\over\eta}
\sum_{n={-\infty}\atop{i=0 {\rm ~to~} 5}}^{\infty}
q^{[{(6n+i)\over R}]^2/2}\,\, .\eqno\eqnlabel{zpart-b}$$\suboff
This suggests the following more succinct
identities:\footnote{These were verified up to $q^{1300}$ using mathematica.}
\subon
$$\eqalignno
{\eta c^2_2
&={1\over\eta}q^{1\over12}
\sum_{n=-\infty}^{\infty}
q^{3n^2+n}
&\eqnlabel{id4-a}
\cr
\eta c^2_0
&={1\over\eta}q^{1\over3}
\sum_{n=-\infty}^{\infty}
q^{3n^2+2n}
&\eqnlabel{id4-b}
\cr
\eta c^4_2 = \eta c^4_{-2}
&={1\over 2\eta}q^{3\over4}
\sum_{n=-\infty}^{\infty}
q^{3n^2+3n}
&\eqnlabel{id4-c}
\cr
\eta (c^0_0+c^4_0)
&={1\over\eta}
\sum_{n=-\infty}^{\infty}
q^{3n^2}\,\, .
&\eqnlabel{id4-d}
\cr}$$
\suboff
The corresponding free boson representations of the parafermion primary
fields are given in Table 4.3:
\centertext{\hbox to 1cm{\hfill}}
\centertext{Table 4.3 Primary Field Representation From $R= {{\sqrt{6}}}$
Bosonization}
$$\vbox{\settabs 4 \columns
\+ {{$\Z_4$ Parafermion}}\hskip .5cm {{$R={\sqrt 6}$ Boson Rep.}}
&&\hbox to 2.65cm{\hfill Verma Module\hfill}
&\hbox to 4cm{\hfill Boson Rep.\hfill}\cr
\+ {\overline{\phantom{$\Z_4$ Parafermion}}}\hskip .5cm
{\overline{\phantom{\hbox to 3.63cm{$R={6}$ Boson Rep.\hfill}}}}
&&\hbox to .6cm{\hfill}{\overline{\phantom{\hbox to 2.65cm{\hfill Verma
Module\hfill}}}}
&\overline{\phantom{\hbox to 4cm{\hfill Boson Rep.\hfill}}}\cr
\+ \hbox to 2.8cm{\hfill $\phi_0$\hfill}\hskip .5cm
\hbox to 3.6cm{\hfill 1\hfill}
&&\hbox to 2.65cm{\hfill $[\phi_0]$\hfill}
&$\{ 1,\, \e^{i{6n\over\sqrt 6}X} + \e^{-i {6n\over\sqrt 6}}\, ;$\cr
\+ &&& \hskip 2.7cm $n>0 \} $\cr
\+ \hbox to 2.8cm{\hfill $\phi_1$\hfill}\hskip .5cm
\hbox to 3.6cm{\hfill $\e^{i{3\over\sqrt6} X}$\hfill}
&&\hbox to 2.65cm{\hfill [$\phi_1$]\hfill}
&\hbox to 4cm{\hfill $\{\e^{i{6n+3\over\sqrt6} X}\}$\hfill}\cr
\+ \hbox to 2.8cm{\hfill $\phi_{-1}$\hfill}\hskip .5cm
\hbox to 3.6cm{\hfill $\e^{-i{3\over\sqrt6} X}$\hfill}
&&\hbox to 2.65cm{\hfill [$\phi_{-1}$]\hfill}
&\hbox to 4cm{\hfill $\{\e^{-i{6n+3\over\sqrt6} X}\}$\hfill}\cr
\+ \hbox to 2.8cm{\hfill $\phi_2$\hfill}\hskip .5cm
\hbox to 3.6cm{\hfill $i\partial X$\hfill}
&&\hbox to 2.65cm{\hfill [$\phi_{-2}$]\hfill}
&\hbox to 4cm{\hfill $\{\alpha_{-n}\}$\hfill}\cr
\+ \hbox to 2.8cm{\hfill $\epsilon$\hfill}\hskip .5cm
\hbox to 3.6cm{\hfill $\e^{i{2\over\sqrt6} X}$,
$\e^{-i{2\over\sqrt6} X}$\hfill}
&&\hbox to 2.65cm{\hfill [$\epsilon$]\hfill}
&\hbox to 4cm{\hfill $\{\e^{i{6n+2\over\sqrt6} X}\}$,
$\{\e^{-i{6n+2\over\sqrt6} X}\}$\hfill}\cr
\+ \hbox to 2.8cm{\hfill $\phi^1_1$\hfill}\hskip .5cm
\hbox to 3.6cm{\hfill $\e^{i{1\over\sqrt6} X}$,
$\e^{-i{1\over\sqrt6} X}$\hfill}
&&\hbox to 2.65cm{\hfill [$\phi^1_1$]\hfill}
&\hbox to 4cm{\hfill $\{\e^{i{6n+1\over\sqrt6} X}\}$,
$\{\e^{-i{6n+1\over\sqrt6} X}\}$\hfill}\cr}$$\vskip .2cm
In this representation
$\phi_1$ and $\phi_{-1}$ need not be paired together.
Also, $\epsilon$ and $\phi^1_1$ have double representations.
For $\epsilon$, this allows the fractional supercurrent,
$J_{\rm FSC}$, to be
expressed as
$J^+_{{\rm FSC}}+ J^-_{{\rm FSC}}$.
For $\phi^1_1$, this should correspond to the two
spin modes, call them $(+)$ and $(-)$. The zero mode of one representation
of $\epsilon$
should act as a raising operator between these spin states and the other as
a lowering operator:
$${\eqalign{ \epsilon^+_0 (+) &= (-)\cr
\epsilon^+_0 (-) &= 0\cr
\epsilon^-_0 (-) &= (+)\cr
\epsilon^-_0 (+) &= 0\,\, .\cr}}\eqno\eqnlabel{spinmodes}$$

The free boson/orbifold representations of the $\Z_4$ parafermion
CFT should be a valuable tool for better understanding the $K=4$ FSC model,
especially its associated partition function.
\hfill\vfill\eject
{\hc{4.3.b}{\sl Generalized Commutation Relations and the GSO Projection}}
\vhalf

One of the major complications of
generalizing from the $K=2$ fermion case to $K>2$ is that the parafermions
(and bosonic field representations) do not have simple commutation
relations.\markup{[\putref{zamol87}]}
What are the commutation relations for non-(half) integral spin particles?
Naively,
the first possible generalization of standard (anti-)commutation
relations for two
fields $A$ and $B$ with fractional spins seems to be:
$$AB-\e^{[i4\pi\, {\rm spin}(A)\,  {\rm spin}(B)]}BA=0\eqno\eqnlabel{wrong}$$
(which reduces to the expected result for bosons and fermions). This is too
simple a generalization, however.\markup{[\putref{wilczek90}]}
Fractional spin particles must be representations
of the braid group.\mpr{wilczek90}
Zamolodchikov and Fateev\markup{[\putref{zamol87}]} have
shown that worldsheet parafermions (of fractional spin) have complicated
commutation relations that
involve an infinite number of modes of a given field. For example:
\subon
$$\eqalignno{\sum_{l=0}^{\infty} C^{(l)}_{(-1/3)}&\left[
A_{n+(1-q)/3-l} A^{\dagger}_{m-(1-q)/3 +l} +
A^{\dagger}_{m- (2-q)/3-l}A_{n-(2-q)/3 +l}
\right]=\cr
&-{1\over 2}\left( n-{q\over 3}\right) \left( n+1 -{q\over 3}\right)
\delta_{n+m,0} + {8\over 3c}L_{n+m}&\eqnlabel{commut-a}}$$
and
$$\sum_{l=0}^{\infty} C^{(l)}_{(-2/3)}
\left[
A_{n-q/3-l}A_{m+(2-q)/3+l}-
A_{m-q/3-l}A_{n+(2-q)/3+l}\right]
= {\lambda\over 2}\left( n-m \right)A^{\dagger}_{(2-2q)/3 +n+m}\,\, ,
\eqno\eqnlabel{commut-b}$$
\suboff
where $A$ is a parafermion field, and
$L_n$ are the generators of the
Virasoro algebra.
$\lambda$ is a real coefficient,
$n$ is integer, and $q=0,1,2 \pmod{3}$ is a $\Z_3$ charge of
Zamolodchikov and Fateev that can be
assigned to each primary field in the $K=4$ model.
The coefficients, $C^{(l)}_{(\alpha)}$, are determined by
the power expansion
$$(1-x)^{\alpha}= \sum_{l=0}^{\infty} C^{(l)}_{(\alpha)}x^l\,\,
.\eqno\eqnlabel{cla}$$ As usual,
$c\equiv {2(K-1)\over K+2}$ is the central charge of the level-K PCFT.
These commutation relations were derived from the OPE of the related
fields.\markup{[\putref{zamol87}]}
(Hence more terms in a given OPE result in more complicated
commutation relations.)
Similar relations between the modes of two different primary
fields can also be derived from their OPE's.
The significance of these commutation relations is that they
severely reduce the number of
distinct physical states in parafermionic models. There are
several equivalent ways of creating a given physical state from the
vacuum using different mode excitations from different parafermion
primary fields
in the same CFT. Thus, the actual Hilbert space of states for this $K=4$
model will be much reduced compared to the space prior to moding
out by these equivalences.\footnote{These equivalences
have subsequently been explicitly shown and the distinct low ${\rm mass}^2$
fields determined in Argyres {\it et al.}\markup{[\putref{argyres91e}]}}

Although the fields in the PCFT do not (anti-)commute,
but instead have complicated commutation relations, some insight can be
gained by comparing the $D=6$, $K=4$ FSC model to the standard $D=10$
superstring. We can, in fact, draw parallels between $\epsilon$ and the
standard fermionic superpartner, $\psi$, of an uncompactified boson X.
In the free fermion approach, developed simultaneously
 by Kawai, Lewellen and Tye
and by
Antoniadis, Bachas and Kounas, generalized GSO projections based on boundary
conditions of the world sheet fermions are
formed.\mpr{kawai87a,antoniadis87,antoniadis88}
Fermions with
half-integer modes (NS-type) are responsible for $\Z_1$ (trivial)
projections; fermions
with integer modes (R-type) induce $\Z_2$ projections. In the non-Ramond
sectors these $\Z_2$ projections remove complete states, while in the
Ramond sector itself,
eliminate half of the spin modes, giving chirality. Fermions
with general complex boundary conditions,
$$\psi (\sigma_1=2\pi)
=-\e^{i{\pi x}}\psi (\sigma_{1}=0)\,\, ,\eq{fbc}$$
where $x\equiv {a\over b}$ is rational with
$a$ and $b$ coprime and chosen in the range
$-1\leq x < 1$,
form in the non-Ramond sector $\Z_{2b}$ projections if $a$ is odd and $\Z_b$
projections if $a$ is even.
For free-fermion models, the GSO operator, originating
from a sector where the world sheet
fermions $\psi^i$ have  boundary conditions
\subon
$$\psi^i(2\pi)= -\e^{i\pi x^i}\psi^i(0)\, ,\eqno\eqnlabel{fbc-a}$$
and that acts on a physical state $\vert {\rm phys}\rangle_{\vec y}$
in a sector where the same
fermions have boundary conditions
$$\psi^i(2\pi)= -\e^{i\pi y^i}\psi(0)\, ,\eqno\eqnlabel{fbc-b}$$
\suboff
takes the form,
$$\left\{\e^{i\pi {\vec x}\cdot {\vec F}_{\vec y}}=\delta_{\vec y}
C({\vec y}\vert {\vec x})\right\}\vert {\rm phys}\rangle_{\vec y}
\eqno\eqnlabel{gso}$$
for states surviving the projection.
Those states not satisfying the demands of the GSO operator for at least
one sector $\vec x$ will not appear in the partition function of the
corresponding model.\footnote{In eq.~(\puteqn{gso}),
${\vec F}_{\vec y}$ is the (vector) fermion number
operator for states in sector $\vec y$. $\delta_{\vec y}$
is $-1$ if either the left-moving or right-moving $\psi^{\rm sfpacetime}$ are
periodic and 1 otherwise.
$C(\vec y\vert\vec x)$ is a phase with value chosen from an allowed set
of order $g_{\vec y,\vec x} = GCD(N_{\vec y},N_{\vec x})$, where
$N_{\vec y}$ is the lowest positive integer such that
$$N_{\vec y}\times \vec y = \vec 0 \pmod{2}\,\, .$$}

The boundary conditions are encoded in
the mode expansions of the complex fermion field, $\psi^+$,
and its complex conjugate, $\psi^-$, on a torus.
These have the following form for a general twist by $x\equiv {a\over b}$:
\subon
$$ \eqalignno{
\psi^+ (\sigma_1, \sigma_2) &=
\sum_{n= 1}^{\infty}
[
\psi_{n-1/2-x/2}^{\alpha}\,
\exp\left\{ -i(n - 1/2 - x/2) (\sigma_1 +\sigma_2) \right\} \cr
&\phantom{\sum_{n= 1}^{\infty}}\hbox to .2truecm{\hfill}
 + \psi_{1/2 - n - x/2}^{\beta}\,
\exp\left\{-i(1/2 - n - x/2)(\sigma_1+\sigma_2 )\right\}
]
&\eqnlabel{psimodes-a}\cr
\psi^- (\sigma_1, \sigma_2) &=
\sum_{n= 1}^{\infty}[
 \psi_{1/2 - n + x/2}^{\alpha}\,
\exp\left\{-i(1/2 - n + x/2 )(\sigma_1+\sigma_2)
\right\}\cr
&\phantom{\sum_{n= 1}^{\infty}}\hbox to .2truecm{\hfill}
+ \psi_{n - 1/2 +x/2}^{\beta}\,
\exp\left\{-i( n - 1/2 +x/2)(\sigma_1 +\sigma_2)\right\}]
&\eqnlabel{psimodes-b}}$$
\suboff
(where ${\psi^{\alpha}_r}^{\dag}\equiv\psi^{\alpha}_{-r}$ and
${\psi^{\beta}_r}^{\dag}\equiv\psi^{\beta}_{-r}$).
$\psi_r^{\alpha}$ and $\psi_r^{\beta}$ are independent modes. Thus,
\subon
$$\eqalignno{
\psi^+(\sigma_1+2\pi)&=
e^{+i 2\pi (1/2)}\,  e^{i\pi x}\, \psi^+(\sigma_1)
&\eqnlabel{twistbcnspsi-a}\cr
\psi^-(\sigma_1+2\pi)&=
e^{-i 2\pi(1/2)}\,  e^{-i\pi x}\, \psi^-(\sigma_1).
&\eqnlabel{twistbcnspsi-b}}$$
\suboff
The specification of the fields is completed by stating the commutation
relation  that the modes obey,
\subon
$$\eqalignno{
\left\{{\psi^{\alpha}_c}^{\dag},\psi^{\alpha}_d\right\}
&=\left\{ {\psi^{\beta}_c}^{\dag},\psi^{\beta}_d\right\}=
\delta_{cd}\,\,  , &\eqnlabel{anticomm-a}\cr
\left\{{\psi^{\alpha}_c}^{\dag},\psi^{\beta}_d\right\}
&=\left\{ {\psi^{\beta}_c}^{\dag},\psi^{\alpha}_d\right\}= 0
\,\,  . &\eqnlabel{anticomm-b}}
$$
\suboff
A similar analysis can be done with the $\epsilon = \phi^1_0$
fields of the $K=4$ parafermion theory.
The normal untwisted ({\it i.e.,} Neveu-Schwarz) modes of $\epsilon$ are
$\epsilon^+_{-{1\over3}-n}$ and $\epsilon^-_{{1\over3}-n}$
where $n\in\Z$. That is, untwisted $\epsilon= (\epsilon^+, \epsilon^-)$
has the following normal-mode expansions.
{\settabs 3\columns
\+ \cr
\+ N-S Sector:\cr}
\subon
$$ \eqalignno{
\epsilon^+ (\sigma_1, \sigma_2) &=
\sum_{n= 1}^{\infty}
[
\epsilon_{n-1/3}^{\alpha}
\,
\exp\left\{ -i(n - 1/3) (\sigma_1 +\sigma_2) \right\} \cr
&\phantom{\sum_{n= 1}^{\infty}}\hbox to .2truecm{\hfill}
 + \epsilon_{2/3 - n }^{\beta}
\,
\exp\left\{-i(2/3 - n )(\sigma_1+\sigma_2 )\right\}
]
&\eqnlabel{emodes-a}\cr
\epsilon^- (\sigma_1, \sigma_2) &=
\sum_{n= 1}^{\infty}[
 \epsilon^{\alpha}_{1/3 - n }\, \exp\left\{-i(1/3 - n
)(\sigma_1+\sigma_2)\right\}\cr
&\phantom{\sum_{n= 1}^{\infty}}\hbox to .2truecm{\hfill}
+ \epsilon^{\beta}_{n - 2/3 }\,
\exp\left\{-i( n - 2/3 )(\sigma_1 +\sigma_2)\right\}]
&\eqnlabel{emodes-b}}$$
\suboff
(where ${\epsilon^{\alpha}_r}^{\dag}=\epsilon^{\alpha}_{-r}$ and
${\epsilon^{\beta}_r}^{\dag}=\epsilon^{\beta}_{-r}$).
Similarly, the associated boundary conditions in this sector are
\subon
$$ \eqalignno{
\epsilon^+(\sigma_1+2\pi)&=
   \e^{+i 2\pi/3}\, \epsilon^+(\sigma_1)
&\eqnlabel{bcns-a}\cr
\epsilon^-(\sigma_1+2\pi)&=
   \e^{-i 2\pi/3}\, \epsilon^-(\sigma_1)\,\, .
&\eqnlabel{bcns-b}}$$
\suboff

Like the standard fermion, the $\epsilon$ operators at $K=4$
can be in twisted sectors,
where the normal-mode
expansions have the following form.
\hfill\vfill\eject
{\settabs 3\columns
\+ General Twisted Sector:\cr}
\subon
$$\eqalignno{
\epsilon^+ (\sigma_1, \sigma_2) &= \sum_{n= 1}^{\infty}[
\epsilon^{\alpha}_{n - {1/3} - {x/2} }\,
    \exp\left\{ -i(n - {1/3} - {x/2} )(\sigma_1 +\sigma_2)\right\}\cr
&\phantom{\sum_{n= 1}^{\infty}}\hbox to .2truecm{\hfill}
+ \epsilon^{\beta}_{{2/3} - n - {x/2}  }\,
\exp\left\{-i({2/3} - n - {x/2} )(\sigma_1 +\sigma_2)\right\} ]
&\eqnlabel{twistmodes-a}\cr
\epsilon^- (\sigma_1, \sigma_2) &=
\sum_{n= 1}^{\infty}[
\epsilon^{\alpha}_{{1/3} - n +  {x/2} }\,
\exp\left\{-i({1/3} - n + {x/2} )(\sigma_1 +\sigma_2)\right\}\cr
&\phantom{\sum_{n= 1}^{\infty}}\hbox to .2truecm{\hfill}
+ \epsilon^{\beta}_{n - {2\over 3} + {x/2}}\,
\exp\left\{ -i( n - {2/3} + x/2)(\sigma_1 +\sigma_2)\right\} ]\,\, .
&\eqnlabel{twistmodes-b}}$$
\suboff
The associated boundary conditions are
\subon
$$\eqalignno{
\epsilon^+(\sigma_1+2\pi)&=
\e^{+i 2\pi (1/3)}\,  \e^{i\pi {x}}\, \epsilon^+(\sigma_1)
&\eqnlabel{twistbcns-a}\cr
\epsilon^-(\sigma_1+2\pi)&=
\e^{-i 2\pi(1/3)}\,  \e^{-i\pi {x}}\, \epsilon^-(\sigma_1)\,\, .
&\eqnlabel{twistbcns-b}}$$
\suboff
The complicated
commutation relations of the modes of $\epsilon$ have already been discussed.
(See eq.~\pe{commut-a}.)

{}From the analogy of free-fermion models, we suggest that in $K=4$ parafermion
models the presence of a sector containing twisted $\epsilon$ fields
with boundary conditions
(\puteqn{twistbcns-a}) or (\puteqn{twistbcns-b}) will result in
$\Z_{b}$ or $\Z_2\times\Z_{b}$ GSO projections,
depending on whether $a$ is even or odd respectively. (We assume as before
that $a$ and $b$ are relative primes but now use the range
 $-2/3\leq x\equiv a/b< 4/3$.)
Zero-modes correspond to a twist by $x= -2/3$.
Whatever the GSO projection is, states resulting from $D$ factors
of $\epsilon_0$ acting on the fermionic
vacuum must survive, in order to have spacetime fermions. Thus, we
conjecture that the presence of these (twisted) zero-modes
$\epsilon_n$, $n\in\Z$ in a model, results in a
generalized $\Z_3$ GSO projection.
Likewise for $K=8$ and $16$,
one might expect $\Z_5$ and
$\Z_9$ projections, respectively. Such projections for
$K=8$ and $16$ could be significantly altered though, by the effects of the
non-Abelian braiding of the non-local interactions.

One other aspect to notice is that within the range $-2/3\leq x<4/3$
there are actually two distinct N-S sectors, corresponding not just to
$x=0$, but also to $x= 2/3$.  This is associated with the
$\Z_2$ symmetry that interchanges
$\epsilon^+$ and $\epsilon^-$.  This symmetry may
explain the origin of the additional $\Z_2$ GSO-type projection we will
shortly discuss.

For the $K=4$ FSC model, one expects a GSO projection to depend on a
generalization of fermion number. However, the naive generalization to
parafermion number, $F(\phi^1_0)$, is insufficient. We find that we must also
consider the multiplicities of the twist field, $\phi^1_1$, and  the field
$\phi^0_1$, which increases the $m$ quantum number by one while keeping
$j$ constant.
In order to derive the MIPF we discovered that, indeed, a  $\Z_3$ projection
must be applied to both the left-moving modes
(LM) and right-moving modes (RM) independently.
Survival of a physical state, $\vert {\rm phys}\big>$, in the Hilbert space
under this $\Z_3$ projection requires\footnote{Note, this projection alone
does not prevent mixing holomorphic $A_4$--sector and
antiholomorphic $B_4$--sector terms.
This is prevented by the standard requirement
$M^2_{\rm LM}=M^2_{\rm RM}$, {\rm i.e.,} $L_0 = \bar L_0$, which here results
in
the RM factors in the partition function being the complex conjugates of
the LM,  giving only mod-squared terms in the partition functions.
This allows us to examine only the left-movers in detail in the
following.}
\subon
$$\left\{\e^{\left\{i\pi \vec {2\over 3}\cdot\left[ {\vec F}_{{\rm
LM}\,{\rm (RM)}}(\phi^1_0)
+ {\vec F}_{{\rm LM}\, {\rm (RM)}}(\phi^1_1)\right]\right\}}
= \e^{i\pi {2\over3}}\right\}
\vert {\rm phys}\big>\,\, ,\eqno\eqnlabel{gsoz3-a}$$
or equivalently
$$\left\{Q_{3,\, {\rm LM}\,{\rm (RM)}}\equiv\sum_i F_{i,\, {\rm LM} \,
{\rm (RM)}}(\phi^1_0)
+ \sum_i F_{i,\, {\rm LM} \,{\rm  (RM)}}(\phi^1_1) = 1 \pmod{3}
\right\} \vert {\rm phys}\big>\,\, , \eqno\eqnlabel{gsoz3-b}$$
where $F_i(\phi^j_m)_{{\rm LM}\,{\rm  (RM)}}$ is the number operator for the
field
$\phi^j_m$ along
the $i$ direction for left-moving (right-moving) modes.
Prior to projection by this extended GSO operator,
we consider all physical states associated with
the LM partition function terms in the expansion of
$(c^0_0 + c^4_0 + c^2_0)^4$ or $(c^2_2 + c^4_2)^4$
to be in the $A_4$--sector.
Similarly, we initially place in the $B_4$--sector
all the LM physical states associated with the partition function terms in the
expansion of $(c^2_2 + c^4_2)^2(c^0_0 + c^4_0 +c^2_0)^2$ or
$(c^0_0 + c^4_0 + c^2_0)^2(c^2_2 + c^4_2)^2$.
There is however, a
third class of states; let us call this the ``$D_4$'' class. This last
class would be present in the original Hilbert space if not for an additional
$\Z_2$ GSO projection.  Left-moving states in $D$ class, would have partition
functions that are terms in the expansion of
$(c^0_0 + c^4_0 + c^2_0)^3(c^2_2 + c^4_2)$ or
$(c^2_2 + c^4_2)^3(c^0_0+c^4_0+c^2_0)$. The thirty-two $D_4$ terms in
the expansions are likewise divisible into subclasses based on their
associated $\Z_3$ charges, $Q_3$.\footnote{For clarity, we are
always pairing $c^0_0$ and $c^4_0$ in these partition functions, rather than
expanding $(c^0_0 +c^4_0)^n\,$, as is done in Table 4.4.}
Twelve have charge $0 \pmod{3}$, twelve have charge $1 \pmod{3}$
and eight have charge $2 \pmod{3}$. Without the $\Z_2$ projection, it
is impossible to
to keep only the wanted terms in the $A_4$-- and $B_4$--sectors,
while projecting away all of the $D_4$--sector terms.
Simple variations of the projection (\puteqn{gsoz3-a}) cannot accomplish this.
All $D_4$ terms can be eliminated, without further projections on the $A$ and
$B$ terms, by a $\Z_2$ projection defined by
$$\left\{\sum_i F_{i,\, {\rm LM} \,{\rm  (RM)}}(\phi^1_1)
+ \sum_i F_{i,\,{\rm  LM} \,{\rm  (RM)}}(\phi^0_{\pm 1}) = 0 {\rm ~mod~} 2
\right\} \vert {\rm phys}\big>\,\, . \eqno\eqnlabel{gsoz3-c}$$
\suboff
(Note that for $K=2$,
$\phi^1_1$ is equivalent to the identity field and $\phi^1_0$ is
indistinguishable from the usual fermion, $\phi^1_0$. Thus for $K=2$
there is no additional $\Z_2$ GSO projection.)

Consideration
of these $D_4$ class states reveals some physical meaning to
our particular $\Z_3$ charge and the additional $\Z_2$ projection.
First, in all sectors the charge $Q_3$ commutes with
$(\phi^0_{K/4})^{D-2}$, which transforms between non-projection and
projection states of opposite spacetime statistics in the $A_4$-- and
$B_4$--sectors.
Second, the values of this charge are also associated with specific
${\rm mass}^2\pmod{1}$ levels. Third, only for the $A_4$-- and
$B_4$--sector
states does ${\rm mass}^2\pmod{1}$ commute with the same
twist operator $(\phi^0_{K/4})^{D-2}$.
Recall, in section 4.2 we suggested that
twisting by this latter field was the key
to spacetime SUSY. Without any of our projections the
${\rm mass}^2$ levels $\pmod{1}$
of states present would be
${\rm mass}^2= 0,~ {1\over 12},~ {2\over 12},~\dots {11\over 12}$.
When acting on  $D_4$--sector fields,
$(\phi^0_{K/4})^{D-2}$ transforms
\hbox {${\rm mass}^2={i\over 12} \pmod{1}$} states into
\hbox {${\rm mass}^2={ i +6\over 12} \pmod{1}$} states.
Thus, if present in a model, states in the $D_4$--sector paired by
the supersymmetry operator $(\phi^0_{K/4})^{D-2}$
would have to be associated with
different mod-squared terms of the
partition function, in order to preserve $T$ invariance. As a result, the
paired contributions to the partition function could not cancel,
proving that $D$
terms cannot be part of any supersymmetric theory.
Although ${\rm mass}^2 \pmod 1$ commutes with $(\phi^0_{K/4})^{D-2}$ in the
$A(Q_3=0)$,
$A(Q_3=-1)$, $B(Q_3=0)$, $B(Q_3=-1)$  subsectors, within these subsectors
(1) there is either a single bosonic state or fermionic state of lowest
mass without superpartner of equal mass, and/or
(2) the lowest mass states are tachyonic.
(See Table 4.4.)\footnote{Our assignments of states as spacetime
bosons or fermions in the $B$-sector,
uses an additional projection that we believe distinguishes between the
two.  Following the pattern in
eq.~\pe{part4-b}
with bosonic/fermionic assignment of related states defined in
eqs.~(4.2.33a-c), we suggest that for these states the two primary fields,
$\phi^{j_3}_{m_3}$ and $\phi^{j_4}_{m_4=m_3}$
(implicitely) assigned compactified spacetime
indices must be the same, {\it i.e.,} $j_3= j_4$,
or else must form  a term in the expansion of
$(\phi^0_0 + \phi^2_0)^2$. This second case is related to $\phi^0_0$ and
$\phi^2_0$ producing the same spacetime fermion field, $\phi^2_1$, when
separately twisted by $\phi^0_{K/4}$.
(Note however that $\phi^2_1\otimes\phi^0_{K/4}=\phi^0_0$ only.)
Following this rule, neither the states
corresponding to
$(c^2_0)(c^0_0)(c^2_2)(c^4_2)$ and $(c^2_2)(c^4_2)(c^2_0)(c^0_0)$,
(which transform between each other under twisting by
$\phi^0_{K/4}\phi^0_{K/4}\phi^0_{K/4}\phi^0_{K/4}$)
nor those associated with
$(c^2_0)(c^4_0)(c^2_2)(c^4_2)$ and $(c^2_2)(c^4_2)(c^2_0)(c^4_0)$,
survive the projections as either spacetime bosons or fermions.
However, for completeness we include these partition functions
in the B-sector columns of Table 4.4. We define the associated states as
either spacetime bosons or fermions based on the value of $m_3=m_4$.
This is academic, though, because the states do not survive the
{\tenBbb Z}$_3$ projection.}
Thus, our specific GSO
projections in terms of
our $\Z_3$ charge projection and our $\Z_2$ projection equate to spacetime
SUSY, uniquely so.
\hfill\vfill\eject
\def\cbox#1{\hbox to 0 pt{\hss#1\hss}}
\centertext{Table 4.4 Masses of $K=4$ Highest Weight States\\
(Represented by Their Associated Characters)\\
\hbox to 1cm{\hfill}}
 \halign to \hsize{%
#\hfil \tabskip=1em& \hfil#\tabskip=0em& #\hfil \tabskip=1em plus 0.2em&
#\hfil \tabskip=0.9em plus 0.1em& \hfil#\tabskip=0.2em& \hfil#\hfil&
#\hfil \tabskip=0.1em& \hfil#\tabskip=0em& #\hfil \tabskip=1em plus 0.2em&
#\hfil \tabskip=0em\cr
& \cbox{$A$-Sector}&&&&&& \cbox{$B$-Sector}&& \cr
\multispan4\hrulefill& & Survives & \multispan4\hrulefill \cr
\ Boson& \cbox{Mass$^2$}&&
\ Fermion& $Q_3$& GSO&
\ Boson ?& \cbox{Mass$^2$}&&
\ Fermion ?\cr
$(c^4_0)^2(c^4_0)^2$& 3& ${2\over3}$&
& 0& No&
$(c^4_0)^2(c^4_2)^2$& 3& ${1\over6}$&
$(c^4_2)^2(c^4_0)^2$ \cr
$c^2_0\>\> c^4_0\>\>(c^4_0)^2$ & 3&&
& 1& Yes&
$c^2_0\>\> c^4_0\>\> (c^4_2)^2$& 2& ${1\over2}$&
$c^2_2\>\> c^4_2\>\>(c^4_0)^2$ \cr
$c^0_0\>\> c^4_0\>\> (c^4_0)^2$& 2& ${2\over 3}$&
$(c^4_2)^2(c^4_2)^2$& 0& No&
$c^0_0\>\> c^4_0\>\> (c^4_2)^2$& 2& ${1\over6}$&
$(c^4_2)^2c^0_0\>\> c^4_0\>\>$ \cr
$(c^4_0)^2(c^2_0)^2$& 2& ${1\over 3}$&
& $-1$& No
& $(c^4_0)^2(c^2_2)^2$& 1& ${5\over 6}$&
$(c^4_2)^2(c^2_0)^2$ \cr
$(c^2_0)^2(c^4_0)^2$&&&
&&&
$(c^2_0)^2(c^4_2)^2$&&&
$(c^2_2)^2(c^4_0)^2$ \cr
&&&
&&&
$c^2_0\>\> c^4_0\>\> c^2_2\>\> c^4_2\>\>$&&&
$c^2_2\>\> c^4_2\>\> c^2_0\>\> c^4_0\>\>$ \cr
$c^2_0\>\> c^0_0\>\> (c^4_0)^2$& 2&&
$c^2_2\>\>c^4_2\>\>(c^4_2)^2$& 1& Yes&
$c^2_0\>\> c^0_0\>\>  (c^4_2)^2$& 1& ${1\over 2}$&
$c^2_2\>\> c^4_2\>\> c^0_0\>\> c^4_0\>\>$ \cr
$c^2_0\>\> c^4_0\>\> (c^2_0)^2$& 1& ${2\over 3}$&
& 0& No&
$c^2_0\>\> c^4_0\>\> (c^2_2)^2$& 1& ${1\over6}$&
$c^2_2\>\> c^4_2\>\> (c^2_0)^2$ \cr
$(c^0_0)^2(c^4_0)^2$&&&
&&&
$(c^0_0)^2(c^4_2)^2$&&&
 \cr
$(c^4_0)^2(c^0_0)^2$&&&
&&&
&&&
$(c^4_2)^2(c^0_0)^2$ \cr
$c^0_0\>\> c^4_0\>\> (c^2_0)^2$& 1& ${1\over 3}$&
$(c^2_2)^2(c^4_2)^2$& $-1$& No&
$c^0_0\>\> c^4_0\>\> (c^2_2)^2$&& ${5\over6}$&
 \cr
&&&
$(c^4_2)^2(c^2_2)^2$&&&
$c^0_0\>\> c^2_0\>\> c^2_2\>\> c^4_2\>\>$&&&
$c^2_2\>\> c^4_2\>\> c^0_0\>\> c^2_0\>\>$ \cr
$(c^2_0)^2(c^2_0)^2$& 1&&
& 1& Yes&
$(c^2_0)^2(c^2_2)^2$&& ${1\over 2}$&
$(c^2_2)^2(c^2_0)^2$ \cr
$c^2_0\>\> c^4_0\>\> (c^0_0)^2$&&&
&&&
&&&
$c^2_2\>\> c^4_2\>\>  (c^0_0)^2$ \cr
$c^2_0\>\> c^0_0\>\> (c^2_0)^2$&& ${2\over 3}$&
$c^2_2\>\>c^4_2\>\>(c^2_2)^2$& 0& No&
$c^2_0\>\> c^0_0\>\> (c^2_2)^2$&& ${1\over6}$&
 \cr
$c^0_0\>\> c^4_0\>\> (c^0_0)^2$&&&
&&&
&&&
 \cr
$(c^0_0)^2(c^2_0)^2$&& ${1\over 3}$&
& $-1$& No&
$(c^0_0)^2(c^2_2)^2$& $-$& ${1\over6}$&
 \cr
$(c^2_0)^2(c^0_0)^2$&&&
&&&
&&&
 $(c^2_2)^2(c^0_0)^2$ \cr
$c^2_0\>\> c^0_0\>\> (c^0_0)^2$& 0&&
$(c^2_2)^2(c^2_2)^2$& 1& Yes&
&&&
 \cr
$(c^0_0)^2(c^0_0)^2$& $-$& ${1\over 3}$&
& 0& No&
&&&
 \cr
}

\begin{ignore}
{\settabs 8 \columns
\+ \cr
\+ & \hso $A$-Sector &&\hso $Q_3$ &Survives&&\hso $B$-Sector\cr
\+ \underbar{\hbox to 6.6cm{\hfill}}&&&\hso
{\underbar{\phantom{$Q_3$}}}&
{\underbar{\hbox to 1.5cm{GSO ?\hfill}}} &\underbar{\hbox to 7.7cm{\hfill}}
\cr
\+ {Boson} &\hso\hskip .25cm {${\rm Mass}^2$}
&\hso {Fermion} &&&{Boson ?} &\hsf\hskip .25cm {${\rm Mass}^2$}
&\hsf {Fermion ?}\cr
\+ $(c^4_0)^2(c^4_0)^2$ &\hso\h $3{2\over3}$ &&\hso\phantom{$-$}0
&\hskip .45cm No & $(c^4_0)^2(c^4_2)^2$
&\hsf\h $3{1\over6}$ &\hsf $(c^4_2)^2(c^4_0)^2$\cr
\+ $(c^2_0)\hof (c^4_0)(c^4_0)^2$ &\hso\h 3 &&\hso\phantom{$-$}1
&\hskip .45cm Yes
& $(c^2_0)\hof (c^4_0)\hof (c^4_2)^2$
&\hsf\h $2{1\over2}$ &\hsf $(c^2_2)\hof (c^4_2)(c^4_0)^2$\cr
\+ $(c^0_0)\hof (c^4_0)\hof (c^4_0)^2$ & $\hso\h 2{2\over 3}$
&\hso $(c^4_2)^2(c^4_2)^2$
&\hso\phantom{$-$}0 &\hskip .45cm No
& $(c^0_0)\hof (c^4_0)\hof (c^4_2)^2$ &\hsf\h $2{1\over6}$ &\hsf
$(c^4_2)^2(c^0_0)\hof (c^4_0)$ \cr
\+ $(c^4_0)^2(c^2_0)^2$ &\hso\h $2{1\over 3}$ && \hso $-1$
&\hskip .45cm No
& $(c^4_0)^2(c^2_2)^2$ &\hsf\h $1{5\over 6}$ &\hsf $(c^4_2)^2(c^2_0)^2$\cr
\+ $(c^2_0)^2(c^4_0)^2$ &&&&& $(c^2_0)^2(c^4_2)^2$ &&\hsf
$(c^2_2)^2(c^4_0)^2$\cr
\+ &&&&& $(c^2_0)\hof (c^4_0)\hof (c^2_2)\hof (c^4_2)$ & &\hsf $(c^2_2)\hof
(c^4_2)\hof (c^2_0)\hof (c^4_0)$\cr
\+ $(c^2_0)\hof (c^0_0)\hof (c^4_0)^2$ &\hso\h 2
&\hso $(c^2_2)(c^4_2)(c^4_2)^2$ &\hso\phantom{$-$}1 &\hskip .45cm Yes
& $(c^2_0)\hof(c^0_0)\hof  (c^4_2)^2$ &\hsf\h
$1{1\over 2}$ &\hsf $(c^2_2)\hof (c^4_2)\hof (c^0_0)\hof (c^4_0)$\cr
\+ $(c^2_0)\hof (c^4_0)\hof (c^2_0)^2$ &\hso\h $1{2\over 3}$
&&\hso\phantom{$-$}0 &\hskip .45cm No &
$(c^2_0)\hof (c^4_0)\hof (c^2_2)^2$
&\hsf\h $1{1\over6}$ & $\hsf (c^2_2)\hof (c^4_2)\hof (c^2_0)^2$\cr
\+ $(c^0_0)^2(c^4_0)^2$&&&&& $(c^0_0)^2(c^4_2)^2$\cr
\+ $(c^4_0)^2(c^0_0)^2$&&&&&&&\hsf $(c^4_2)^2(c^0_0)^2$\cr
\+ $(c^0_0)\hof (c^4_0)\hof (c^2_0)^2$ &\hso\h $1{1\over 3}$
&\hso $(c^2_2)^2(c^4_2)^2$ &\hso $-1$ &\hskip .45cm No
& $(c^0_0)\hof (c^4_0)\hof (c^2_2)^2$ &\hsf\h\hskip .22cm ${5\over6}$\cr
\+ &&\hso $(c^4_2)^2(c^2_2)^2$
&&& $(c^0_0)\hof (c^2_0)\hof (c^2_2)\hof (c^4_2)$
&&\hsf $(c^2_2)\hof (c^4_2)\hof (c^0_0)\hof (c^2_0)$\cr
\+ $(c^2_0)^2(c^2_0)^2$ &\hso\h 1 & &\hso\phantom{$-$}1 &\hskip .45cm Yes
& $(c^2_0)^2(c^2_2)^2$ &
$\hsf\h\hskip .22cm {1\over 2}$ &\hsf $(c^2_2)^2(c^2_0)^2$\cr
\+ $(c^2_0)\hof(c^4_0)\hof (c^0_0)^2$ &&&&&&&\hsf $(c^2_2)\hof (c^4_2)\hof
 (c^0_0)^2$\cr
\+ $(c^2_0)\hof (c^0_0)\hof (c^2_0)^2$ &\hso\h\hskip .22cm ${2\over 3}$
&\hso $(c^2_2)(c^4_2)(c^2_2)^2$
&\hso\phantom{$-$}0 &\hskip .45cm No
& $(c^2_0)\hof (c^0_0)\hof (c^2_2)^2$
&\hsf\h\hskip .22cm ${1\over6}$\cr
\+ $(c^0_0)\hof (c^4_0)\hof (c^0_0)^2$&\cr
\+ $(c^0_0)^2(c^2_0)^2$
&\hso\h\hskip .22cm ${1\over 3}$
&&\hso $-1$ &\hskip .45cm No & $(c^0_0)^2(c^2_2)^2$
&\hskip .57cm\h $-{1\over6}$ \cr
\+ $(c^2_0)^2(c^0_0)^2$&&&&&&&\hsf $(c^2_2)^2(c^0_0)^2$\cr
\+ $(c^2_0)\hof (c^0_0)\hof (c^0_0)^2$ &\hso\h\hskip .22cm 0
&\hso $(c^2_2)^2(c^2_2)^2$
&\hso\phantom{$-$}1 &\hskip .45cm Yes \cr
\+ $(c^0_0)^2 (c^0_0)^2$
&\hskip .9cm\h $-{1\over 3}$ &&\hso\phantom{$-$}0 &\hskip .45cm No\cr
\+ \cr}
\end{ignore}
\hfill\vfill\eject
\centertext{Table 4.5 Mass Sectors as Function of $\Z_3$ Charge}
{\settabs 7\columns
\+\cr
\+ Lowest $M^2$  & $M^2$ mod 1 & Sector & $\Z_3$ Charge & Sector
& $M^2$ mod 1 & Lowest $M^2$\cr
\+ \overline{\hbox to 5.9cm{\hfill}}&&
&\overline{\hbox to 1.8cm{\hfill}}
&\overline{\hbox to 6.7cm{\hfill}}\cr
\+\hskip .65cm \htf 0
&\hskip .8cm 0
&\hskip .4cm $A$
&\hskip .15cm $Q_3={\hbox to .3cm{\hfill}}1$
&\hskip .4cm $B$
&\hskip .8cm ${6\over 12}$
&\hskip .65cm ${6\over 12}$\cr
\+\cr
\+\hskip .65cm $-{1\over 12}$
&\hskip .8cm ${11\over 12}$
&\hskip .4cm $D$
&\hskip .15cm $Q_3= {\hbox to .3cm{\hfill}}0$
&\hskip .4cm $D$
&\hskip .8cm ${5\over 12}$
&\hskip .65cm ${5\over12}$\cr
\+\cr
\+\hskip .65cm $-{2\over 12}$
&\hskip .8cm ${10\over 12}$
&\hskip .4cm $B$
&\hskip .15cm $Q_3= -1$
&\hskip .4cm $A$
&\hskip .8cm ${4\over 12}$
&\hskip .65cm ${4\over12}$\cr
\+\cr
\+\hskip .65cm $-{3\over 12}$
&\hskip .8cm ${9\over 12}$
&\hskip .4cm $D$
&\hskip .15cm $Q_3= {\hbox to .3cm{\hfill}}1$
&\hskip .4cm $D$
&\hskip .8cm ${3\over 12}$
&\hskip .65cm ${3\over 12}$\cr
\+\cr
\+\hskip .65cm $-{4\over 12}$
&\hskip .8cm ${8\over 12}$
&\hskip .4cm $A$
&\hskip .15cm $Q_3={\hbox to .3cm{\hfill}}0$
&\hskip .4cm $B$
&\hskip .8cm ${2\over 12}$
&\hskip .65cm ${2\over 12}$\cr
\+\cr
\+\hskip .65cm\htf ${7\over 12}$
&\hskip .8cm ${7\over 12}$
&\hskip .4cm $D$
&\hskip .15cm $Q_3= -1$
&\hskip .4cm $D$
&\hskip .8cm ${1\over 12}$
&\hskip .65cm ${1\over 12}$\cr
\+ \cr}

\n (In Table 4.5, columns one and seven give the lowest ${\rm mass}^2$  of a
state with center
column $\Z_3$ charge in the appropriate sector. For the $D$ sector states,
under $(\phi^0_{K/4})^{D-2}$ twistings, ${\rm mass}^2$ values in column two
transform into ${\rm mass}^2$ values in column six of the same row and
vice-versa.)

Unlike in the $K=2$ case, for $K=4$
the $\Z_3$ projection in the Ramond sector wipes out complete spinor fields,
not
just some of the modes within a given spin field. This type of projection
does not occur in the Ramond sector for $K=2$ since there are no fermionic
states with fractional ${\rm mass}^2$ values in the $D=10$ model.
Note also that
our $\Z_3$ GSO projections relate to the $\Z_3$ symmetry
pointed out in
[\putref{zamol87}] and briefly commented on following
eqs.~(\puteqn{commut-a}-b).

For $K=8$, a more generalized $\Z_5$ projection
holds true for all
sectors.  For the $K=16$ theory, there are too few terms and products of
string functions to determine if a $\Z_9$ projection
is operative.  In the
$K=4$ case, the value of our LM (RM) $Q_3$ charges for states surviving the
projection is set by demanding that the massless spin-2 state
$\epsilon^{\mu}_{-{1\over3}}\bar\epsilon^{\bar\nu}_{-{1\over3}}\left\vert 0
\right\rangle$ survives.  In the $A_K$--, $B_K$--,
(and $C_K$-- for $K=8,16$)  sectors, these
projections result in states with squared masses of $0+$ integer,
${1\over 2} +$ integer, and ${3\over 4} +$ integer, respectively.

\vhalf
{\hc{4.3.c}{\sl The Unique Role of the Twist Field, $\phi_{K/4}^{K/4}$.}}
\vhalf

In this subsection
we examine whether other consistent models are possible if one
generalizes from the twist
field, $\phi^{K/2}_{K/2}$ to another that could fulfill its role.
When it is demanded that the standard
twist and $\epsilon\equiv\phi^1_0$ fields of reference
[\putref{dienes92b,argyres91b,argyres91d}], be used
we can derive
the critical dimensions of possible models simply by observing that
$K=2,\, 4,\, 8,$ and $16$  are  the only levels for which
\subon
$$h(\phi^1_0)/h(\phi^{K/4}_{K/4})\in\Z\,\, . \eqno\eqnlabel{stdim-a}$$
If we assume (as in [\putref{dienes92b}]) that the
operator $(\phi_{K/4}^{K/4})^{\mu} $ acting on the (tachyonic) vacuum produces
a massless spacetime
spinor vacuum along the direction $\mu$, and $(\phi^1_0)^{\mu}$ produces a
massless spin-1 state, then for spacetime supersymmetry
(specifically $N=2$ SUSY for fractional type II theories and $N=1$
for fractional heterotic)
$h(\phi^1_0)/h(\phi^{K/4}_{K/4})$  must equal the number
of  transverse
spin modes, {\it i.e.,}
$$\eqalignno{ h(\phi^1_0) &= (D-2) h(\phi^{K/4}_{K/4})\cr
{2\over K+2}&= (D-2) {K/8\over K+2}\,\, .&\eqnlabel{stdim-b}\cr}$$
Hence,
$$ D= 2 + {16\over K}\in\Z\,\, .\eqno\eqnlabel{stdim-c}$$
\suboff
Thus, from this one assumption, the possible integer spacetime
dimensions are determined along with theassociated levels.
Perhaps not coincidentally, the allowed dimensions are precisely the ones
in which classical supersymmetry is possible. This is clearly a
complementary method to the approach for determining $D$ followed
in refs.~[\putref{dienes92b,argyres91b,argyres91d}].

Demanding eq.~(\puteqn{stdim-a}) guarantees
\hbox{spin-1} and \hbox{spin-1/2} superpartners in the open string a
(\hbox{spin-2} and \hbox{spin-3/2} in the closed string) with
$${\rm mass}^2= {\rm mass}^2({\rm vacuum}) + h(\phi^1_0) = {\rm mass}^2({\rm
vacuum}) +
(D-2)* h(\phi^{K/4}_{K/4})\, .\eqno\eqnlabel{mass1}$$
(Double the total mass$^2$ for the closed string.)
{\it A priori} simply demanding the ratio
be integer in eq.~(\puteqn{stdim-a})   is not
sufficient to guarantee local spacetime supersymmetry in the closed string.
However, in the previous
subsections it proved to be;
for the $K=4$ model the masslessness of the open string
\hbox{(spin-1, spin-1/2)} pair occurred automatically
and hence also in the closed string for the \hbox{(spin-2, spin-3/2)}
pair.
\centertext{\hbox to 1cm{\hfill}}
\centertext{Figure 4.1 Supersymmetry of Lowest Mass States of Fractional
Open String}
\vskip .5cm
{\centertext{\underline{\hbox to 4in{$m^2({\rm spin}-1)\hfill =\hfill
m^2({\rm spin}-1/2)$}}}}\\
{\centertext{\hbox to 1in{\hfill}}}\\
{\centertext{\hbox to 4in{$h(\phi^1_0)$ $\big\Uparrow$\hfill
$(D-2)\times h(\phi^{K/4}_{K/4})$ $\big\Uparrow$}}}\\
{\centertext{\hbox to 1 in{\hfill}}}\\
{\centertext{\underline{\hbox to 4 in{\hfill $m^2({\rm vacuum})$\hfill}}}}

\vskip 1cm
In fractional superstrings, the primary field
$\phi^{K/4}_{K/4}\equiv\phi^{K/4}_{-K/4}$ for $K=4,\, 8,\, {\rm and~} 16$,
and its associated character
\hbox{$Z^{K/4}_{K/4}=\eta c^{K/2}_{K/2}$}, are
 viewed as the generalizations of
$\phi^{1/2}_ {1/2}$ at $K=2$ and $(\vartheta_2/\eta)^{1/2}$.
Are there any other parafermion operators at
additional levels $K$ that could be used to transform the bosonic vacuum
into a massless fermionic vacuum
and bring about local spacetime supersymmetric models? The
answer is that by demanding masslessness\footnote{Masslessness of
at least the left- (right)-moving spin-1 spacetime fields (whose
tensor product forms the massless spin-2 graviton in a closed string)
is of course required
for a consistent string theory.  Consistent two-dimensional field
theories with
$$\eqalignno{
{\rm ~lowest~mass}&{\rm ~of~left-~(right-)moving~spacetime~spin-1~fields}
=\cr
& {\rm ~lowest~mass~of~left-~(right-)moving~spacetime~spin-1/2~fields}
\equiv M_{\rm min} > 0}$$
may exist (as we discuss below) but, the physical interpretation of such
models is not clear, (other than to say they would not be theories with
gravity.}
of the (spin-1, spin-1/2) pair,
there is clearly no other choice for  $K<500$. (We believe this will
generalize to $K<\infty$.)

The proof is short. We do not start from the assumption that the
massless spin-1 fields are a result of the $\phi^1_0$ fields.  Rather,
to the contrary, the necessity of choosing $\phi^1_0$
appears to be the result of the uniqueness of
$\phi^{K/4}_{K/4}$.
{\settabs 2\columns
\+ \cr}

Proof:  Assume we have a consistent (modular invariant) closed fractional
superstring theory
at level-$K$ with supersymmetry in $D$ dimensional spacetime,
($N=2$ for type-II and $N=1$ for heterotic).
Let the massless left- (right-)moving spin-1 field be
$(\phi^{j_1}_{m_1})^{\mu} \vert {\rm vacuum}>$. This requires that
$\phi^{j_1}_{m_1}$  have conformal dimension
$$h(\phi^{j_1}_{m_1})= c_{\rm eff}/24= (D-2) {K\over 8(K+2)}\,\,
.\eqno\eqnlabel{spin1cd}$$
Thus, the twist field $\phi^{j_2}_{m_2}$ that produces the spinor vacuum along
one of the $D-2$
transverse dimensions must have conformal dimension
$$h(\phi^{j_2}_{m_2})= {K\over 8(K+2)}\,\, . \eqno\eqnlabel{spin1half}$$
For $K<500$ the only primary fields with this dimension are the series
of $\phi^{K/4}_{K/4}$ for $K\in 2\Z$, and the accidental solutions
$\phi^2_0$ for $K=48$, $\phi^3_0$ for $K=96$,
and $\phi^{9/2}_{7/2}$ for $=98$. Being fields with $m=0$, neither
$\phi^2_0$ nor $\phi^3_0$ at any level
cannot be used to generate spacetime fermions. The $\phi^{9/2}_{7/2}$
alternative is not acceptable either because at $K=98$ there is not an
additional field to replace $\epsilon$. In other words, there is not
a field to be paired with $\phi^{9/2}_{7/2}$ whose
conformal dimension is an integer multiple of $\phi^{9/2}_{7/2}$'s.
(A proof of the uniqueness of $\phi^{K/4}_{K/4}$ for all
$K$ is being prepared by the author.)

Confirmation of $\phi^{K/4}_{K/4}$ as the spin-1/2 operator, though, does not
immediately lead one to conclude that $\epsilon$ is the only possible
choice for producing massless boson fields. Table 4.6 shows alternative
fields at new levels $K\not= 2,\, 4,\, 8,$ or $16$ whose conformal
dimension is one, two, or four times the conformal dimension of
$\phi^{K/4}_{K/4}$.
(Note that successful alternatives to
$\epsilon$ would lead to a  relationship between level and
spacetime dimension differing from eq.~(\puteqn{stdim-c}).) However, nearly
all alternatives are of the form $\phi^{j>1}_0$ and we would expect that
modular invariant models using
$\phi^{j>1}_0$ to create massless bosons, would necessarily include
(at least) the tachyonic state, $(\phi^1_0)^{\mu}\vert {\rm vacuum}\rangle$.
That is,  we do not believe valid GSO projections
exist which can project away these tachyons while simultaneously
keeping the massless graviton and gravitino and giving modular invariance.
Further, the remaining fields on the list have $m\not= 0\pmod{K}$.
Each of these
would not have the correct fusion rules with itself, nor with
$\phi^{K/4}_{K/4}$ to be a spacetime boson.
\centertext{\hbox to 1cm{\hfill}}
\centertext{Table 4.6 Fields $\phi^{j_1}_{m_1}\neq \phi^1_0$ with Conformal
Dimensions in Integer Ratio with $h(\phi^{K/4}_{K/4})$}
{\settabs 5 \columns
\+\cr
\+ & {$K$} & {$\phi^j_m$} &
{$h(\phi^j_m)/h(\phi^{K/4}_{K/4})$}\cr
\+ & $\overline{\hbox to 9.3cm{\hfill}}$\cr
\+ & 12 & $\phi^2_0$ & \hskip 1.3cm 4\cr
\+ & 24 & $\phi^2_0$ & \hskip 1.3cm 2\cr
\+ &   & $\phi^3_0$ & \hskip 1.3cm 4\cr
\+ & 36 & $\phi^7_6$ & \hskip 1.3cm 4\cr
\+ & 40 & $\phi^4_0$ & \hskip 1.3cm 4\cr
\+ & 48 & $\phi^2_0$ & \hskip 1.3cm 1\cr
\+ &    & $\phi^3_0$ & \hskip 1.3cm 2\cr
\+ & 60 & $\phi^5_0$ & \hskip 1.3cm 4\cr
\+ & 80 & $\phi^4_0$ & \hskip 1.3cm 2\cr
\+ & 84 & $\phi^6_0$ & \hskip 1.3cm 4\cr
\+ & 96 & $\phi^3_0$ & \hskip 1.3cm 1\cr
\+ & 112 & $\phi^7_0$ & \hskip 1.3cm 4\cr
\+ & 120 & $\phi^5_0$ & \hskip 1.3cm 2\cr
\+ & $\vdots$ & $\vdots$ & \hskip 1.3cm $\vdots$\cr
\+\cr}

Lastly, we want to consider the possibility that there is meaning
to (non-stringy) two-dimensional field theories that
contain neither supergravity nor even gravity. Instead let a
model of this type contain only a global supersymmetry. The
lowest mass spin-1 ($(\phi^1_0)^{\mu}\vert {\rm vacuum}>$)
and spin-1/2 ($(\phi^{j_3}_{m_3})^{D-2}\vert {\rm vacuum}>$)
left- or right-moving fields would be related by
$${\rm mass}^2({\rm vacuum}) + h(\phi^1_0) = {\rm mass}^2({\rm vacuum}) +
(D-2)\times h(\phi^{j_3}_{m_3})\, .\eqno\eqnlabel{massive}$$
In PCFT's there
is only a very small number (12) of potential candidates for
$\phi^{j_3}_{m_3}$.
(Like $\phi^{K/4}_{K/4}$ these twelve are all of the form
$\phi^{j_3}_{\pm j_3}$.) We are
able to reduce the number of candidates down to this finite number very
quickly by proving no possible candidate could have $j_3>10$, independent of
the level. We demonstrate this as follows:

Any potential level-$K$ candidate $\phi^{j_3}_{m_3}$ must satisfy the
condition of
$${K\over K+2}\left[ j_3(j_3+1) -2\right] \leq (m_3)^2 \leq (j_3)^2 \leq K^2/4
\,\, .\eqno\eqnlabel{constraint}$$
By parafermion equivalences (\puteqn{cid-a}-b),
$\vert m\vert\leq j\leq K/2$ can be required for any level-$K$
fields.
The other  half of the inequality,
${K\over K+2}\left[ j_3(j_3+1) -2 \right]\leq m^2$
results from the weak requirement that the
conformal dimension of the candidate (spacetime) spin-1/2 field,
$\phi^{j_3}_{m_3}$, creating the fermion ground state
along one spacetime direction cannot be greater than the conformal
dimension of $\epsilon$, {\it i.e.,} $h(\phi^{j_3}_{m_3})\leq h(\phi^1_0)$.

{}From eq.~(\puteqn{constraint}),
we can determine both the minimum and maximum values
of $K$, for a given $j_3$, (independent of $m_3$).
These limits are \hbox{$K_{\rm min}= 2j_3$} and
\hbox{$K_{\rm max}= {\rm ~int}\left( {2(j_3)^2\over j_3-2} \right)$.}
Thus, the number of different levels that can correspond to the field
$\phi^{j_3}_{m_3}$ is int$\left({{5j_3-2}\over {j_3-2}}\right)$.
This number quickly decreases to six as $j_3$ increases to ten and
equals five for $j_3$ greater than ten.
For a given $j_3$, we will express the levels under consideration as
\hbox{$K_i= 2j_3 +i$}.
Also, since \hbox{$K_{\rm min}=2j_3$},
the weak constraint on $m_3$
implies that we need only consider $\phi^{j_3}_{m_3= \pm j_3}$ fields.
Thus, our search reduces to finding fields $\phi^{j_3}_{\pm j_3}$ whose
conformal dimensions satisfy
$${{h(\phi^1_0)\over h(\phi^{j_3}_{\pm j_3})} =
  {{2\over K_i +2}\over {j_3(j_3+1)\over K_i+2} - {(j_3)^2 \over K_i}}\in\Z}
\,\, .
  \eqno\eqnlabel{ratioj}$$
Clearly, there are no solutions to
eq.~(\puteqn{ratioj}) for $i=0 {\rm ~~to~~} 4$ and $j_3>10$.
Hence, our range of possible alternative sources for fermionic
ground states reduces
to only cnsidering
those $\phi^{j_3}_{\pm j_3}$ with $0<j_3\leq 10$.  Within this
set of $j_3$'s, a computer
search reveals the complete set of fields that
obey eq.~(\puteqn{ratioj}), as shown in Table 4.7.
\hfill\vfill\eject
\centertext{Table 4.7 Potential Alternatives, $\phi^{j_3}_{\pm m_3}$,
 to $\phi^{K/4}_{K/4}$ for Spin Fields}
{\settabs 7 \columns
\+ \cr
\+ {$j_3$}     & {$\pm m_3$}     & {$K$}
& {$i$} & {$h(\phi^1_0)$} & {$h(\phi^{j_3}_{m_3})$}
& $D$\cr
\+ $\overline{\hbox to 14.1cm{\hfill}}$\cr
\+ 1/2 & 1/2 & 2  & 1  & 1/2 & 1/16 & 10 **\cr
\+     &     & 3  & 2  & 2/5 & 1/15 & 8\cr
\+     &     & 5  & 4  & 2/7 & 2/35 & 7 \cr
\+ \cr
\+ 1   & 1   & 3  & 1  & 2/5 & 1/15 & 8 \cr
\+     &     & 4  & 2  & 1/3 & 1/12 & 6 **\cr
\+     &     & 6  & 4  & 1/4 & 1/12 & 5 \cr
\+ \cr
\+ 3/2 & 3/2 & 9  & 6  & 2/11 & 1/11 & 4 \cr
\+ \cr
\+ 2   & 2   & 5  & 1  & 2/7 & 2/35 & 7\cr
\+     &     & 6  & 2  & 1/4 & 1/12 & 5 \cr
\+     &     & 8  & 4  & 1/5 & 1/10 & 4 **\cr
\+ \cr
\+ 5/2 & 5/2 & 25  & 20 & 2/27 & 2/27 & 3 \cr
\+ \cr
\+ 3   & 3   & 9  & 3  & 2/11 & 1/11 & 4 \cr
\+     &     & 18 & 12 & 1/10 & 1/10 & 3 \cr
\+ \cr
\+ 4   & 4   & 16 & 8  & 1/9  & 1/9  & 3 **\cr
\+ \cr
\+ 6   & 6   & 18 & 6  & 1/10 & 1/10 & 3 \cr
\+ \cr
\+ 10  & 10  & 25 & 5  & 2/27 & 2/27 & 3 \cr
\+ \cr}
\hfill\vfill\eject
The sets of solutions for
$j_3= \half,\, 1,\, {\rm and~} 2$
are related. The existence of a set
\hbox{$\{i=1,\, 2,\, {\rm and~} 4\}$} of solutions for any
one of these $j_3$ implies identical sets \hbox{$\{ i\}$} for the remaining
two $j_3$ as well.
The known $\phi^{K/4}_{K/4}$ solutions (marked with a **)
correspond to the \hbox{$i=1,\, 2,\, {\rm and~} 4$} elements in the
\hbox{$j_3=\half,\, 1,\, {\rm and~}2$} sets respectively. Whether this pattern
suggests anything about the additional related $\phi^{j_3}_{\pm j_3}$ in
these sets, other than explaining their appearance in the above table, remains
to be seen.

The set of distinct fields can be further reduced. There is
a redundancy in the above list. Among this list,
for all but the standard $\phi^{K/4}_{K/4}$ solutions,
there are two fields at each level, with different values of $j_3$.
However, these pairs are related by
the field equivalences (\puteqn{phidents}):
\subon
$$\eqalignno{
\hbox to 1cm{$\phi^{1/2}_{\pm 1/2}$\hfill}
&\equiv\hbox to 1cm{$\phi^1_{\mp 1}$\hfill} {\rm~~at~level~~} K=3
&\eqnlabel{id-a}\cr
\hbox to 1cm{$\phi^{1/2}_{\pm 1/2}$\hfill}
&\equiv\hbox to 1cm{$\phi^2_{\mp 2}$\hfill} {\rm~~at~level~~} K=5
&\eqnlabel{id-b}\cr
\hbox to 1cm{$\phi^1_{\pm 1}$\hfill}
&\equiv\hbox to 1cm{$\phi^2_{\mp 2}$\hfill} {\rm~~at~level~~} K=6
&\eqnlabel{id-c}\cr
\hbox to 1cm{$\phi^{3/2}_{\pm 3/2}$\hfill}
&\equiv\hbox to 1cm{$\phi^3_{\mp 3}$\hfill} {\rm~~at~level~~} K=9
&\eqnlabel{id-d}\cr
\hbox to 1cm{$\phi^3_{\pm 3}$\hfill}
&\equiv\hbox to 1cm{$\phi^6_{\mp 6}$\hfill} {\rm~~at~level~~} K=18
&\eqnlabel{id-e}\cr
\hbox to 1cm{$\phi^{5/2}_{\pm 5/2}$\hfill}
&\equiv\hbox to 1cm{$\phi^{10}_{\mp 10}$\hfill} {\rm~~at~level~~} K= 25
\,\, . &\eqnlabel{id-f}\cr}$$
\suboff
Because $\phi^j_m$ and $\phi^j_{-m}$ have identical  partition functions
and $\phi^j_{-m}\equiv(\phi^j_m)^{\dagger}$
we can reduce the number of possible alternate fields in half, down to
six. (Note that we not been distinguishing between $\pm$ on $m$ anyway.)

If we want models with {\it minimal} (global) super Yang-Mills Lagrangians
we can reduce the number of the fields to investigate further.
Such theories
exist classically only in $D_{\rm SUSY}=10,\, 6,\, 4,\, 3,\, ({\rm and~} 2)$
spacetime. Thus we can consider only
those $\phi^{j_3}_{\pm j_3}$ in the above list that have integer conformal
dimension
ratios of $D_{\rm SUSY}-2 = h(\phi^1_0)/h(\phi^{j_3}_{j_3})= 8,\, 4,\, 2,\,
{\rm and~} 1$.
This would reduce the fields to consider to just the two
new possibilities for $D= 4$, and $3$ since there there are no new additional
for $D=10$ or $6$.\newpage

{\hb{4.4}{\bfs Concluding Discussion}}\vhalf
\sectionnum=4

A consistent generalization of the superstring would be an
important
development. Our work has shown that the fractional
superstring has many intriguing features that merit further study. The
partition functions for these theories have simple
origins when derived systematically
through the factorization approach of Gepner and Qiu.
Furthermore, using this affine/theta-function factorization of the
parafermion partition  functions, we have related the $A_K$--sector
containing the graviton and
gravitino with the massive sectors, $B_K$ and $C_K$. A bosonic/fermionic
interpretation
of the $B_K$--subsectors was given.  Apparent ``self-cancellation'' of the
$C_K$--sector was shown, the meaning of which is under investigation.
A possible GSO projection was found, adding
hope that the partition functions have a natural physical
interpretation.

Nevertheless, fundamental questions
remain concerning the ghost system and current algebra, which prevent a
definite conclusion as to whether or not these are consistent theories.
Perhaps most important are arguments suggesting that fractional superstrings
in $D$ dimensions are not formed from tensor products of $D$ separate
$SU(2)_{K}/U(1)$ CFT's. Rather, a tensor product
CFT may be an illusion of the tensor product form of the partition function.
Instead of having a total conformal anomaly contribution of $c=12$ from
matter, the appearance in the six-dimensional ($K=4$) theory of
extra null states at the
critical value suggests that $c=10$.
This would require a non-tensor product representation of
the fractional superconformal algebra
(\puteqn{FSCalgebra-a}-\puteqn{FSCalgebra-c}).
However, even if the theories
are ultimately shown to be inconsistent, we believe that this
program will at least provide interesting identities and new insight into
the one case that we know is consistent, $K=2$. In other words, viewed in this
more general context, we may understand better what is special about
the usual superstring.

On the other hand, fractional superstrings may eventually prove to
be a legitimate class of solutions to a new string theory. This class would
then join the ranks of bosonic, type-II, and heterotic string theories.
Further, it is claimed
that MIPF's for heterotic fractional superstrings with left-movers at
level-$K_1$ and right-movers at level-$K_2$ are also possible.\mpr{dienes92b}
Let us call these heterotic$_{(K_1,K_2)}$ models. The simplest of this class,
the heterotic$_{(1,K)}$ model,
should have $SO(52-2D)= SO(48- 16/K)$ gauge symmetry.

The dominant view holds that the bosonic, type-II, and (standard)
heterotic theories
are uniquely defined by their underlying extended Virasoro
algebras, with many solutions (models) existing for each
theory.footnote{The
number of uncompactified dimensions is regarded as a parameter in the space of
solutions.} An alternate view\mpr{schellekens89c} suggests that heterotic
and type-II strings are related to subregions in the parameter space of
bosonic strings.
Specifically, the claim is that for {\it any} heterotic (type-II)
string there exists a bosonic string that faithfully represents its
properties.
This means that the classification of heterotic and type-II
strings is contained within that of the bosonic strings,
$${\rm Bosonic~Strings} \supset {\rm Heterotic~Strings}
\supset {\rm Type~II~Strings}\,\, . \eq{subclasses}$$
Thus, theoretically,
once the conditions for modular invariance of bosonic strings
are known, determination of them for heterotic or type-II is transparent.
The basis for this mapping
is that the non-unitary supersymmetric ghost system can be
transformed into a unitary conformal field theory.
This transformation preserves both conformal and modular invariance.
The partition functions of a new
unitary theory satisfies all the consistency conditions to serve
as a partition function for
a bosonic string compactified on a lattice.\mpr{schellekens89c}

Where might (heterotic) fractional superstrings, if proven consistent,
fit in this scheme? If their ghost system is finally understood,
perhaps the same mapping technique can be applied.  If so, can
fractional superstrings be represented by subclasses of bosonic
strings?  We suspect that the answer is ``yes,''
given that fractional (heterotic) strings
are found to be consistent.
Further, we would expect the fractional heterotic superstrings to
correspond specifically to a subset of heterotic (or type-II if $K_1=K_2$)
strings. This is suggested by the apparent spacetime SUSY of
heterotic$_{(K_1,K_2)}$ strings,
even though the local world sheet symmetry is ``fractional.''
\hfill\vfill\eject

\n {\bf Appendix A: Dynkin Diagrams for Lie Algebras and \KMAS}

\vskip 7truecm
\hfill\vfill

\n Figure A.1 Generalized Dynkin diagrams of the finite KM algebras
({\it i.e.}, of the compact simple Lie algebras)

\vskip 1.6truecm

Each simple root of a Lie algebra or a \KM algebra is represented by a
circle. Two circles not directly connected by at least one line imply two
orthogonal roots. One line connecting circles signifies a $120^{\circ}$
angle between corresponding roots, two lines  $135^{\circ}$, three lines
$150^{\circ}$, and four lines $180^{\circ}$.
For non-simply-laced algebras, arrows point toward the shorter of two
connected roots. The co-marks associated with the simple roots of a Lie
algebra appear inside the circles.

\eject

\hbox to 1cm{\hfill}

\hfill\vfill

\centertext{Figure A.2 Generalized Dynkin diagrams of the untwisted
affine KM algebras}

\eject
\hbox to 1cm{\hfill}

\hfill\vfill

\centertext{Figure A.3 Generalized Dynkin diagrams of the twisted
affine KM algebras}

\eject

\n{{\bf Appendix B:} Proof that completeness of the A-D-E classification of
modular\hfill\\}
\n{\phantom{\bf Appendix B:} invariant partition functions for $SU(2)_K$ is
unrelated to uniqueness\hfill\\}
\n{\phantom{\bf Appendix B:} of the vacuum\hfill}
\vhalf

In this appendix we prove that relaxing the condition of uniqueness of
the vacuum does not allow new solutions to ${SU}(2)_K$ MIPFs.  The
allowed solutions are still just the A-D-E classes.  We prove this through a
review of  Cappelli, Itzykson, and Zuber's (CIZ's) derivation of the A-D-E
classification.\mpr{capelli87} We treat the coefficients
$N_{l\bar l}$ in the partition function of eq.~(B.5) as the components of a
symmetric matrix, $\bmit N$, that operates between vectors of characters
$\vec\chi$:
$$\eqalign {
Z &= \sum_{l\bar l} N_{\bar l l}\bar\chi_{\bar
l}^{(K)}(\bar\tau)\chi_{l}^{(K)}(\tau)\cr
  &\equiv \bigl\langle\vec\chi\vert{\bmit
N}\vert{\vec\chi}\bigr\rangle}\,\, . \eqno ({\rm B}.1)$$
In this notation, under  ${\bmit V}\in {\bmit S},{\bmit T}\}$,
${\bmit V}^{\dag}{\bmit V}=1$,
the partition function
transforms as
$$ Z_{\bmit V}= \bigl\langle\vec\chi\vert{\bmit V^{\dag}}{\bmit N}{\bmit
V}\vert{\vec\chi}\bigr\rangle\,\, , \eqno ({\rm B}.2)$$
where $\bmit S$ and $\bmit T$ are the matrix representations of the standard
$S$ and $T$ modular transformations for a specific conformal field theory
(CFT).
Thus, a partition function $Z$ for a specific CFT is modular invariant
iff $\bmit N$ commutes with $\bmit S$ and $\bmit T$ in the given CFT.
CIZ proved that for a general $SU(2)_K$ algebra the commutant
of ${\bmit S}$ and ${\bmit T}$ is generated by a
set of linearly independent symmetric matrix operators, $\Om$, labeled by
$\delta$, a divisor of $n\equiv K+2$. Thus,  for $(B.1)$ to be a MIPF
$\bmit N$
must be formed from the basis set of $\Om$.
In \pr{capelli87}, CIZ showed that the
additional requirements of (1) uniqueness of the vacuum ($N_{11}=1$) and
(2) absence of $\bar\chi_{\bar l}\chi_l$ with coefficients
$N_{\bar l l}<0$
constrain MIPFs for $SU(2)_K$ to the A-D-E classification.
This is apparent from the
limited possibilities
for $\sum_{\delta} c_{\delta}\Om$ with $c_{\delta}\in Z$
that produce MIPF's satisfying both (1) and (2). (See Table B.1 below.)
\vskip .4cm
\n Table B.1 A--D--E Classification in Terms of $\Omega_{\delta}$ Basis Set
$$\vbox{\settabs 3 \columns
\+ \cr
\+  {\underbar {$n\equiv$ level $+2$}}\hskip 1.35cm
{\underbar {Basis Elements For MIPF's}}
&& {\underbar {A--D--E Classification}}\cr
\+ \hbox to 2.3cm{\hfill $n\geq
2$\hfill}\hskip 1.35cm\hbox to 4.95cm{\hfill $\Omega_n$\hfill}
&& \hbox to 3.9cm{\hfill $(A_{n-1})$\hfill}\cr
\+ \hbox to 2.3cm{\hfill $n {\rm ~even}
$\hfill}\hskip 1.35cm\hbox to 4.95cm{\hfill $\Omega_n + \Omega_2$\hfill}
&& \hbox to 3.9cm{\hfill $(D_{n/2 +1})$\hfill}\cr
\+ \hbox to 2.3cm{\hfill $n = 12$\hfill}\hskip 1.35cm\hbox to 4.95cm{\hfill $
\Omega_{n=12} + \Omega_3 + \Omega_2$\hfill}
&& \hbox to 3.9cm{\hfill $(E_6)$\hfill}\cr
\+ \hbox to 2.3cm{\hfill $n =
18$\hfill}\hskip 1.35cm\hbox to 4.95cm{\hfill $\Omega_{n=18}
+ \Omega_3 + \Omega_2$\hfill}
&& \hbox to 3.9cm{\hfill $(E_7)$\hfill}\cr
\+ \hbox to 2.3cm{\hfill $n = 30$\hfill}\hskip 1.35cm\hbox to 4.95cm{\hfill $
\Omega_{n=30} + \Omega_5 + \Omega_3 + \Omega_2$\hfill}
&& \hbox to 3.9cm{\hfill $(E_8)$\hfill}\cr}$$
In actuality, relaxing requirement (1) to
$N_{11}\geq 1$ does not enlarge this solution
set. Rather, the solutions in column two of Table B.1
are simply multiplied by an
overall constant, $N_{1 1}=c_n$. Our proof proceeds along the lines of
\pr{capelli87}:
\vhalf

Let
$\alpha (\delta)= {\rm ~GCF}(\delta,\bar \delta \equiv n/\delta)$ and
$N=2(K+2)$.
Next we define $p\equiv \bar \delta/\alpha$ and
$p'\equiv \delta/\alpha$. Then we choose a pair of
integers $\rho$, $\sigma$ such that $\rho p - \sigma p' = 1$. From these
we form $\omega(\delta) = \rho p + \sigma p' \pmod{N/\alpha^2}$,
which leads to the following equations:\footnote{Note for future reference that
interchanging the roles of $\delta$ and $\bar \delta$ in these equations
amounts to replacing $\omega$ by $-\omega$.}
$$\eqalign{\omega^2 -1 &= 4\rho\sigma p p' = 0 \pmod{2 N/\alpha^2};\cr
           \omega + 1  &= 2\rho p \pmod{N/\alpha^2};\cr
           \omega - 1  &= 2\sigma p' \pmod{N/\alpha^2}\,\, .\cr}
\eqno ({\rm B}.3)$$

An  $N\times N$-dimensional matrix $\Om$
operates on an enlarged set of
characters $\chi_{\lambda}$, with $\lambda$ defined mod $N$.
The ``additional'' characters carry indices in the range
$-(K+2)<\lambda \pmod{N}<0$ and
are trivially related to the customary $SU(2)_K$ characters,
carrying positive
indices $\pmod{N}$, by
$\chi_{\lambda}= -\chi_{-\lambda}$ and $\chi_{\xi (K+2)}=0$
for $\xi \in \Z$.\footnote{The
character $\chi_l$, for $0<l<K+1$, is associated here with
the primary field that is an $l$-dimensional representation
of $SU(2)$. This notation differs from that used
in the rest of the thesis.
Outside of this appendix,
the character corresponding to the primary field
in the $l$-dimensional representation is denoted by $\chi_{l-1}$.
In particular, while here the character for the singlet representation is
denoted by $\chi_1$, elsewhere it is denoted by $\chi_0$.}
This means the overall sign coefficient   in the $q$-expansion of
$(\chi_{\lambda})$ is positive for $0<\lambda \pmod{N} <n $ and negative
for  $n<\lambda \pmod{N} <N $.
The components of a matrix, $\Om$, are defined to be:
$$({\Om})_{\lambda,\lambda'}=\cases{
0,&if $\alpha\not\vert\lambda$ or $\alpha\not\vert\lambda'$;\cr
\sum_{\xi \pmod{\alpha}}\delta_{\lambda',\omega\lambda + \xi N/\alpha},
&otherwise.\cr}\eqno({\rm B}.4)$$
Thus, $\Omega_n$ is the $N\times N$-dimensional identity matrix.

A general MIPF for  $SU(2)$ at level $K$ can be written as
$$ Z(\tau,\, \bar\tau) = {1\over 2}\sum_{\lambda,\lambda'
{\pmod{N}}}\bar\chi_{\lambda}
(\bar \tau)(\sum_{\delta} c_{\delta}\Om)_{\lambda \lambda'} \chi_{\lambda'}
(\tau)~~.\eqno({\rm B}.5)$$
We divide the integers $\lambda\neq 0\pmod{N}$ into two disjoint sets $U$
and $L$ with $\lambda \in U$ if $1\leq\lambda\leq n-1$ and $\lambda\in L$
if $n+1\leq\lambda\leq 2n-1$. Therefore, $L\equiv -U\pmod{N}$
and we choose U as the
fundamental domain over which $\lambda$ is varied for $\chi_{\lambda}$.

The matrices $\Om$ have the following properties between their elements:
$$\eqalignno{(\Om)_{\lambda,\lambda'} &= (\Om)_{-\lambda,-\lambda'} &(B.6)\cr
(\Om\chi)_{\lambda} &= (\Omega_{n/\delta}\chi)_{-\lambda}\cr
                 &= -(\Omega_{n/\delta}\chi)_{\lambda}\,\, .&(B.7)\cr}$$
We use these relationships
to reexpress the  partition function $(B.5)$ as
$$\eqalignno{Z &= \sum_{\lambda\in U, \lambda' \pmod{N}}
\bar\chi_{\lambda}(\bar \tau)\sum_{\delta\mid n}(c_{\delta}\Om)_{\lambda,
\lambda'}
\chi_{\lambda'}(\tau)&(B.8)\cr
               &= \sum_{\lambda,\lambda'\in U}\bar\chi_{\lambda}(\bar \tau)
\bigl\{\sum_{\delta\mid
n}c_{\delta}[(\Om)_{\lambda,\lambda'}-(\Om)_{\lambda,-\lambda'}\bigr\}\chi_{\lambda'}(\tau)&(B.9)\cr
&= {1\over 4}\sum_{\lambda,\lambda'\in U,L}
\bar\chi_{\lambda}(\bar\tau)\bigl\{\sum_{(\delta,\bar \delta)
{\rm ~pairs},\,\delta\mid n} (c_{\delta} - c_{\bar
\delta})[(\Om)_{\lambda,\lambda'}
- (\Omega_{\bar
\delta})_{\lambda,\lambda'}]\bigr\}\chi_{\lambda'}(\tau)&(B.10)\cr}$$
with $c_{\delta}\geq 0$ and $c_{\delta}> 0$ implying $c_{\bar
\delta=n/\delta}=0$ by
convention. Two properties of these partition functions become apparent:
(i) either $\Omega_n$ or $\Omega_1$ contribute to the coefficient of
the vacuum state $\bar\chi_1 \chi_1$ but not both,
and (ii) the $\Om$  corresponding to
$\delta^2=n$ makes   no net contribution
to the partition function.\footnote{CIZ
shows (ii) from a different approach.}

The coefficient of $\bar\chi_1$ is (choosing $c_n\geq 1$ a
nd, therefore, $c_1=0$)
$$c_n\chi_1 + \sum_{\delta \not= n,1;~\alpha(\delta)=1}c_{\delta}
\chi_{\omega(\delta)}\eqno ({\rm B}.11)$$
(since $(\Om)_{\lambda,\lambda'}=0 {\rm ~~unless~~} \alpha\vert
\lambda^{(')}$). For $1<\delta<n$, $\omega(\delta)\in (\Z /N\Z)^*$
are all distinct from
$\pm 1$.\footnote{Here
({\tenBbb Z}$/N${\tenBbb Z})$^*\equiv$  integers (mod $N$) that are prime to
$N$. We also define
\hbox {$ U^* \equiv $({\tenBbb Z}$/N${\tenBbb Z})$^*\cap U$}
and\break
\hbox {$ L^* \equiv $({\tenBbb Z}$/N${\tenBbb Z})$^*\cap L$.}}
Additionally, $\omega(\delta)= -\omega(\delta')$
implies $\delta' = n/\delta$ for $\alpha(\delta')=\alpha(\delta)$.
This forces $\omega(\delta)$ to
 belong to $U$ for $c_{\delta}>0$ and
$\alpha(\delta)=1$. Otherwise, if $\omega(\delta)\in L$, then
$c_{\delta}\chi_{\omega(\delta)}= -c_{\delta}\chi_{-\omega(\delta)}$ would
contribute  negatively to the coefficient
of $\bar\chi_1$. This would require a positive
$c_{\delta'}\chi_{\omega'(\delta')}$ contribution from a $\delta'$,
such that $\alpha(\delta')=1$
and $\omega(\delta')= -\omega(\delta)$. But this implies that
$\delta'=n/\delta$. Hence,
$c_{\delta}$ and $c_{n/\delta}$ must both be positive definite, which
is excluded.

\hfill\break
\noindent {\it Lemma 1}:
\rm  (i) $\alpha_{\rm min}= 1$ or $2$, (ii) if
$\alpha_{\rm min}=2$, then the unique partition function (other than the
diagonal A-type) is
$$ \Omega_n + \Omega_2 {\rm ~with~} n= 0 \pmod{4}\,\, .\eqno ({\rm B}.12)$$

\vhalf
\hskip .6cm Proof:
$\alpha_{\rm min}$ is defined as the lowest $\alpha(\delta)$ of those
$\delta \not= n,1$ with $c_{\delta}>0$. The coefficient of
$\bar\chi_{\lambda=\alpha_{\rm min}}$ is
$$ c_n\chi_{\lambda'=\alpha_{\rm min}} + \sum_{\delta\neq
1;\alpha(\delta)=\alpha_{\rm min}} \sum_{\xi= 0}^{\alpha_{\rm
min}-1}c_{\delta}\chi_{\lambda'=\omega(\delta)
\alpha_{\rm min} + \xi N/\alpha(\delta)}.\eqno ({\rm B}.13)$$
For $\alpha(\delta)>1$ the
$\lambda' = \omega(\delta)\lambda + \xi N/\alpha(\delta)$ of $\Om$ in
eq.~(B.8) correspond to vertices of an $\alpha$-sided
polygon.  (Eq.~(B.8) is the form of the
partition function that we use from here on.
These $\lambda'$ can be viewed as points on a circle
of radius $N/2\pi$, where one half of the circle corresponds to the region $U$
and the opposite half to the region $L$. Any $\lambda'$ in $L$ must be
compensated by a vertex point
$-\lambda' \pmod{N}$ in $U$ either (1) from the same polygon, (2) a
different polygon, or (3) from the point corresponding to
$\chi_{\lambda'=\lambda}$ from $\Omega_n$.\newpage

\centertext{Figure B.1 The integers (mod $N$) mapped to a circle of radius
$N/2\pi$}
\vskip 9cm
Independent of the value of $c_n\geq 1$,
only the negative contribution to the coefficient of
$\bar\chi_{\lambda}$
from at most one
point, $\lambda'\in L$, on the $\alpha$-gon of a generic $\Om$
can be compensated
by the contribution from the single point $-\lambda'>0 \pmod{N}$
from $\Omega_n$.
If $\lambda'<0 \pmod{N}$ from $\Om$ must be cancelled by a
$\lambda'>0 \pmod{N}$ from
$\Omega_n$ then $c_n\geq c_{\delta}$.

Any $\Om$ with $\alpha\geq 4$ must have at least one point of its related
$\alpha$-gon in $L$. The case of only one point in $L$ corresponds to
$\alpha(\delta) = 4$ and a set of indices
$$ \lambda' = \omega\alpha + \xi N/\alpha\in 0,{n\over 2},n,{3n\over 2}
\pmod{N}\,\, .\eqno ({\rm B}.14)$$
This set of indices
only exists at $n=\alpha^2$, {\it i.e.,} $n=\delta^2$.
Therefore, the corresponding $\Om=\Omega_{\sqrt{n}}$ cannot
contribute to the partition function. Consider then partition functions
with $\alpha_{\rm min}\geq 4$.
Based on the above, the coefficient
of $\bar\chi_{\alpha_{\rm min}}$ in any of
these models will contain at least two different negative $\chi_{\lambda'\in
L}$ terms. Hence at least one of these terms must be
cancelled by methods (1) or (2).  First consider method (2).
Assume $\Omega_{\delta'}$ can compensate a negative term from $\Om$.
This requires that
$$\omega'(\delta')\alpha_{\rm min}\equiv -\omega(\delta)\alpha_{\rm min}
\pmod{N/\alpha_{\rm min}}.\eqno ({\rm B}.15)$$
Equivalently,
$\omega'(\delta')\equiv -\omega(\delta) \pmod{N/\alpha^2_{\rm min}}$,
which again implies $\delta\delta'=n$, in contradiction to
$c_{n/\delta}=0$
if
$c_{\delta}>0$. Hence method (2) cannot be used. Similarly,
cancelling the negative terms from $\Om$ with positive ones from the same
$\alpha$-gon of $\Om$ implies that $2\omega\equiv 0 \pmod{N/\alpha_{\rm min}}$,
which again is only possible for $n=\alpha_{\rm min}^2 = \alpha^2(\delta)$.
Thus,
neither methods (1) nor (2) can be used in the cancellation of negative terms
in partition functions. Thus,
MIPFs with $\alpha_{\rm min}\geq 4$ cannot exist for any value of $c_n$.

We now consider potential MIPFs with $\alpha_{\rm min}=3$. In this case
it is possible for just one point, $\lambda'$, from a $\Om$ to be in $L$.
This can be cancelled by a $\chi_3$ term
from $\Omega_n$. However, then $\omega\equiv -1 \pmod{N/9}$ and the terms
from $\Om$ would be $\chi_{-3} + \chi_{-3+N/3} + \chi_{-3-N/3}$, where
$\pm N/3-3\in U\cup \{0,n\}$. This implies $n\leq 9$ while
$9\vert n$. Hence, we have another case of $n=\alpha_{\rm min}^2$.
Therefore,  MIPFs cannot have $\alpha_{\rm min}>2$, independent of the
coefficient
$c_n$. The $2-$gon (line) resulting from a $\Om$ with $\alpha=2$ has as
its vertices the two points $2\omega$ and $2\omega + n$. The coefficient of
$\chi_2$ for a model with $\alpha_{\rm min}=2$ is
$$\chi_2 + \sum_{\alpha(\delta)=2} c_{\delta}(\chi_{\lambda'=2\omega}+
\chi_{\lambda'=2\omega+n}).\eqno ({\rm B}.16)$$
Three sets of values are possible for
$\lambda'=2\omega, 2\omega+n$. One corresponds to the excluded case
$n=\alpha^2_{\rm min}$, the second to another excluded by  contradiction
(requiring both $2\omega=0 \pmod{n}$ and $\omega^2 = 1 \pmod{n}$).
The remaining case corresponds to a unique $\delta$ with $\alpha(\delta)=2$.
Choosing $\omega\pmod{N/4}$) $2\omega=-2 \pmod{N}$,
$2\omega+n\in U$,
we find the negative term, $\chi_{2\omega}$, is compensated by
$\chi_2$ from $\Omega_n$. This requires $n\equiv 4m\in 4\Z$, and
$\omega(\delta)\equiv
-1$ mod $2m$, which is only satisfied by $\delta=2$. Since there is
one-to-one cancellation between $\chi_2$ and
$\chi_{2\omega}$, $c_n>c_2$ is also mandatory. If $c_n=c_2$ and all other
$c_{\delta}=0$ for even $n$, then the resulting partition function is
simply an integer multiple of the D-type solution. Further,
$c_n>c_2$, with all other $c_{\delta}=0$, can be expressed as
$(c_n - c_2)Z(A) + (c_2)Z(D)$.
Thus, for $\alpha_{\rm min}=2$, the freedom to have
$c_n\geq 1$ for even $n$ only increases the solution set of MIPFs beyond the
A-D-E
class if and only if (iff)
additional $\Om$'s (not in E class for $n=12,18,30$) with $\alpha(\delta)\geq
3$ can be included
with $c_n(\Omega_n +\Omega_2)$. We now show that is not possible to include
such terms and keep the partition function positive.

For $\lambda\in U$, $(\Omega_n + \Omega_2)\chi_{\lambda}$ equals
$\chi_{\lambda}$ if $\lambda$ is odd and $\chi_{n-\lambda}$ if $\lambda$
is even. That no other $\Om$'s are allowed in MIPFs
with $\alpha_{\rm min}=2$ is evident
by repetition of prior arguments after replacing
$\chi_{\lambda}$ with $\chi_{n-\lambda}$ if $\lambda$ is even.
The only remaining candidate for giving non-zero $c_{\delta}$ are
those $\Om$ corresponding to $\alpha$-gons with exactly 2 vertices in $L$. This
limits consideration to $\Om$ with $\alpha(\delta)= 3,4,5,6$. The negative
contributions from any such  $\Om$ must be cancelled by positive from
$\Omega_n$ and $(\Omega_n + \Omega_2)$.
First we consider the coefficients of
$\bar\chi_3$ ($\bar\chi_5$) coming
from a $\Om$ with $\alpha(\delta)= 3\,(5)$,
respectively. In these cases $(\Omega_n + \Omega_2)$ acts identically to
$\Omega_n$ since $\lambda$ is odd.  The prior arguments that eliminated
partition functions with $\alpha_{\rm min}= 3\, ,\, 5$ likewise exclude any
$\Om$ with $\alpha(\delta)=3\, ,\, 5$. The two  $\delta'\in L$ for
a $\Om$ with $\alpha(\delta)=4$ that contribute
negative coefficients $\chi_{\delta'\in L}$ to $\bar\chi_4$ can be
cancelled by a combination
of $\Omega_n$ and $(\Omega_n + \Omega_2)$ iff
$\delta'= N-4\, ,~ n+4$. However, this only occurs for $N=32$, {\it i.e.,}
$n=\alpha^2$. By similar argument the last possibility,
which is including a $\Om$ with $\alpha(\delta)=6$,
requires $N=36$, which by factorization of $n$ only allows $\alpha=1\, ,~
3$. Thus no additional $\Omega_{\delta\neq n,2}$ are allowed for
$\alpha_{\rm min}=2$ even when $c_n>1$.
So {\it lemma} 1 of
CIZ is independent of the restriction of $c_n = 1$.

\hfill\break
\noindent {\it Lemma 2}:
If $n$ is odd, then the unique possibility is $\Omega_n$.

\vhalf
\hskip .6cm Proof:
We show this by contradiction.  Assume, contrary to {\it lemma} 2, that
MIPFs can be formed from additional combinations of $\Om$'s.
Recall that
{\it lemma 1} requires that $\alpha_{\rm min}=1$ for odd $n$ and
consider specifically
the coefficient of $\bar\chi_{2^{\gamma}}$ for $2^{\gamma}<n$.
Since an off $n$ limits the $\Om$'s, that contribute terms to the coefficient
of $\bar\chi_{2^{\gamma}}$,
 to those with $\alpha(\delta)=1$, this coefficient is:
$$ c_n\chi_{2^{\gamma}} + \sum_{\alpha(\delta)=1}
c_{\delta}\chi_{\omega(\delta)
2^{\gamma}}~~.\eqno ({\rm B}.17)$$
By prior argument all of these $\omega(\delta)\in U$.
That is, $(0<\omega<n)$.
Consider the case of $2^{\gamma}\omega\in L$. The resulting
negative contribution to the coefficient would require
an additional $\omega'(\delta')$ from some $\Omega_{\delta'}$
(including the possibility $\omega'=1$) such that
$2^{\gamma}(\omega + \omega')\equiv 0 \pmod{N}$.
For $\gamma =1$, $2(\omega +\omega')\equiv 0 \pmod{N}$. However,
$\omega + \omega'<N$ implies
 $2(\omega +\omega')<2N$. This leads to
$2(\omega + \omega')= N$, {\ie} $\omega + \omega'=n$;
but $\omega$ and $\omega'$ are both odd by
$(\omega^{(')})^2\equiv 1 \pmod{2N}$ (from $\omega^2 - 1\equiv 0
\pmod{2N/\alpha^2}$). Thus $\omega +\omega'$ is even
while $n$ is by assumption odd.
To resolve  this potential contradiction requires
$2\omega \in U$ and
$\omega < n/2$.  This argument can be iterated to prove that any $\gamma$,
with the property that
$2^{\gamma} \in U$, requires $2^{\gamma}\omega\in U$ with
$0<\omega<n/2^{\gamma}$.

Now consider $\gamma_{\rm max}$ defined by
$2^{\gamma_{\rm max}}<n<2^{\gamma_{\rm max}+1}$.
Based on above arguments, we require that
$0<\omega<n/2^{\gamma_{\rm max}}$. From the defining eq.~of
$\omega(\delta)$ we can show that
$\omega(\delta)\bar\delta\equiv \bar\delta \pmod{N}$
for $\bar\delta=n/\delta$. Since $n$ has been chosen odd, $\delta>2$.
So $\bar\delta<2^{\gamma_{\rm max}}{1\over{\delta/2}}<2^{\gamma_{\rm max}}$
while
$\omega\bar\delta<{n\over{2^{\gamma}}}2^{\gamma_{\rm max}}$.
Thus $\omega\bar\delta\equiv\bar\delta \pmod{N}$ implies $\omega(\delta)=1$,
in contradiction to the assumption that $1<\delta<n$ (for which
$\vert\omega(\delta)\vert >1$). The only solution is that
$c_{\delta\neq n}=0$ in eq.~(B.17).
Thus, {\it lemma} 2 is also true independent of whether
$c_n=1$.

Since {\it lemmas} 1 and 2 are still valid for $c_n>1$, the only remaining
possibility for new types of MIPFs corresponds to $n$ even and
$\alpha_{\rm min}=1$ (the latter implying $\omega^2\equiv 1 \pmod{2N}$
and $\omega\in U$).
Henceforth
we assume these values for $n$ and $\alpha_{\rm min}$ and
consider the last {\it lemma} of CIZ.

\hfill\break
\noindent{\it Lemma 3}:
For $n$ even, $n\not=12,30$,
 $\omega\in U^*$, $\omega^2=1 \pmod{2N}$, $\omega\not=1$,
and $\omega\not= n-1$ if $n=2 \pmod{4}$, there exists
$\lambda\in U^*$ such that $\omega\lambda\in L^*$.
\vhalf

The $\{\delta,\bar\delta\}$ pairs excluded from the claims of {\it lemma 3}
by the conditions imposed within it are $\{n,1\}$ for any even $n$ since
$w(n)=1$, $w(1)= -1\not\in U^*$;
$\{2,n/2\}$ for $n=2 \pmod{4}$ where $\alpha=1$ and $\omega(2)= n-1$;
$\{3,4\}$ for $n=12$; $\{2,15\}$, $\{3,10\}$, $\{5,6\}$ for $n=30$.
As {\it lemma 3} in no way involves the coefficients $c_{\delta}$,
generalizing $c_n=1$ to
{\hbox{$c_n\geq 1$ clearly does not invalidate it.}}
So we assume {\it lemma} 3 and show that the conclusions based upon it
do not alter if $c_n>1$.

Consider the coefficient of $\bar\chi_{\lambda\in U^*}$ (for $n$ odd and
$\alpha_{\rm min}=1$).
Only $\Om$ with
$\alpha(\delta)=1$ can contribute and
the $\omega(\delta)$ corresponding to $c_{\delta}\not= 0$ must
have the property that
$\omega\lambda\in U$, for all $\lambda\in U$.
This can be shown by contradiction:  Assume instead
that $\omega\lambda\in L^*$. Cancellation of this term requires another
$\omega'(\delta')$ such that $\omega'\lambda\in U^*$ and
$(\omega' + \omega)\lambda \equiv 0 \pmod{N}$, which requires that
$\omega' + \omega \equiv 0 \pmod{N}$, since $\lambda$ is invertible
$\pmod{N}$. However,
$\omega'\equiv -\omega \pmod{N}$ implies $\delta'=n/\delta$ in
contradiction  to $c_{\delta}>0$ implying $c_{n/\delta}=0$.

Now apply {\it lemma} 3:
For $n\equiv 0 \pmod{4}$ and $n\not= 12$, all $\Om$,
with $\delta\not= n$ are $\alpha(\delta)=1$, are
excluded by  this {\it lemma} since no $\omega$ exists that
always giving $\omega\lambda\in U$ for
all $\lambda$.
For $n\equiv 2 \pmod{4}$, if $n\not= 30$, the only possible allowed $\Om$
with $\alpha(\delta)=1$ is $\Omega_2$ ($\omega=n-1$). From prior arguments
$c_2$ must be less than $c_n$. The matrix, $\bmit N$,
 for MIPFs with even $n$,
$\alpha_{\rm min}=1$ and $c_n$, $c_2\not=0$
can be expressed as:
$$ {\bmit N} = c^A_n\Omega_n + c^D_n(\Omega_n + \Omega_2)
+  [c^E_n(\Omega_n +\Omega_2) +      \sum_{{\rm odd~}\alpha(\delta)\geq
3}c_{\delta\not= n,2}\Omega_{\delta}]\,\, .\eqno ({\rm B}.18)$$
The first term on the RHS is just a multiple of the (diagonal) A-type
and the second term a multiple of the D-type.
In $(B.18)$, all $c_{\delta\not= n,2}$ for
$n\not= 12, 18, 30$ must vanish. The arguments demanding this parallel those
for {\it lemma 1}. Let $\hat\alpha_{\rm min}$ be the smallest  of odd
$\alpha(\delta)\geq 3$ associated
with $c_{\delta}>0$ and consider the coefficients of
$\bar\chi_{\hat\alpha_{\rm min}}$.\footnote{We consider
only odd $\alpha$ since $4{\not\vert}(n\equiv 2 \pmod{4})$.}
$c^E_n(\Omega_n+\Omega_2)_{\lambda,\lambda'}\chi_{\lambda'}$ contributes
the coefficients
$ c^D_n(\chi_{\hat\alpha_{\rm min}} + \chi_{n-\hat\alpha_{\rm min}})$
while $c^A\Omega_n$ contributes additional
$ c^A\chi_{\hat\alpha_{\rm min}}$.

Any $\lambda'\in L$ from an $\Om$ with $c_{\delta}>0$ and
$\alpha(\delta)=\hat\alpha_{\rm min}$ can only be compensated by the positive
$\lambda'= \hat\alpha_{\rm min}, n-\hat\alpha_{\rm min}$ coming from
$\Omega_{\delta=n}$ or $\Omega_{\delta=2}$.
Odd $\alpha$-gons with just one or two
vertices in $L$ are limited to $\alpha= 3, 5$. In the event
of two vertices in $L$, $\hat\alpha_{\rm min}=3$  requires
${2n\over 3}= (n+\hat\alpha_{\rm min}) - {n\over 2}$,
{\it i.e.,} $n=18$.
Thus $\delta=3$ and $c_{\delta=3}\leq c^E$. $c_3=c^E$ for $n=18$ forms the
standard $E_7$ invariant.\footnote{We let $c_3= c^E$ since
$c^E - c_3$ can be redefined as a contribution to $c^D$.}
By the same logic, $\hat\alpha_{\rm min}= 5$ requires
${4n\over 5}= (n-\hat\alpha_{\rm min}) - {n\over 2}$,
{\ie} $n= {50\over 3}$,
which is not allowed since $n\in \Z$).
The last possibility, that
of a 3-gon having only one vertex in $L$, that is
compensated
by either $\lambda'= 3$ from $\Omega_n$ or $\lambda'= n-3$ from $\Omega_2$
was shown not possible in discussion of {\it lemma} 1.

The only remaining cases not covered by any of the {\it lemmas} are
$n= 12, 30$ with $\alpha_{\rm min}=1$. It is straightforward to show that
in these cases too,
$c_n>1$ does not lead to new MIPFs, only to multiples of A, D, or E classes.
Therefore, we conclude that
relaxing the condition of uniqueness of the vacuum does not enlarge
the solution space of MIPFs beyond the A-D-E classification of Cappelli,
Itzykson, and Zuber.
Whether this rule can be applied to
MIPFs of other Ka\v c-Moody algebras, we do not know.
\hfill\vfill\eject

\chapternumstyle{blank}
\n {\bf\chapter{References:}}

\begin{putreferences}

\ref{abbott84}{L.F.~Abbott and M.B.~Wise, {\it Nucl.~Phys.~}{\bf B244}
(1984) 541.}

\ref{alvarez86}{L.~Alvarez-Gaum\' e, G.~Moore, and C.~Vafa,
{\it Comm.~Math.~Phys.~}{\bf 106} (1986) 1.}

\ref{antoniadis86} {I.~Antoniadis and C.~Bachas, {\it Nucl.~Phys.~}{\bf B278}
 (1986) 343;\\
M.~Hama, M.~Sawamura, and H.~Suzuki, RUP-92-1.}
\ref{li88} {K.~Li and N.~Warner, {\it Phys.~Lett.~}{\bf B211} (1988)
101;\\
A.~Bilal, {\it Phys.~Lett.~}{\bf B226} (1989) 272;\\
G.~Delius, preprint ITP-SB-89-12.}
\ref{antoniadis87}{I.~Antoniadis, C.~Bachas, and C.~Kounnas,
{\it Nucl.~Phys.~}{\bf B289} (1987) 87.}
\ref{antoniadis87b}{I.~Antoniadis, J.~Ellis, J.~Hagelin, and D.V.~Nanopoulos,
{\it Phys.~Lett.~}{\bf B149} (1987) 231.}
\ref{antoniadis88}{I.~Antoniadis and C.~Bachas, {\it Nucl.~Phys.~}{\bf B298}
(1988) 586.}

\ref{ardalan74}{F.~Ardalan and F.~Mansouri, {\it Phys.~Rev.~}{\bf D9} (1974)
3341; {\it Phys.~Rev.~Lett.~}{\bf 56} (1986) 2456;
{\it Phys.~Lett.~}{\bf B176} (1986) 99.}

\ref{argyres91a}{P.~Argyres, A.~LeClair, and S.-H.~Tye,
{\it Phys.~Lett.~}{\bf B235} (1991).}
\ref{argyres91b}{P.~Argyres and S.~-H.~Tye, {\it Phys.~Rev.~Lett.~}{\bf 67}
(1991) 3339.}
\ref{argyres91c}{P.~Argyres, J.~Grochocinski, and S.-H.~Tye, preprint
CLNS 91/1126.}
\ref{argyres91d}{P.~Argyres, K.~Dienes and S.-H.~Tye, preprints CLNS 91/1113;
McGill-91-37.}
\ref{argyres91e} {P.~Argyres, E.~Lyman, and S.-H.~Tye
preprint CLNS 91/1121.}
\ref{argyres91f}{P.~Argyres, J.~Grochocinski, and S.-H.~Tye,
{\it Nucl.~Phys.~}{\bf B367} (1991) 217.}
\ref{dienes92a}{K.~Dienes, Private communications.}
\ref{dienes92b}{K.~Dienes and S.~-H.~Tye, {\it Nucl.~Phys.~}{\bf B376} (1992)
297.}

\ref{athanasiu88}{G.~Athanasiu and J.~Atick, preprint IASSNS/HEP-88/46.}

\ref{atick88}{J.~Atick and E.~Witten, {\it Nucl.~Phys.~}{\bf B2 }
(1988) .}

\ref{axenides88}{M.~Axenides, S.~Ellis, and C.~Kounnas,
{\it Phys.~Rev.~}{\bf D37} (1988) 2964.}

\ref{bailin92}{D.~Bailin and A.~Love, {\it Phys.~Lett.} {\bf B292}
(1992) 315.}

\ref{barnsley88}{M.~Barnsley, {\underbar{Fractals Everywhere}} (Academic
Press, Boston, 1988).}

\ref{bouwknegt87}{P.~Bouwknegt and W.~Nahm,
{\it Phys.~Lett.~}{\bf B184} (1987) 359;\\
F.~Bais and P.~Bouwknegt, {\it Nucl.~Phys.~}{\bf B279} (1987) 561;\\
P.~Bouwknegt, Ph.D.~Thesis.}

\ref{bowick89}{M.~Bowick and S.~Giddings, {\it Nucl.~Phys.~}{\bf B325}
(1989) 631.}
\ref{bowick92}{M.~Bowick, SUHEP-4241-522 (1992).}
\ref{bowick93}{M.~Bowick, Private communications.}

\ref{brustein92}{R.~Brustein and P.~Steinhardt, preprint UPR-541T.}

\ref{capelli87} {A.~Cappelli, C.~Itzykson, and
J.~Zuber, {\it Nucl.~Phys.~}{\bf B280 [FS 18]} (1987) 445;
{\it Commun.~Math.~Phys.~}113 (1987) 1.}

\ref{carlitz}{R.~Carlitz, {\it Phys.~Rev.~}{\bf D5} (1972) 3231.}

\ref{candelas85}{P.~Candelas, G.~Horowitz, A.~Strominger, and E.~Witten,
{\it Nucl.~Phys.~}{\bf B258} (1985) 46.}

\ref{cateau92}{H.~Cateau and K.~Sumiyoshi,
{\it Phys.~Rev.~}{\bf D46} (1992) 2366.}

\ref{christe87}{P.~Christe, {\it Phys.~Lett.~}{\bf B188} (1987) 219;
{\it Phys.~Lett.~}{\bf B198} (1987) 215; Ph.D.~thesis (1986).}

\ref{clavelli90}{L.~Clavelli {\it et al.}, {\it Int.~J.~Mod.~Phys.~}{\bf A5}
(1990) 175.}

\ref{cleaver92a}{G.~Cleaver. {\it ``Comments on Fractional Superstrings,''}
To appear in the Proceedings of the International Workshop on String
Theory, Quantum Gravity and the Unification of Fundamental Interactions,
Rome, 1992.}
\ref{cleaver93a}{G.~Cleaver and D.~Lewellen, {\it Phys.~Rev.~}{\bf B300}
(1993) 354.}
\ref{cleaver93b}{G.~Cleaver and P.~Rosenthal, preprint CALT 68/1756.}
\ref{cleaver93c}{G.~Cleaver and P.~Rosenthal, preprint CALT 68/18__.}
\ref{cleaver}{G.~Cleaver, Unpublished research.}

\ref{cornwell89}{J.~F.~Cornwell, {\underbar{Group Theory in Physics}},
{\bf Vol. III}, (Academic Press, London, 1989).}

\ref{deo89a}{N.~Deo, S.~Jain, and C.~Tan, {\it Phys.~Lett.~}{\bf
B220} (1989) 125.}
\ref{deo89b}{N.~Deo, S.~Jain, and C.~Tan, {\it Phys.~Rev.~}{\bf D40}
(1989) 2626.}
\ref{deo92}{N.~Deo, S.~Jain, and C.~Tan, {\it Phys.~Rev.~}{\bf D45}
(1992) 3641.}
\ref{deo90a}{N.~Deo, S.~Jain, and C.-I.~Tan,
in {\underbar{Modern Quantum Field Theory}},
(World Scientific, Bombay, S.~Das {\it et al.} editors, 1990).}

\ref{distler90}{J.~Distler, Z.~Hlousek, and H.~Kawai,
{\it Int.~Jour.~Mod.~Phys.~}{\bf A5} (1990) 1093.}
\ref{distler93}{J.~Distler, private communication.}

\ref{dixon85}{L.~Dixon, J.~Harvey, C.~Vafa and E.~Witten,
 {\it Nucl.~Phys.~}{\bf B261} (1985) 651; {\bf B274} (1986) 285.}
\ref{dixon87}{L.~Dixon, V.~Kaplunovsky, and C.~Vafa,
{\it Nucl.~Phys.~}{\bf B294} (1987) 443.}

\ref{drees90}{W.~Drees, {\underbar{Beyond the Big Bang},}
(Open Court, La Salle, 1990).}

\ref{dreiner89a}{H.~Dreiner, J.~Lopez, D.V.~Nanopoulos, and
D.~Reiss, preprints MAD/TH/89-2; CTP-TAMU-06/89.}
\ref{dreiner89b}{H.~Dreiner, J.~Lopez, D.V.~Nanopoulos, and
D.~Reiss, {\it Phys.~Lett.~}{\bf B216} (1989) 283.}

\ref{ellis90}{J.~Ellis, J.~Lopez, and D.V.~Nanopoulos,
{\it Phys.~Lett.~}{\bf B245} (1990) 375.}

\ref{fernandez92}{R.~Fern\' andez, J.~Fr\" ohlich, and A.~Sokal,
{\underbar{Random Walks, Critical Phenomena, and Triviality in}}
{\underbar{Quantum Mechanics}}, (Springer-Verlag, 1992).}

\ref{font90}{A.~Font, L.~Ib\'a\~ nez, and F.~Quevedo,
{\it Nucl.~Phys.~}{\bf B345} (1990) 389.}

\ref{frampton88}{P.~Frampton and M.~Ubriaco, {\bf D38} (1988) 1341.}

\ref{francesco87}{P.~di Francesco, H.~Saleur, and J.B.~Zuber,
{\it Nucl.~Phys.~} {\bf B28 [FS19]} (1987) 454.}

\ref{frautschi71}{S.~Frautschi, {\it Phys.~Rev.~}{\bf D3} (1971) 2821.}

\ref{gannon92}{T.~Gannon, Carleton preprint 92-0407.}

\ref{gasperini91}{M.~Gasperini, N.~S\'anchez, and G.~Veneziano,
{\it Int.~Jour.~Mod.~Phys.~}{\bf A6} (1991) 3853;
{\it Nucl.~Phys.~}{\bf B364} (1991) 365.}

\ref{gepner87}{D.~Gepner and Z.~Qiu, {\it Nucl.~Phys.~}{\bf B285} (1987)
423.}
\ref{gepner87b}{D.~Gepner, {\it Phys.~Lett.~}{\bf B199} (1987) 380.}
\ref{gepner88a}{D.~Gepner, {\it Nucl.~Phys.~}{\bf B296} (1988) 757.}

\ref{ginsparg88}{P.~Ginsparg, {\it Nucl.~Phys.~}{\bf B295 [FS211]}
(1988) 153.}
\ref{ginsparg89}{P.~Ginsparg, in \underbar{Fields, Strings and Critical
Phenomena}, (Elsevier Science Publishers, E.~Br\' ezin and
J.~Zinn-Justin editors, 1989).}

\ref{gross84}{D.~Gross, {\it Phys.~Lett.~}{\bf B138} (1984) 185.}

\ref{green53} {H.~S.~Green, {\it Phys.~Rev.~}{\bf 90} (1953) 270.}

\ref{hagedorn68}{R.~Hagedorn, {\it Nuovo Cim.~}{\bf A56} (1968) 1027.}

\ref{kac80}{V.~Ka\v c, {\it Adv.~Math.~}{\bf 35} (1980) 264;\\
V.~Ka\v c and D.~Peterson, {\it Bull.~AMS} {\bf 3} (1980) 1057;
{\it Adv.~Math.~}{\bf 53} (1984) 125.}
\ref{kac83}{V.~Ka\v c, {\underbar{Infinite Dimensional Lie Algebras}},
(Birkh\" auser, Boston, 1983);\\
V.~Ka\v c editor, {\underbar{Infinite Dimensional Lie Algebras and Groups}},
(World Scientific, Singapore, 1989).}

\ref{kaku91}{M.~Kaku, \underbar{Strings, Conformal Fields and Topology},
(Springer-Verlag, New York, 1991).}

\ref{kawai87a} {H.~Kawai, D.~ Lewellen, and S.-H.~Tye,
{\it Nucl.~Phys.~}{\bf B288} (1987) 1.}
\ref{kawai87b} {H.~Kawai, D.~Lewellen, J.A.~Schwartz,
and S.-H.~Tye, {\it Nucl.~Phys.~}{\bf B299} (1988) 431.}

\ref{kazakov85}{V.~Kazakov, I.~Kostov, and A.~Migdal,
{\it Phys.~Lett.~}{\bf B157} (1985) 295.}

\ref{khuri92}{R.~Khuri, CTP/TAMU-80/1992; CTP/TAMU-10/1993.}

\ref{kikkawa84}{K.~Kikkawa and M.~Yamasaki, {\it Phys.~Lett.~}{\bfB149}
(1984) 357.}

\ref{kiritsis88}{E.B.~Kiritsis, {\it Phys.~Lett.~}{\bf B217} (1988) 427.}

\ref{langacker92}{P.~Langacker, preprint UPR-0512-T (1992).}

\ref{leblanc88}{Y.~Leblanc, {\it Phys.~Rev.}{\bf D38} (1988) 38.}

\ref{lewellen87}{H.~Kawai, D.~Lewellen, and S.-H.`Tye,
{\it Nucl.~Phys.~}{\bf B288} (1987) 1.}
\ref{lewellen}{D.~C.~Lewellen, {\it Nucl.~Phys.~}{\bf B337} (1990) 61.}

\ref{lizzi90}{F.~Lizzi and I.~Senda, {\it Phys.~Lett.~}{\bf B244}
(1990) 27.}
\ref{lizzi91}{F.~Lizzi and I.~Senda, {\it Nucl.~Phys.~}{\bf B359}
(1991) 441.}

\ref{lust89}{D.~L\" ust and S.~Theisen,
{\underbar{Lectures on String Theory,}} (Springer-Verlag, Berlin, 1989).}

\ref{maggiore93}{M.~Maggiore, preprint IFUP-TH 3/93.}

\ref{mansouri87} {F.~Mansouri and X.~Wu, {\it Mod.~Phys.~Lett.~}{\bf A2}
(1987) 215; {\it Phys.~Lett.~}{\bf B203} (1988) 417;
{\it J.~Math.~Phys.~}{\bf 30} (1989) 892;\\
A. Bhattacharyya {\it et al.,} {\it Mod.~Phys.~Lett.~}{\bf A4} (1989)
1121; {\it Phys.~Lett.~}{\bf B224} (1989) 384.}

\ref{narain86} {K.~S.~Narain, {\it Phys.~Lett.~}{\bf B169} (1986) 41.}
\ref{narain87} {K.~S.~Narain, M.H.~Sarmadi, and C.~Vafa,
{\it Nucl.~Phys.~}{\bf B288} (1987) 551.}

\ref{obrien87}{K.~O'Brien and C.~Tan, {\it Phys.~Rev.~}{\bf D36} (1987)
1184.}

\ref{parisi79}{G.~Parisi, {\it Phys.~Lett.~}{\bf B81} (1979) 357.}

\ref{polchinski88}{J.~Polchinski, {\it Phys.~Lett.~}{\bf B209} (1988)
252.}
\ref{polchinski93}{J.~Polchinski, Private communications.}

\ref{pope92}{C.~Pope, preprint CTP TAMU-30/92  (1992).}

\ref{raiten91}{E.~Raiten, Thesis, (1991).}

\ref{roberts92}{P.~Roberts and H.~Terao, {\it Int.~J.~Mod.~Phys.~}{\bf A7}
(1992) 2207;\\
P.~Roberts, {\it Phys.~Lett.~}{\bf B244} (1990) 429.}

\ref{sakai86}{N.~Sakaii and I.~Senda, {\it Prog.~Theo.~Phys.~}
{\bf 75} (1986) 692.}

\ref{salomonson86}{P.~Salomonson and B.-S.~Skagerstam, {\it
Nucl.~Phys.~}{\bf B268} (1986) 349.}

\ref{schellekens89} {A.~N.~Schellekens and S.~Yankielowicz,
{\it Nucl.~Phys.~}{\bf B327} (1989) 3;\\
A.~N.~Schellekens, {\it Phys.~Lett.~}{\bf 244} (1990) 255;\\
B.~Gato-Rivera and A.~N.~Schellekens, {\it Nucl.~Phys.~}{\bf B353} (1991)
519; {\it Commun.~Math.}
{\it Phys.~}145 (1992) 85.}
\ref{schellekens89b}{B.~Schellekens, ed. \underbar{Superstring Construction},
 (North-Holland Physics, Amsterdam, 1989).}
\ref{schellekens89c}{B.~Schellekens, CERN-TH-5515/89.}

\ref{schwarz87}{M.~Green, J.~Schwarz, and E.~Witten,
\underbar{Superstring Theory}, {\bf Vols. I \& II},
(Cambridge University Press, New York, 1987).}

\ref{turok87a}{D.~Mitchell and N.~Turok, {\it Nucl.~Phys.~}{\bf B294}
(1987) 1138.}
\ref{turok87b}{N.~Turok, Fermilab 87/215-A (1987).}

\ref{verlinde88}{E.~Verlinde, {\it Nucl.~Phys.~}{\bf B300}
(1988) 360.}

\ref{warner90}{N.~Warner, {\it Commu.~Math.~Phys.~}{\bf 130} (1990) 205.}

\ref{wilczek90} {F.~Wilczek, ed. \underbar {Fractional Statistics and Anyon
Superconductivity}, (World Scientific, Singaore, 1990) 11-16.}

\ref{witten92}{E.~Witten, preprint IASSNS-HEP-93-3.}

\ref{vafa1}{R.~Brandenberger and C.~Vafa, {\it Nucl.~Phys.}
{\bf B316} (1989) 391.}
\ref{vafa2}{A.A.~Tseytlin and C.~Vafa, {\it Nucl.~Phys.}
{\bf B372} (1992) 443.}

\ref{zamol87}{A.~Zamolodchikov and V.~Fateev, {\it Sov.~Phys.~}JETP
{\bf 62} (1985)  215; {\it Teor.~}{\it Mat.}
{\it Phys.~}{\bf 71} (1987) 163.}

\end{putreferences}


\bye